\documentclass[aps,twocolumn,showkeys,showpacs,preprintnumbers,prd,superscriptaddress,nofootinbib,10pt]{revtex4-1}
\usepackage{graphicx,epsf,bm,amsmath,amsfonts,amssymb,epstopdf,natbib,hyperref,xcolor,appendix,verbatim,multirow,diagbox,booktabs,siunitx}
\usepackage[normalem]{ulem}

\hypersetup{colorlinks=true,urlcolor=blue,citecolor=blue,linkcolor=blue,menucolor=blue,anchorcolor=blue,filecolor=blue}
\renewcommand{\arraystretch}{1.5}
\newcolumntype{?}{!{\vrule width 3pt}}
\newcommand{\thickhline}{\noalign{\hrule height 3pt}}

\date{\today}

\makeatletter
\newcommand{\fmarki}{\ensuremath{\alpha}}
\newcommand{\fmarkii}{\ensuremath{\beta}}
\newcommand{\fmarkiii}{\ensuremath{\gamma}}
\newcommand{\fmarkiv}{\ensuremath{\delta}}
\newcommand{\fmarkv}{\ensuremath{\epsilon}}
                
\def\@fnsymbol#1{{\ifcase#1\or \fmarki\or \fmarkii\or \fmarkiii\or \fmarkiv\or \fmarkv\or \else\@ctrerr\fi}}
\makeatother
\definecolor{darkred}{rgb}{0.8, 0.0, 0.0}
\definecolor{darkgreen}{rgb}{0, 0.8, 0.0}

\definecolor{valecol}{rgb}{0,0.5, 1.}

\begin{document}

\title{BAO miscalibration cannot rescue late-time solutions to the Hubble tension}

\author{Davide Pedrotti}
\email{davide.pedrotti-1@unitn.it}
\affiliation{Department of Physics, University of Trento, Via Sommarive 14, 38123 Povo (TN), Italy}
\affiliation{Trento Institute for Fundamental Physics and Applications (TIFPA)-INFN, Via Sommarive 14, 38123 Povo (TN), Italy}

\author{Luis A. Escamilla}
\email{luis.escamilla@icf.unam.mx}
\affiliation{Department of Physics, Istanbul Technical University, 34469 Maslak, Istanbul, Turkey \looseness=-1}
\affiliation{School of Mathematics and Statistics, University of Sheffield, Hounsfield Road, Sheffield S3 7RH, United Kingdom \looseness=-1}

\author{Valerio Marra}
\email{valerio.marra@me.com}
\affiliation{Departamento de F\'{i}sica, Universidade Federal do Esp\'{i}rito Santo, 29075-910, Vit\'{o}ria, ES, Brazil}
\affiliation{INAF -- Osservatorio Astronomico di Trieste, Via Tiepolo 11, 34131 Trieste (TS), Italy \looseness=-1}
\affiliation{IFPU -- Institute for Fundamental Physics of the Universe, Via Beirut 2, 34151 Trieste (TS), Italy \looseness=-1}

\author{Leandros Perivolaropoulos}
\email{leandros@uoi.gr}
\affiliation{Department of Physics, University of Ioannina, 45110 Ioannina, Greece \looseness=-1}

\author{Sunny Vagnozzi}
\email{sunny.vagnozzi@unitn.it}
\affiliation{Department of Physics, University of Trento, Via Sommarive 14, 38123 Povo (TN), Italy}
\affiliation{Trento Institute for Fundamental Physics and Applications (TIFPA)-INFN, Via Sommarive 14, 38123 Povo (TN), Italy}

\begin{abstract}
\noindent Baryon Acoustic Oscillation (BAO) measurements play a key role in ruling out post-recombination solutions to the Hubble tension. However, because the data compression leading to these measurements assumes a fiducial $\Lambda$CDM cosmology, their reliability in testing late-time modifications to $\Lambda$CDM has at times been called into question. We play devil's advocate and posit that fiducial cosmology assumptions do indeed affect BAO measurements in such a way that low-redshift acoustic angular scales (proportional to the Hubble constant $H_0$) are biased low, and test whether such a rescaling can rescue post-recombination solutions. The answer is no. Firstly, strong constraints on the shape of the $z \lesssim 2$ expansion history from unanchored Type Ia Supernovae (SNeIa) prevent large deviations from $\Lambda$CDM. In addition, unless $\Omega_m$ is significantly lower than $0.3$, the rescaled BAO measurements would be in strong tension with geometrical information from the Cosmic Microwave Background. We demonstrate this explicitly on several dark energy (DE) models ($w$CDM, CPL DE, phenomenologically emergent DE, holographic DE, $\Lambda_s$CDM, and the negative cosmological constant model), finding that none can address the Hubble tension once unanchored SNeIa are included. We argue that the $\Lambda_s$CDM sign-switching cosmological constant model possesses interesting features which make it the least unpromising one among those tested. Our results demonstrate that possible fiducial cosmology-induced BAO biases cannot be invoked as loopholes to the Hubble tension ``no-go theorem'', and highlight the extremely important but so far underappreciated role of unanchored SNeIa in ruling out post-recombination solutions.
\end{abstract}

\maketitle

\section{Introduction}
\label{sec:introduction}

Over the past three decades the $\Lambda$CDM model, the prevailing standard model of cosmology, has evolved hand-in-hand with the increasing wealth of high-precision observational data~\cite{Scott:2018adl}. Key pillars in this sense include anisotropies in the temperature and polarization of the Cosmic Microwave Background (CMB). Another arguably equally important probe is the Baryon Acoustic Oscillation (BAO) feature in the clustering of tracers of the Large-Scale Structure (LSS), such as galaxies. While these and several other probes initially cemented the $\Lambda$CDM paradigm, later data collected over the past decade have begun exposing potential cracks, in the form of cosmological tensions. The best known among these discrepancies, the Hubble tension, concerns the Hubble constant $H_0$, and is possibly the most convincing evidence for new physics beyond $\Lambda$CDM (see e.g.\ Refs.~\cite{Verde:2019ivm,DiValentino:2021izs,Perivolaropoulos:2021jda,Schoneberg:2021qvd,Shah:2021onj,Abdalla:2022yfr,DiValentino:2022fjm,Hu:2023jqc,Verde:2023lmm,CosmoVerseNetwork:2025alb} for reviews, and Refs.~\cite{Anchordoqui:2015lqa,DiValentino:2016hlg,Karwal:2016vyq,Benetti:2017juy,Mortsell:2018mfj,Kumar:2018yhh,Guo:2018ans,Graef:2018fzu,Poulin:2018cxd,Agrawal:2019lmo,Yang:2019nhz,Escudero:2019gvw,Niedermann:2019olb,Sakstein:2019fmf,Ballesteros:2020sik,Jedamzik:2020krr,Ballardini:2020iws,Niedermann:2020dwg,Gonzalez:2020fdy,Sekiguchi:2020teg,Braglia:2020auw,Karwal:2021vpk,Cyr-Racine:2021oal,Niedermann:2021ijp,Saridakis:2021xqy,Herold:2021ksg,Heisenberg:2022lob,Heisenberg:2022gqk,Sharma:2022ifr,Ren:2022aeo,Nojiri:2022ski,Schoneberg:2022grr,Reeves:2022aoi,Joseph:2022jsf,Gomez-Valent:2022bku,Moshafi:2022mva,Rezazadeh:2022lsf,Banerjee:2022ynv,Alvarez:2022wef,Ge:2022qws,Gangopadhyay:2022bsh,Schiavone:2022wvq,Brinckmann:2022ajr,Khodadi:2023ezj,Ben-Dayan:2023rgt,deCruzPerez:2023wzd,Ballardini:2023mzm,Poulin:2023lkg,Gangopadhyay:2023nli,SolaPeracaula:2023swx,Gomez-Valent:2023hov,Odintsov:2023cli,Greene:2023cro,Frion:2023xwq,Petronikolou:2023cwu,Sharma:2023kzr,Ramadan:2023ivw,Ben-Dayan:2023htq,Fu:2023tfo,Efstathiou:2023fbn,Montani:2023ywn,Lazkoz:2023oqc,Khalife:2023qbu,Jusufi:2024ifp,Erdem:2024vsr,Moshafi:2024guo,Capozziello:2024stm,Co:2024oek,Nozari:2024wir,Montani:2024xys,Escamilla:2024xmz,Chatrchyan:2024xjj,deJesus:2024zny,Hu:2024big,Jiang:2024nha,Simon:2024jmu,Toda:2024uff,DeSimone:2024lvy,Montani:2024ntj,Ye:2024zpk,Fikri:2024klc,Odintsov:2024woi,Teixeira:2024qmw,Li:2025owk,Tsilioukas:2025dmy,Jiang:2025ylr,Pang:2025jtk,Stahl:2025mdu,Jiang:2025hco,Luciano:2025fox,Lee:2025yah,Poulin:2025nfb,Boiza:2025xpn,Carloni:2025jlk,Montani:2025jkk,Simpson:2025kfn,Adhikary:2025khr,Benetti:2025bsl,Hu:2025fsz,Wang:2025dzn,Erdem:2025xtr,Adi:2025hyj,Briscese:2025kiz,Sohail:2025mma,Efstratiou:2025iqi} for examples of proposed solutions based on new physics beyond $\Lambda$CDM, with no claims as to completeness).

Constraints on the BAO feature can be obtained through tests such as the Alcock-Paczynski (AP) one. This exploits the fact that the assumption of an incorrect background cosmology distorts radial and angular scales in a different way when redshifts are converted into distances. The resulting BAO measurements play a key role in the Hubble tension discourse, particularly when it comes to arbitrating possible solutions thereto. These measurements are sensitive to the angular size of the sound horizon at baryon drag, $\theta_d \propto r_d/D$, where $r_d$ is the sound horizon at baryon drag, and $D$ an appropriate type of cosmological distance (e.g.\ $D_V$, $D_H$, or $D_A$). Since distances scale as $1/H_0$, the net sensitivity is to the product $r_dH_0$, and it is this $r_d$-$H_0$ degeneracy which makes BAO a key player in the Hubble tension. It is in fact clear that a solution to the Hubble tension, i.e.\ one which raises $H_0$, necessarily needs to lower $r_d$. This, in turn, calls for new physics operating prior to recombination, since the integral determining $r_d$ is only sensitive to the pre-recombination epoch~\cite{Bernal:2016gxb,Addison:2017fdm,Lemos:2018smw,Aylor:2018drw,Schoneberg:2019wmt,Knox:2019rjx,Arendse:2019hev,Efstathiou:2021ocp,Cai:2021weh,Keeley:2022ojz}. More generally, once $r_d$ is externally calibrated to the $\Lambda$CDM value $r_d \sim 147\,{\text{Mpc}}$, e.g.\ through a Big Bang Nucleosynthesis (BBN)-informed prior on the physical baryon density $\omega_b$, BAO can be combined with unanchored Type Ia Supernovae (SNeIa) to build an inverse distance ladder, from which a ``low'' value of $H_0 \sim 68\,{\text{km}}/{\text{s}}/{\text{Mpc}}$ is inferred, in agreement with the $\Lambda$CDM-based CMB value, but potentially without adopting any CMB data. This makes it very clear that the ``no-go theorem'' precluding a post-recombination solution to the Hubble tension hinges in a crucial manner on BAO measurements.~\footnote{The term ``no-go theorem'' here is of course not meant in the strict mathematical sense, but rather as a data-driven physics result. With this in mind, we nevertheless choose to adopt this terminology throughout the work to reflect its importance in the Hubble tension literature as excluding post-recombination solutions, while retaining the quotation marks to stress that it is not a strict mathematical theorem.}

A possible objection to the above arguments, and therefore to the ``no-go theorem'', is related to the role of fiducial cosmology assumptions in the process of extracting BAO measurements from raw data. This data, in practice, consists of maps of LSS tracers given in terms of two angular components (typically the right ascension and declination angles) and redshifts. The standard pipeline which goes from these maps to the actual BAO measurements, i.e.\ the BAO scaling parameters $\alpha_{\perp,\parallel}$ against which one ultimately fits theoretical model predictions, requires assuming fiducial cosmologies within at least four different steps:
\begin{enumerate}
\item transforming redshifts to radial distances in order to construct the map in comoving coordinates, from which summary statistics such as $n$-point correlators can be computed;
\item constructing a template for the BAO oscillations in these summary statistics, relative to which the BAO features observed in the data are extracted/quantified (in passing, it is worth noting that most state-of-the-art BAO pipelines are template-based);
\item transforming redshifts to radial distances in the mock catalogs used to estimate the covariance matrix of the data;
\item performing reconstruction to sharpen the BAO signature and mitigate non-linear effects in order to improve the accuracy of cosmological inference: this requires explicit knowledge of the logarithmic growth rate parameter.
\end{enumerate}
While in principle the above four fiducial cosmological models (which we could refer to as \textit{grid cosmology}, \textit{template cosmology}, \textit{covariance cosmology}, and \textit{reconstruction cosmology}) need not be the same, in practice they are usually taken to be one and the same, and we shall generically refer to it as \textit{fiducial cosmology} in what follows. It is worth stressing that the choice of fiducial cosmology does not merely amount to the choice of the values of parameters within a given model (e.g.\ the values of $H_0$ and $\Omega_m$ within $\Lambda$CDM), but crucially involves the choice of cosmological model itself (e.g.\ $w$CDM or CPL dynamical dark energy as opposed to $\Lambda$CDM).

The choice of fiducial cosmology typically falls on the $\Lambda$CDM model, which has itself raised questions on the potential circularity of BAO analyses: could the initial assumption of a fiducial $\Lambda$CDM cosmology lead to biases in the extracted $\alpha$s, which might therefore inadvertently reinforce the fiducial cosmology? A few works have looked at whether this model-dependence is truly under control (see e.g.\ Refs.~\cite{Xu:2012hg,BOSS:2016sne,Sherwin:2018wbu,Carter:2019ulk,Heinesen:2019phg,Bernal:2020vbb,eBOSS:2020uxp,Pan:2023zgb,Sanz-Wuhl:2024uvi,DESI:2024ude,Nadal-Matosas:2024dun,DESI:2025qqy}), with mixed conclusions. For instance, Ref.~\cite{Sanz-Wuhl:2024uvi} finds that for surveys such as DESI and Euclid, appreciable systematic shifts in the $\alpha$s can be introduced if the grid cosmology differs significantly from the true cosmology, because the assumption of the distance-redshift relation between true and fiducial cosmology matching a linear scaling breaks down (see also Ref.~\cite{Gsponer:2025sox} for a recent study focusing on full-shape measurements, which under certain conditions findings comparatively large shifts in the inferred cosmological parameters). Moreover, Ref.~\cite{Heinesen:2019phg} finds that in models which deviate significantly from $\Lambda$CDM at late times, particularly in the presence of large metric gradients, the AP scaling underlying standard BAO measurements breaks down. It was in fact suggested that these measurements should not be used to constrain cosmologies which deviate significantly from $\Lambda$CDM at late times~\cite{Heinesen:2019phg}.~\footnote{More specifically, Ref.~\cite{Heinesen:2019phg} argues that the two key assumptions which underlie the AP test can break down in cosmologies where the late-time distance-redshift relation differs more than a few percent from $\Lambda$CDM. These two assumptions are that \textit{a)} differences in the distance-redshift relationship between true and fiducial cosmology obey a linear scaling, and \textit{b)} differences in comoving clustering between true and fiducial cosmology can be nulled by the same free parameters which null the non-BAO signal in the galaxy correlation function or power spectrum.}

These concerns constitute grounds for caution when interpreting conclusions on beyond-$\Lambda$CDM models based on BAO data. These concerns have been invoked, albeit somewhat hand-wavingly, as a potential loophole in the ``no-go theorem'' precluding post-recombination solutions to the Hubble tension. In other words, the following question remains completely open:\\ \textit{Should one assume from the start a fiducial cosmology \textbf{vastly} different from} $\mathit{\Lambda}$\textit{CDM, would the recovered $\alpha$s and associated cosmological inference be strongly affected?} \\ To the best of our knowledge, such an analysis has never been performed in full generality. In fact, it is worth noting that the aforementioned studies~\cite{Xu:2012hg,BOSS:2016sne,Sherwin:2018wbu,Carter:2019ulk,Heinesen:2019phg,Bernal:2020vbb,eBOSS:2020uxp,Pan:2023zgb,Sanz-Wuhl:2024uvi,DESI:2024ude,Nadal-Matosas:2024dun,DESI:2025qqy} have only tested the stability of the BAO pipeline against the assumption of individual fiducial cosmological models, in all cases considering small ``perturbations'' away from the best-fit $\Lambda$CDM model. For obvious reasons of practicality, no attempt has been made to cover the much larger space of general fiducial cosmologies potentially deviating significantly from $\Lambda$CDM in the post-recombination Universe. This self-consistency problem was first strongly highlighted in Ref.~\cite{Anselmi:2022exn}, which showed that standard BAO analyses may rely on ``untested extrapolations'' to explore the parameter space far from their fiducial flat $\Lambda$CDM assumption. The same work cautions that these extrapolations may not be entirely self-consistent, recommending that traditional BAO methodologies be validated across the full range of models and parameters over which their results are quoted~\cite{Anselmi:2022exn}. Therefore, the question of whether or not the assumption of a fiducial cosmology in BAO analyses constitutes a loophole to the Hubble tension ``no-go theorem'', remains an open one to which a definitive answer (or an explicit counterexample) is yet to be provided.

In this work, we take a completely different approach to this issue: we shall not attempt to provide an answer to the previous question, but instead choose to play devil's advocate. We shall assume, for the sake of argument, that the adoption of a fiducial $\Lambda$CDM cosmology in BAO analyses does indeed result in a misdetermination of the $\alpha$s, in such a way that the inferred low-redshift acoustic angular scale is biased low. In other words, that the ``true'' $\theta_d$ (itself related to $\alpha_{\perp,\parallel}$) is larger than what is inferred from the standard BAO pipeline. Recall in fact that $\theta_d \propto r_dH_0$: hence, with $r_d$ calibrated, a determination of $\theta_d$ which is biased low would lead to $H_0$ being biased low as well. For concreteness and simplicity, we assume that current BAO measurements are subject to a redshift-independent bias which, if accounted for, should shift the inferred value of $H_0$ towards the local SH0ES value. We then ask ourselves whether such a bias would be sufficient to rescue a number of popular, existing post-recombination proposals for addressing the Hubble tension. In short, even under our extreme and rather generous assumptions, we find the answer to be negative, because of two important effects:
\begin{enumerate}
\item unanchored SNeIa tightly constrain the shape of the late-time expansion history, preventing it from deviating significantly from $\Lambda$CDM (a conclusion in qualitative agreement with the one recently drawn in Ref.~\cite{Zhou:2025kws});
\item a severe tension would be introduced between BAO measurements and geometrical CMB information, unless $\Omega_m$ is significantly (and implausibly) lower than $0.3$.
\end{enumerate}
While the impact on BAO analyses of assuming a fiducial cosmology radically different from the best-fit $\Lambda$CDM one remains a valid and interesting question, a potential BAO miscalibration is not sufficient to constitute a loophole to the ``no-go theorem'' (with the only possible exception being that of a highly contrived and fine-tuned hypothetical scenario we will discuss in Sec.~\ref{sec:discussion}, involving multiple systematics in BAO and unanchored SNeIa, as well as a non-$\Lambda$CDM late-time expansion history). Our results therefore further strengthen the case for pre-recombination new physics.

The rest of this paper is then organized as follows. In Sec.~\ref{sec:bao} we review the role of BAO in arbitrating possible solutions to the Hubble tension, and the rescaling/miscalibration we invoke to play devil's advocate, whereas in Sec.~\ref{sec:models} we introduce the post-recombination new physics models against which we test the possible loophole in the ``no-go theorem''. In Sec.~\ref{sec:datasets} we discuss the datasets and methodology adopted to obtain our results, which are instead discussed in Sec.~\ref{sec:results}. A critical discussion of our findings and their implications for the Hubble tension is carried out in Sec.~\ref{sec:discussion}. Finally, we draw concluding remarks in Sec.~\ref{sec:conclusions}.

\section{BAO and the Hubble tension}
\label{sec:bao}

We now review the key role of BAO measurements in the Hubble tension. To set the stage, in what follows we work within a homogeneous and isotropic cosmology described by the spatially flat Friedmann-Lema\^{i}tre-Robertson-Walker (FLRW) metric. Working in units where $c=1$, comoving angular diameter distances (also referred to as transverse comoving distances) are given by the following expression:
\begin{equation}
D_M(z)=\frac{1}{H_0}\int_0^z\,\frac{dz'}{E(z')}\,,
\label{eq:dm}
\end{equation}
where $H_0$ is the Hubble constant and $E(z) \equiv H(z)/H_0$ is the unnormalized expansion rate.

BAO observations from the clustering of tracers of the large-scale structure such as galaxies, quasars, or the Lyman-$\alpha$ forest carry the imprint of the sound horizon at baryon drag $r_d$, which plays the role of standard ruler, and is determined by the following integral:
\begin{eqnarray}
r_d = \int_{z_d}^{\infty}dz\,\frac{c_s(z)}{H(z)}\,,
\label{eq:rd}
\end{eqnarray}
with $c_s(z)$ being the sound speed of the photon-baryon plasma, and $z_d \approx 1060$ the redshift of the drag epoch. However, this standard ruler is not observed directly: only its angular extent (transverse to the line-of-sight), redshift extent (along the line-of-sight), or a volume-averaged combination of the two are observed. Transverse, line-of-sight, and isotropic (volume-averaged) BAO measurements at an effective redshift $z_{\text{eff}}$ are then sensitive to the transverse angular scale $\theta_d$, redshift span $\delta z_d$, and isotropic angular scale $\theta_v$, given by the following:
\begin{align}
\label{eq:thetad}
\theta_d(z_{\text{eff}}) &= \frac{r_d}{D_M(z_{\text{eff}})}=\frac{r_dH_0}{\displaystyle \int_0^{z_{\text{eff}}}\frac{dz'}{E(z')}}\,, \\
\label{eq:zd}
\delta z_d(z_{\text{eff}}) &= \frac{r_d}{D_H(z_{\text{eff}})} \equiv r_d H(z_{\text{eff}})= r_dH_0 E(z_{\text{eff}})\,,\\
\label{eq:thetav}
\theta_v(z_{\text{eff}}) &= \frac{r_d}{D_V(z_{\text{eff}})} \equiv \frac{r_d}{ \left [ z_{\text{eff}} D_M^2(z_{\text{eff}}) D_H(z_{\text{eff}}) \right ] ^{1/3}}\,,
\end{align}
making it clear that BAO measurements are directly sensitive to the product $r_dH_0$, which controls the apparent extent of the BAO feature. Measuring this feature across a sufficiently wide range of effective redshifts (at present $0.1 \lesssim z_{\text{eff}} \lesssim 2.5$) can (at least partially) disentangle the effects of $r_dH_0$ and $E(z)$. The reason is that the slope describing the $r_dH_0$-$\Omega_m$ correlation slowly changes with $z_{\text{eff}}$ (see e.g.\ Fig.~2 of Ref.~\cite{Lin:2021sfs}), reflecting how the importance of the dark energy (DE) contribution relative to the matter one decreases as $z_{\text{eff}}$ increases. In $\Lambda$CDM, $E(z)$ depends exclusively on $\Omega_m$, and therefore BAO measurements across a wide redshift range can constrain $\Omega_m$ (although not very tightly, unless geometrical CMB information is added, as we will see later, see also Ref.~\cite{Lin:2021sfs}); on the other hand, in models introducing late-time new physics $E(z)$ will generally depend on other parameters as well (typically those related to the DE sector).

To extract $H_0$ from BAO measurements, an external determination/calibration of the sound horizon $r_d$ is clearly required, which in turn depends on the physical baryon density $\omega_b$. Once the physical radiation density parameter $\omega_r$ is fixed from extremely precise measurements of the CMB temperature $T_{\text{CMB}}$, information on $\omega_b$ can be supplied, for instance, by BBN considerations on the primordial abundances of light elements. This explains why the BAO+BBN combination is particularly powerful and important in constraining $H_0$, despite none of the two being directly related to $H_0$~\cite{Addison:2013haa,Addison:2017fdm,Cuceu:2019for,Schoneberg:2019wmt,Schoneberg:2022ggi}. Alternatively, $\omega_b$ can be determined very robustly from the even-to-odd peak height ratio in the CMB, and the resulting determination is highly consistent with the BBN one. In principle, as we will discuss in more detail in Sec.~\ref{subsubsec:omegam}, calibrating $r_s$ requires knowledge of both $\omega_b$ and $\omega_c$, which together control the denominator of the integrand in Eq.~(\ref{eq:rd}). This is somewhat problematic since, as recently argued in Refs.~\cite{Poulin:2024ken,Pedrotti:2024kpn}, $\omega_c$ strongly depends on the assumed cosmological model, especially for those models attempting to solve the Hubble tension. To work around this problem, one can trade information on $\omega_c$ for information on $\Omega_m$, since the latter parameter is much more stable than $\omega_c$. At the same time, $\Omega_m$ is also well constrained by a very wide range of late-time probes (e.g.\ unanchored SNeIa). It can be shown that, within $\Lambda$CDM, the quantity $r_dH_0$ scales as follows~\cite{Schoneberg:2022ggi}:
\begin{equation}
r_dH_0 \propto \Omega_m^{-0.23}h^{0.52}\omega_b^{-0.11}\,,
\label{eq:h0rd}
\end{equation}
where $h \equiv H_0/(100\,{\text{km}}/{\text{s}}/{\text{Mpc}})$ is the reduced Hubble constant. From Eq.~(\ref{eq:h0rd}) we clearly see that, once $\omega_b$ is determined from BBN and $\Omega_m$ is given, $H_0$ can be uniquely determined from inferences of $r_dH_0$. Essentially the reason is that, since DE can be neglected at early times, $r_dH_0$ can be rewritten in the following way:
\begin{equation}
r_dH_0 \approx \int_{z_d}^{\infty}dz\,\frac{c_s(z,\omega_b,\omega_r)}{\sqrt{\Omega_m(1+z)^3+\dfrac{\omega_r}{h^2}(1+z)^4}}\,,
\label{eq:soundhorizonomegam}
\end{equation}
which shows that the unnormalized expansion rate does indirectly depend on $H_0$. While the previous discussion is phrased within the context of $\Lambda$CDM, similar considerations hold for late-time modifications thereof, such as those we will be interested in.

The sensitivity of BAO measurements to the product $r_dH_0$ is what makes them relevant to the Hubble tension. Current BAO measurements are broadly speaking consistent with $r_dH_0 \approx 10000\,{\text{km}}/{\text{s}}$ (or alternatively $r_dh \approx 100$). This product is consistent with the ``$\Lambda$CDM solution'' characterized by a ``high'' sound horizon $r_d \sim 147\,{\text{Mpc}}$ and a ``low'' Hubble constant $H_0 \approx 67\,{\text{km}}/{\text{s}}/{\text{Mpc}}$. In order to remain consistent with this product while accommodating a higher Hubble constant, $H_0 \approx 73\,{\text{km}}/{\text{s}}/{\text{Mpc}}$, as required to fully address the Hubble tension, one must therefore reduce the sound horizon to $r_d \sim 136\,{\text{Mpc}}$ (see e.g.\ Fig.~1 of the ``Hubble hunter's guide''~\cite{Knox:2019rjx} for a very clear visual representation). This inevitably requires new physics operating prior to recombination. All of these considerations rest on the specific product $r_dH_0$ inferred from standard BAO analyses. As noted in the Introduction, this determination involves several fiducial cosmology assumptions built into the standard BAO pipeline, whose output -- the scaling parameters $\alpha_{\perp,\parallel}$ -- scale inversely with $r_dH_0$. The question of whether these fiducial cosmology assumptions may (circularly) bias the resulting BAO measurements is not yet completely settled.~\footnote{In partial response to these issues, BAO analysis methods other than the standard ones have been developed. One possibility is related to the so-called ``linear point'', which can be safely extrapolated far from the assumed fiducial cosmology, while being much less sensitive to non-linear gravitational effects~\cite{Anselmi:2015dha,Anselmi:2017cuq,Anselmi:2018vjz,ODwyer:2019rvi,Parimbelli:2020wyw,Anselmi:2022exn,He:2023xof,Lee:2024uuz,Lee:2024rvh,Paranjape:2024ytm}. A fundamentally different strategy is instead to measure the BAO feature from the 2-point angular correlation function of galaxies in narrow redshift bins. This approach only requires two angular coordinates for each objects, and therefore does not require a grid fiducial cosmology for converting redshifts to distances. The resulting angular BAO measurements, also referred to as ``2D BAO'', have been argued to carry less model dependence. These measurements have been extensively used to constrain cosmological models (see e.g.\ Refs.~\cite{Carvalho:2015ica,deCarvalho:2017xye,Camarena:2019rmj,Nunes:2020hzy,Nunes:2020uex,Menote:2021jaq,Benisty:2022psx,Bernui:2023byc,Liu:2024dlf,Gomez-Valent:2024tdb,Favale:2024sdq,Ruchika:2024lgi,Giare:2024syw,Dwivedi:2024okk,Santos:2025gjf,Sabogal:2025qhz}).} All conclusions drawn from these measurements (including the need for pre-recombination new physics) could of course be affected by such a bias.

Before moving on, we note that recent analyses have further scrutinized the consistency between different cosmological probes by performing a redshift tomography of the Hubble tension. For instance, Ref.~\cite{Bousis:2024rnb} compared calibrated SNeIa to a comprehensive BAO dataset calibrated by the CMB sound horizon, after translating the calibrated BAO into luminosity distances. They found significant discrepancies between the two distance measurements across all redshift bins, with the tension being more pronounced at lower redshifts ($z \in [0.1, 0.8]$) than at higher ones ($z \in [0.8, 2.3]$). This redshift-dependent inconsistency suggests that a simple modification to the late-time expansion history, $H(z)$, is insufficient to resolve the tension, pointing instead towards a potential issue with the underlying calibration methods of the probes themselves (see also Ref.~\cite{Perivolaropoulos:2024yxv}).~\footnote{See e.g.\ Refs.~\cite{Krishnan:2020obg,Krishnan:2020vaf,Dainotti:2021pqg,Dainotti:2022bzg,Colgain:2022nlb,Dainotti:2022rea,Colgain:2022rxy,Colgain:2022tql,Jia:2022ycc,Akarsu:2024qiq,Jia:2024wix,Lopez-Hernandez:2024osv,Dilsiz:2025ucs,Fazzari:2025mww,Kalita:2025jqz} for recent works examining redshift-dependence of inferred cosmological parameters and possible implications for cosmological tensions.}

\begin{figure}[!t]
\centering
\includegraphics[width=0.9\linewidth]{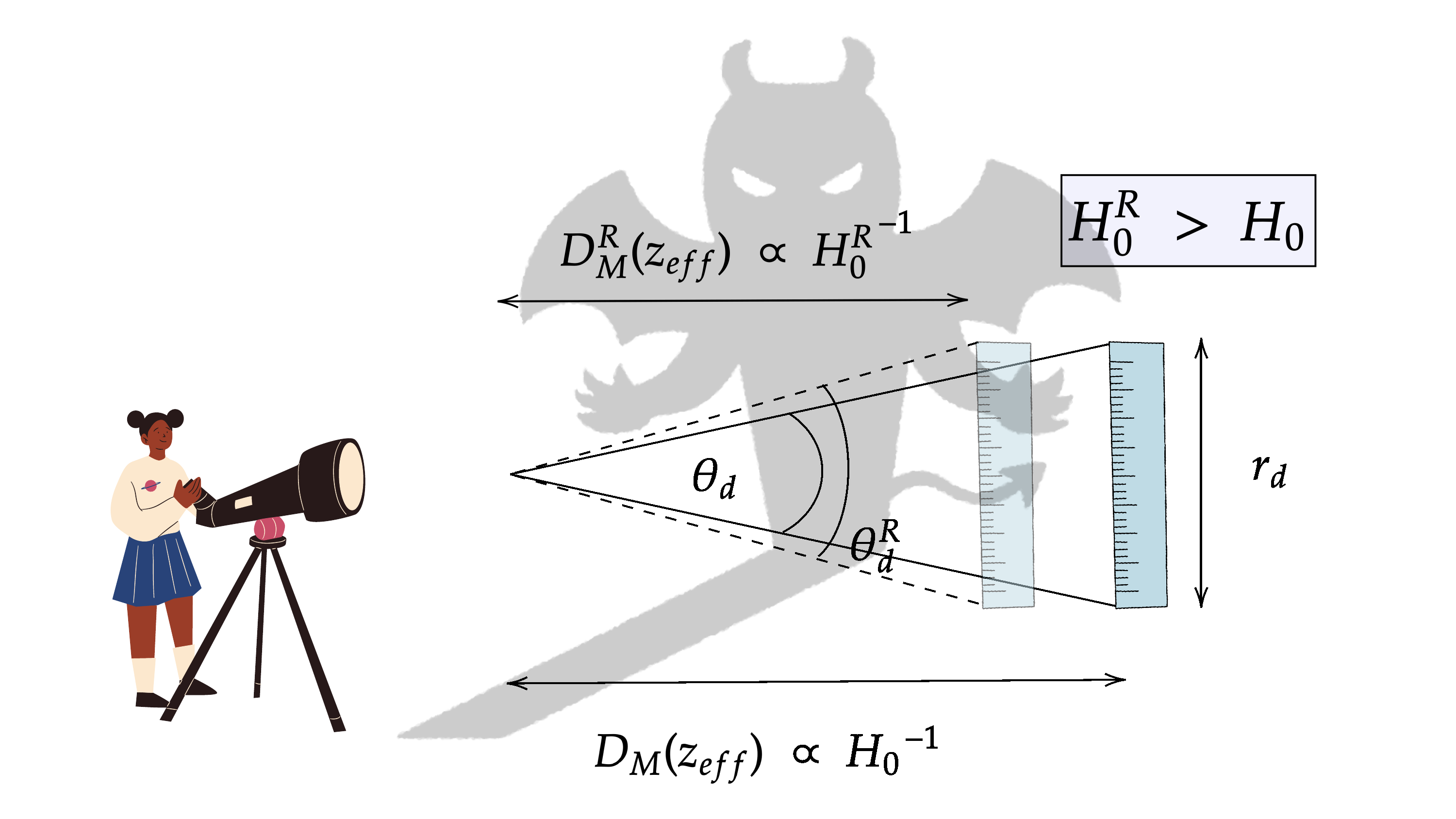}
\caption{Cartoon representation of a biased transverse BAO measurement at the effective redshift $z_{\text{eff}}$. We shall play devil's advocate (in grey) and assume that adopting a fiducial $\Lambda$CDM cosmology in the BAO pipeline biases the resulting angular scales low. This means that the true angular scale is larger than the inferred one, $\theta_d^{\cal R}>\theta_d$, and similarly for the Hubble constant, i.e.\ $H_0^{\cal R}>H_0$, whereas the converse holds for comoving angular diameter distances, i.e.\ $D_M^{\cal R}<D_M$. The superscript $^{\cal R}$ refers to ``bias-corrected'' BAO measurements.}
\label{fig:diagram}
\end{figure}

Returning to the BAO measurements themselves, let us now play devil's advocate and assume that adopting a fiducial $\Lambda$CDM cosmology throughout the standard pipeline does indeed bias the recovered $\alpha$s. We assume that the bias goes in the direction of lowering the inferred value of $H_0$ and, likewise, the inferred acoustic angular scales and redshift span $\theta_d$, $\delta z_d$, and $\theta_v$, since all three are proportional to $H_0$ [see Eqs.~(\ref{eq:thetad},\ref{eq:zd},\ref{eq:thetav})]. For concreteness, we shall model this effect as a redshift-independent shift in the BAO measurements, such that the ``true'' value of $H_0$, inferred from the ``bias-corrected'' BAO measurements, would move closer to the local determination. We will refer to these ``bias-corrected'' BAO measurements as \textit{rescaled BAO measurements} in what follows, given that the assumed redshift-independent shift amounts to an overall rescaling. It is important to stress that this redshift-independent rescaling is a toy model, which we adopt as a minimal assumption to parameterize a potential fiducial cosmology-induced bias. Let us denote by $\theta_d$, $\delta z_d$, and $\theta_v$ the standard (non-rescaled) acoustic angular scales and redshift span, and by $\theta_d^{\cal R}$, $\delta z_d^{\cal R}$, and $\theta_v^{\cal R}$ their rescaled, bias-corrected counterparts. The relation between these two sets of measurements is the following:
\begin{equation}
\theta_d^{\cal R}=\lambda\theta_d\,, \quad \delta z_d^{\cal R}=\lambda\delta z_d\,, \quad \theta_v^{\cal R}=\lambda\theta_v\,,
\label{eq:lambda}
\end{equation}
where $\lambda$ is a constant rescaling parameter which ``undoes'' the fiducial cosmology-associated bias. Following our previous discussion, the ``true'' angular scales and redshift spans are larger than the biased ones, and therefore $\lambda>1$ in Eq.~(\ref{eq:lambda}). In fact, in our subsequent analysis we will set $\lambda=73.0/68.5 \approx 1.06$, i.e.\ approximately the ratio between the SH0ES and $\Lambda$CDM-based BAO+BBN values of $H_0$. We note that the actual BAO measurements are typically reported in terms of $D_V/r_d$, $D_H/r_d$, and $D_M/r_d$, since at a fixed fiducial cosmology it is these quantities which are proportional to the $\alpha$s. These quantities, and the $\alpha$s themselves, are therefore rescaled by $1/\lambda$ as follows:
\begin{align}
\left ( \frac{D_V}{r_d} \right ) ^{\cal R} &= \frac{1}{\lambda} \left ( \frac{D_V}{r_d} \right ) \,, \nonumber \\
\left ( \frac{D_H}{r_d} \right ) ^{\cal R} &= \frac{1}{\lambda} \left ( \frac{D_H}{r_d} \right ) \,, \nonumber \\
\left ( \frac{D_M}{r_d} \right ) ^{\cal R} &= \frac{1}{\lambda} \left ( \frac{D_M}{r_d} \right ) \,.
\label{eq:lambdad}
\end{align}
In our case, $1/\lambda \approx 0.94$. This rescaling is visually conveyed in Fig.~\ref{fig:diagram}, representing an idealized cartoon version of a biased transverse BAO measurement.

In what follows, we rescale existing BAO measurements according to Eqs.~(\ref{eq:lambda},\ref{eq:lambdad}).~\footnote{As discussed above, we fix $\lambda \approx 1.06$. In principle we could have chosen to treat $\lambda$ as a free parameter to be varied in the likelihood, with the data free to choose the preferred value of $\lambda$. However, this would effectively turn our rescaled BAO measurements into unanchored probes of the shape of the expansion rate, albeit with a much weaker constraining power compared to unanchored SNeIa. For this reason, we choose to fix $\lambda$ instead of varying it.} We then explicitly test, against these rescaled BAO measurements, a set of extensively studied post-recombination proposals that have at various times been considered promising candidates to resolve the Hubble tension (these models are presented in Sec.~\ref{sec:models}), assuming instead that the pre-recombination Universe is described by $\Lambda$CDM. As we will show in Sec.~\ref{sec:results}, such models cannot be rescued, at least not by the type of redshift-independent rescaling we have chosen to adopt.

\section{Late-time new physics models}
\label{sec:models}

\begin{figure*} 
\includegraphics[trim = 10mm  10mm 9mm 10mm, clip, width=7.cm, height=4.5cm]{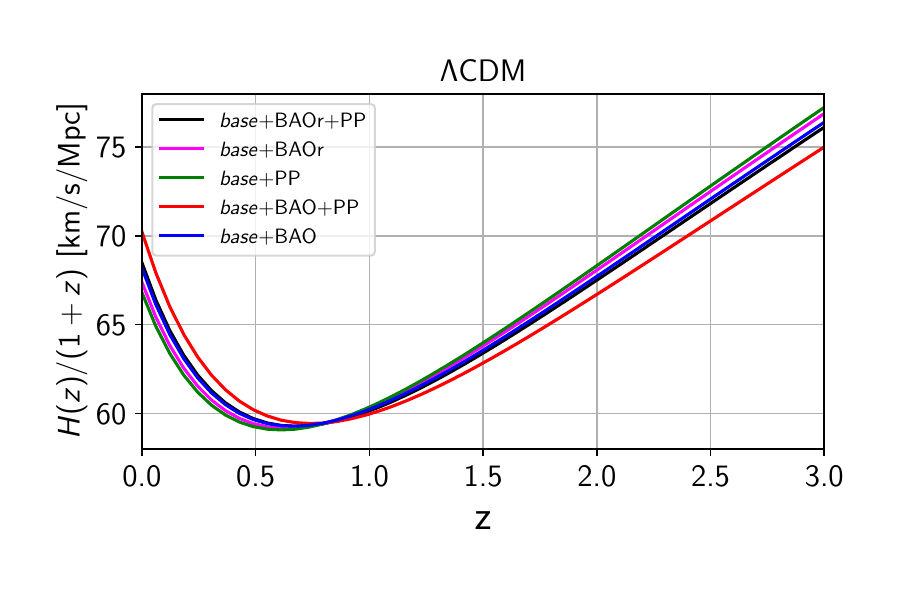}  
\includegraphics[trim = 10mm  10mm 9mm 10mm, clip, width=7.cm, height=4.5cm]{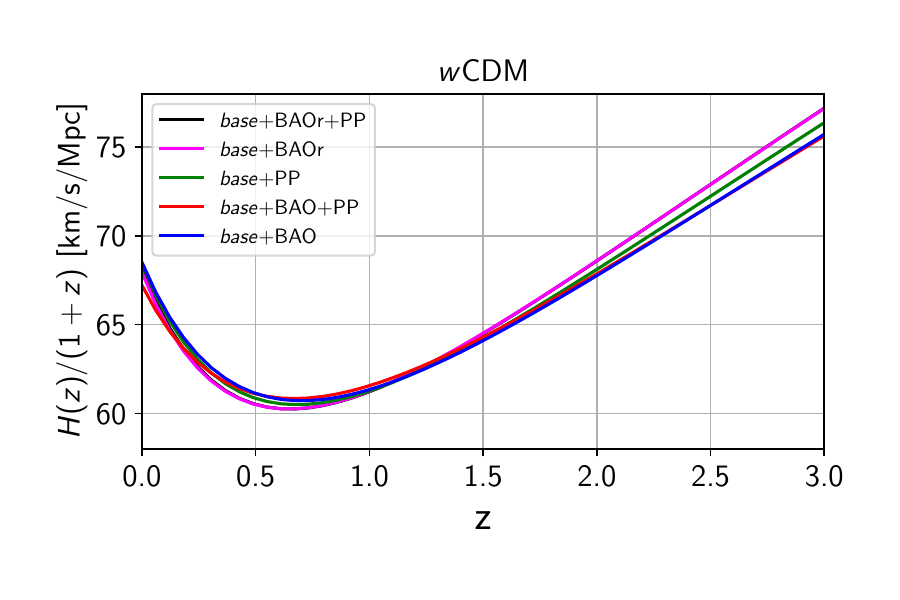} \\
\includegraphics[trim = 10mm  10mm 9mm 10mm, clip, width=7.cm, height=4.5cm]{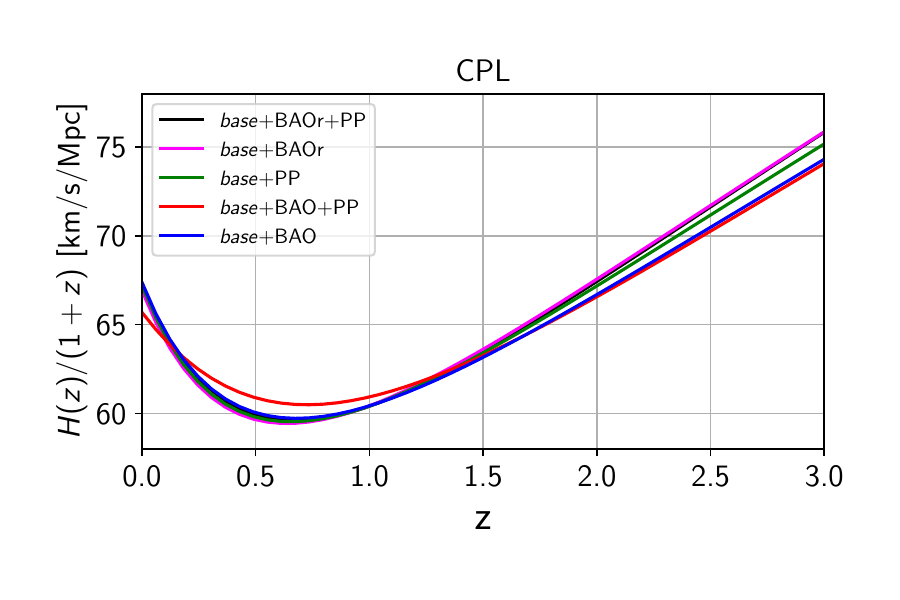} 
\includegraphics[trim = 10mm  10mm 9mm 10mm, clip, width=7.cm, height=4.5cm]{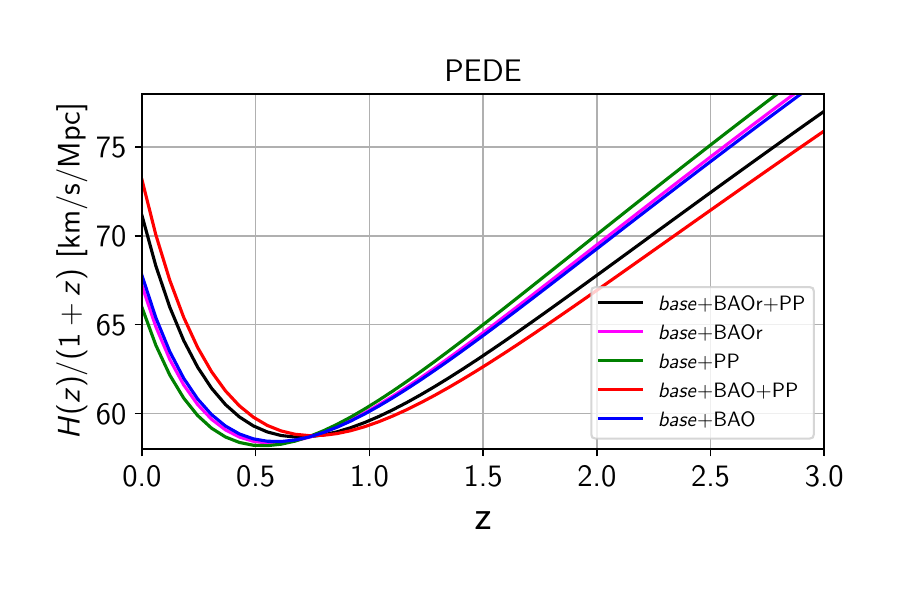}   \\
\includegraphics[trim = 10mm  10mm 9mm 10mm, clip, width=7.cm, height=4.5cm]{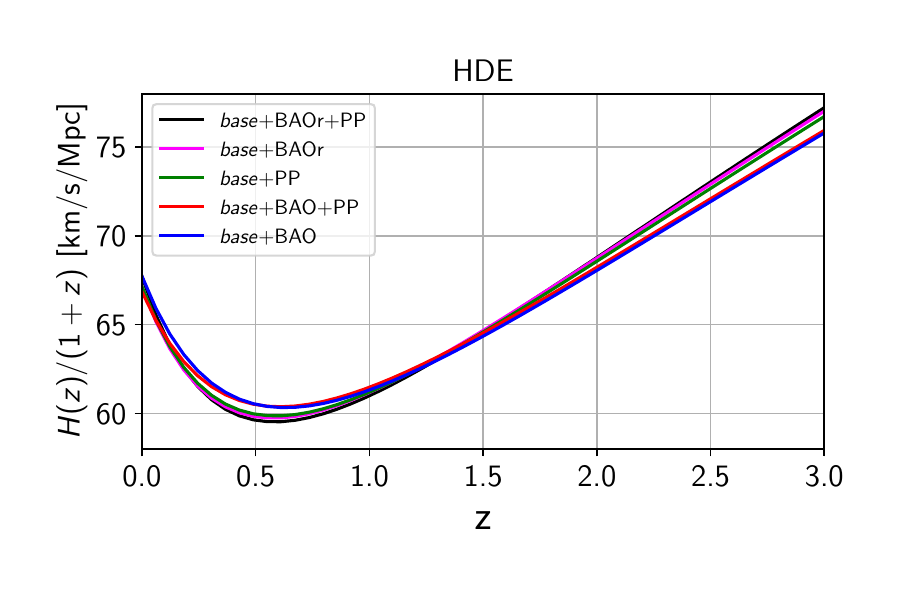} 
\includegraphics[trim = 10mm  10mm 9mm 9mm, clip, width=7.cm, height=4.5cm]{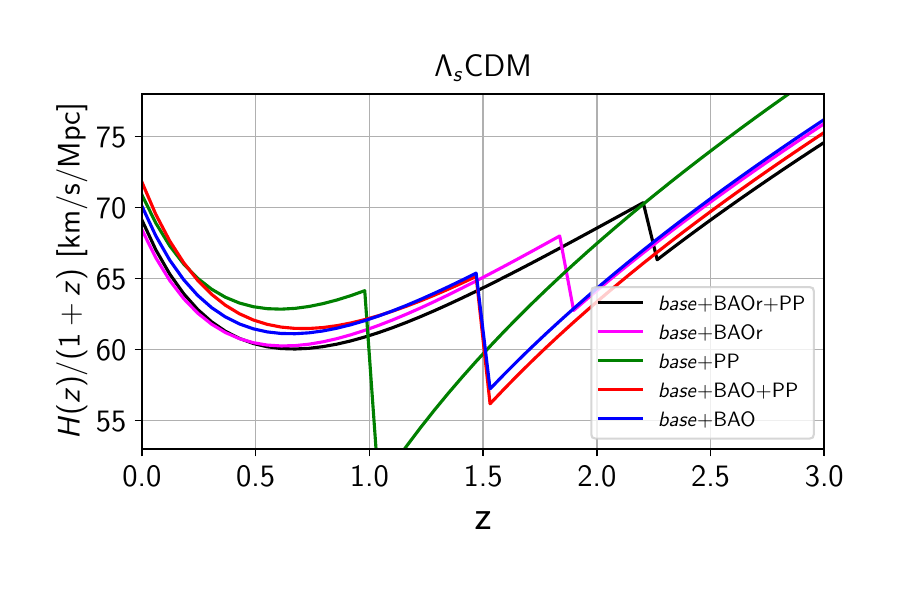}  \\
\includegraphics[trim = 10mm  10mm 9mm 10mm, clip, width=7.cm, height=4.5cm]{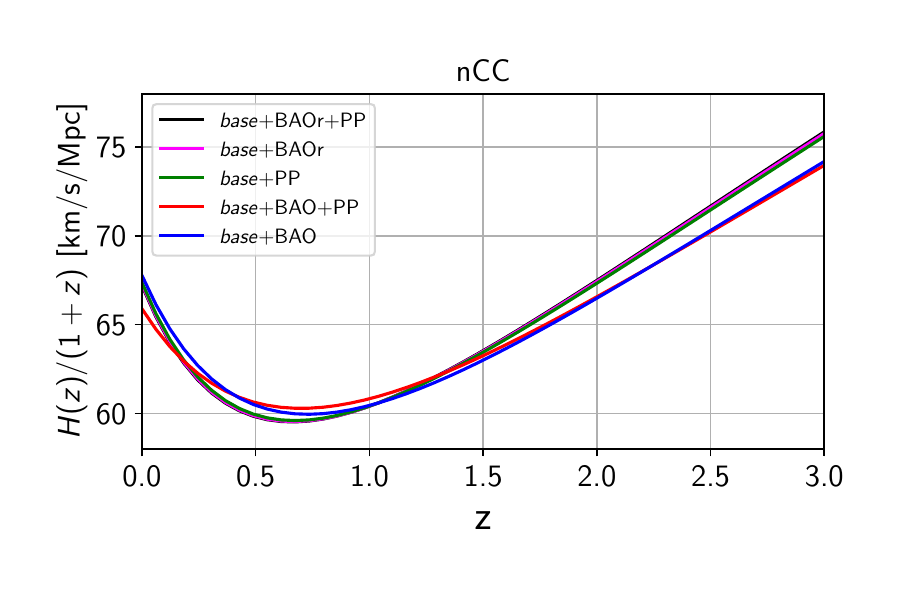} 
\includegraphics[trim = 10mm  10mm 9mm 10mm, clip, width=7.cm, height=4.5cm]{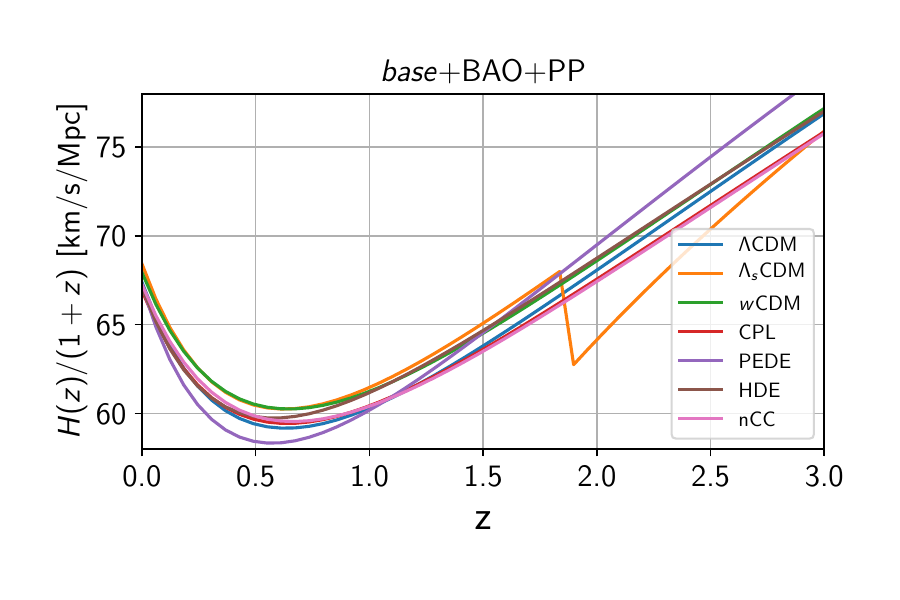} 
\caption{Best-fit reconstructed expansion histories for the seven late-time cosmological models studied in this work. The first seven panels (from left to right, up to down) show the predictions for $H(z)/(1+z)$ within each individual model, with best-fit parameters determined from five different dataset combinations, discussed later in Sec.~\ref{sec:datasets}, and as determined by the color coding. This comparison illustrates the effect of using standard BAO (\textit{BAO}) versus rescaled BAO (\textit{BAOr}) measurements, and the powerful constraints imposed by including unanchored SNeIa (\textit{PP}). A key feature across almost all models is that while the rescaled BAO data (\textit{base}+\textit{BAOr} dataset combination) can lead to significant shifts in the predicted expansion history, the inclusion of unanchored SNeIa data (\textit{base}+\textit{BAOr}+\textit{PP}) largely precludes these shifts by tightly constraining the shape of the expansion history, pulling back the expansion history back towards the $\Lambda$CDM one. Finally, the bottom right panel directly compares the best-fit expansion histories for all seven models, as determined by the color coding, in light of the standard \textit{base}+\textit{BAO}+\textit{PP} dataset combination.}
\label{fig:hz_plots}
\end{figure*}

We now discuss the cosmological models we use to test whether BAO miscalibration could rescue late-time solutions to the Hubble tension. Each model introduces significant post-recombination modifications, typically operative at redshifts $z \lesssim 2$, and reduces to $\Lambda$CDM in the early Universe. Essentially, all these models represent alternative DE scenarios where the cosmological constant $\Lambda$ is replaced by an evolving DE component. Indeed, nearly all ``popular'' (for want of a better word) late-time approaches to alleviating the $H_0$ tension fall within this category. All the models we evaluate have, at one point or another, been considered worth exploring in the context of the Hubble tension. The best-fit reconstructed expansion histories for the seven late-time cosmological models studied in this work are shown in Fig.~\ref{fig:hz_plots}.

We can analyze these models in an unified manner based on their impact on the first Friedmann equation, which we express as follows:
\begin{equation}
E(z) \equiv \frac{H(z)}{H_0} = \sqrt{\Omega_r(1+z)^4 + \Omega_m(1+z)^3 + \Omega_{\text{DE}}f(z)}\,,
\label{eq:fF_equation}
\end{equation}
where $H(z)$ is the expansion rate of the Universe, and $\Omega_r$, $\Omega_m$, and $\Omega_{\text{DE}}$ are the present-day density parameters of radiation, matter, and the DE component. In Eq.~(\ref{eq:fF_equation}) we are leaving out massive neutrinos, whose impact in our discussion is negligible, and explicitly assume a spatially flat Universe, so that $\Omega_r+\Omega_m+\Omega_{\text{DE}}=1$.~\footnote{The latter choice is motivated by the fact that state-of-the-art constraints on the spatial curvature parameter $\Omega_K$ are broadly consistent with the spatially flat case where $\Omega_K=0$ (see e.g.\ Refs.~\cite{Efstathiou:2020wem,Benisty:2020otr,Vagnozzi:2020rcz,Vagnozzi:2020dfn,Yang:2021hxg,Dhawan:2021mel,Gonzalez:2021ojp,Dinda:2021ffa,Zuckerman:2021kgm,Bargiacchi:2021hdp,Glanville:2022xes,Yang:2022kho,Stevens:2022evv,Favale:2023lnp,Qi:2023oxv,Wu:2024faw,Liu:2024yib,Forconi:2025zzu,Benetti:2025ljc,Specogna:2025ufe}), despite previous hints for a potentially closed Universe from \textit{Planck} data alone~\cite{Handley:2019tkm,DiValentino:2019qzk}.} The information on the different DE models we consider is then fully encoded in the redshift-dependent function $f(z)$. The models we consider can typically be phrased in terms of a DE component with time-varying EoS $w_{\text{DE}}(z) \equiv p_{\text{DE}}/\rho_{\text{DE}}$. The relation between $f(z)$ and $w_{\text{DE}}(z)$ is given by the following:
\begin{equation}
f(z) = \exp \left [ 3 \int^{\text{ln(1+z)}}_{0} \text{d ln}(1+z')(1 + w_{\text{DE}}(z')) \right ] \,.
\label{eq:fz}
\end{equation}
Conversely, the effective DE EoS can be expressed as follows:
\begin{equation}
w_{\text{DE}}(z) = \frac{1}{3}\frac{d\ln(\Omega_{\rm DE}f(z))}{d\ln(1+z)}-1 = \frac{1+z}{3f(z)}\frac{df(z)}{dz}-1\,.
\label{eq:w}
\end{equation}
We note that $w_{\text{DE}}(z)$ need not be finite. In fact, models where the energy density of the DE component changes sign, and therefore where $f(z)$ goes through a zero, lead to poles in $w_{\text{DE}}(z)$. This behaviour has been explicitly discussed in the literature~\cite{Ozulker:2022slu,Adil:2023ara,Menci:2024rbq}, and is not in itself pathological, since $w_{\text{DE}}(z)$ is not a directly observable quantity, but only the expansion rate is.

Models which have at some point been (incorrectly) considered viable solutions to the Hubble tension are those for which $f(z>0)<1$, i.e.\ where the past contribution of the DE component is smaller than that of the cosmological constant, when both are assumed to have the same energy density at $z=0$. The reason is that these models allow for a higher value of $H_0$ relative to $\Lambda$CDM, at least when considering CMB data alone (but ignoring at least unanchored SNeIa data). The reason is as follows. Let us consider the best-fit $\Lambda$CDM model, and hold the $\Omega$s fixed. For $f(z>0)<1$, the unnormalized expansion rate $E(z)$ in Eq.~(\ref{eq:fF_equation}) is clearly lower with respect to the reference $\Lambda$CDM model. If $H_0$ were also held fixed, the distance to the last-scattering surface $D_A(z_{\star})$ would then increase relative to $\Lambda$CDM, since $D_A(z_{\star})$ is determined by an integral of $[H_0E(z)]^{-1}$. Since the DE component is negligible at recombination, these modifications do not alter the sound horizon at recombination $r_s(z_{\star})$. The increase in $D_A(z_{\star})$ would therefore imply a reduction in the acoustic angular scale $\theta_s=r_s(z_{\star})/D_A(z_{\star})$, which however is tightly constrained by CMB observations. An increase in $H_0$ is then required to bring down $D_A(z_{\star})$, and in doing so raise $\theta_s$ to a value compatible with observations. Models where $f(z>0)<1$ effectively behave as a phantom DE component, whose contribution going towards earlier times redshifts faster than the cosmological constant, and are therefore often referred to as ``phantom-like''.

Before discussing the seven models considered in this work, we stress that our analysis deliberately excludes one of the most widely studied late-time scenarios for the Hubble tension: interacting dark energy (IDE). In this class of models, non-gravitational interactions between dark matter and dark energy lead to energy-momentum exchange between the two. IDE models have received tremendous attention in recent years, particularly in the context of the Hubble tension, and the subset of such models which can partially alleviate the tension does so via an effective phantom-like behaviour (see e.g. Refs.~\cite{Yang:2018euj,Yang:2018uae,Martinelli:2019dau,Yang:2019vni,Pan:2019gop,DiValentino:2019ffd,Benetti:2019lxu,DiValentino:2019jae,vonMarttens:2019ixw,Vagnozzi:2019kvw,Yang:2020uga,Lucca:2020zjb,Hogg:2020rdp,Li:2020gtk,vonMarttens:2020apn,Gao:2021xnk,Wang:2021kxc,Benetti:2021div,Nunes:2021zzi,Hogg:2021yiz,Guo:2021rrz,Gariazzo:2021qtg,Ferlito:2022mok,Nunes:2022bhn,Yao:2022kub,Pan:2022qrr,Gao:2022ahg,Roy:2023uhc,vanderWesthuizen:2023hcl,Borges:2023xwx,Zhai:2023yny,Forconi:2023hsj,Sebastianutti:2023dbt,Benisty:2024lmj,Giare:2024ytc,Montani:2024pou,Sabogal:2024yha,Ghedini:2024mdu,Benetti:2024dob,Aboubrahim:2024cyk,Sabogal:2025mkp,Tamayo:2025wiy,Rahimy:2025iyj,Feng:2025mlo,Silva:2025hxw,Richarte:2025tkj,Montani:2025rcy,vanderWesthuizen:2025vcb,vanderWesthuizen:2025mnw,vanderWesthuizen:2025rip,Li:2025muv,Zhang:2025dwu,Figueruelo:2026eis}). As discussed later in Sec.~\ref{subsubsec:cmb}, in our analysis we make use of a compressed CMB likelihood. This is appropriate for models affecting only the late-time expansion history, with primarily geometrical effects on the CMB. However, the effects of IDE on the evolution of matter perturbations, not only at low redshifts but also at much earlier times, can be significant, and cannot be captured solely through their impact on the geometrical observables of the CMB. As such, for IDE the choice between compressed and full CMB likelihoods drastically alters the resulting parameter constraints and degeneracy directions, as some of us have explicitly tested in work in progress~\cite{compressedinpreparation}. In order to be as conservative as possible, we have therefore opted not to include IDE in the present work.

\subsection{$\Lambda$CDM}
\label{subsec:lcdm}

As a warm-up in our subsequent analysis, we will consider the impact of BAO miscalibration on parameter inference within the standard $\Lambda$CDM model. In this case, the redshift-dependent function $f(z)$ entering Eq.~(\ref{eq:fF_equation}) is of course a constant $f(z)=(1+z)^{1-1}=1$, so that $\Omega_{\text{DE}}=\Omega_{\Lambda}$ is the present-day density parameter of the cosmological constant, whose EoS is $w_{\Lambda}=-1$.

\subsection{$w$CDM}
\label{subsec:wcdm}

The simplest possibility for going beyond $\Lambda$CDM is to consider a DE component whose EoS $w \neq -1$ is constant in time, while satisfying $w<-1/3$ in order to drive cosmic acceleration. We refer to the resulting cosmological model as $w$CDM. In this case, the function $f(z)$ which appears in Eq.~(\ref{eq:fF_equation}) is given by the following:
\begin{equation}
f(z) = (1+z)^{3(1+w)}\,.
\label{eq:fwcdm}
\end{equation}
Values $w>-1$ are referred to as being ``quintessence-like'', since this is the range of EoS values which can be realized within the simplest quintessence models featuring a single scalar field with a canonical kinetic term, minimally coupled to gravity, and without higher derivative operators~\cite{Nesseris:2006er}, whereas values $w<-1$ are in the so-called phantom regime, and the $w=-1$ demarcation line is often referred to as the ``phantom divide''. Various works over the past years have examined constraints on $w$ in light of cosmological data. One of the most recent analyses indicates $w = -1.013 ^{+0.038}_{-0.043}$ from a combination of SDSS BAO, \textit{PantheonPlus} unanchored SNeIa, cosmic chronometers, and \textit{Planck} PR3 CMB likelihoods, in perfect agreement with the cosmological constant scenario~\cite{Escamilla:2023oce}. A later analysis replacing SDSS BAO with DESI measurements finds $w=-1.049 \pm 0.024$, still in good agreement with the cosmological constant~\cite{Sammut:2025eik}.

The $w$CDM model has at some point been considered a contender to fully solve the Hubble tension due to its ability to accommodate a phantom DE component. One indeed observes from Eq.~(\ref{eq:fwcdm}) that $w<-1$ implies $f(z>0)<1$, recovering the condition discussed earlier. Therefore, one expects a negative correlation between $w$ and $H_0$, which has been observed and extensively discussed in the literature~\cite{Vagnozzi:2018jhn,Vagnozzi:2019ezj,Alestas:2020mvb,Alestas:2021xes,deSa:2022hsh}. In fact, exploiting this degeneracy, one finds that a value of $w \sim -1.3$ would formally be required to address the Hubble tension: this is of course completely at odds with BAO and unanchored SNeIa data, given the significant modifications introduced to the expansion history.

\subsection{CPL dynamical dark energy}
\label{subsec:cpl}

The next simplest possibility for going beyond $\Lambda$CDM is to consider a simple parametrization for the EoS $w_{\text{DE}}(z)$ of an evolving DE component. While several (typically two-parameter) parametrizations for the EoS of dynamical DE exist in the literature (see e.g.\ Refs.~\cite{Efstathiou:1999tm,Jassal:2004ej,Feng:2004ff,Gong:2005de,Barboza:2008rh,Ma:2011nc,Pantazis:2016nky,Escamilla-Rivera:2016qwv,Yang:2017alx,Pan:2017zoh,Yang:2018qmz,Capozziello:2020ctn,DiValentino:2020naf,Capozziello:2022jbw,Chaudhary:2023vxz,Singh:2023ryd,Adil:2023exv,Chaudhary:2023zzo,Dunsby:2023qpb,Rezaei:2024vtg,Kessler:2025kju,Shlivko:2025fgv,Cheng:2025lod,Lee:2025pzo,Fazzari:2025lzd}), there is no doubt that the most widely adopted and popular one is the so-called Chevallier-Polarski-Linder (CPL) parametrization, with the resulting cosmological model sometimes referred to as $w_0w_a$CDM model. Within the CPL parametrization, the DE EoS is given by the following~\cite{Chevallier:2000qy,Linder:2002et}:
\begin{equation}
w(z) = w_0 + w_a\frac{z}{1+z}\,.
\label{eq:w0wa}
\end{equation}
At face value, the EoS given in Eq.~(\ref{eq:w0wa}) can be interpreted as a truncated Taylor expansion of the DE EoS as a function of the scale factor $a$, around the present time: we then read off $w_0$ as being the present-day DE EoS, and $w_a$ as the coefficient of the first-order term, describing the dynamics of DE. Within the CPL parametrization, the redshift-dependent function $f(z)$ is then given by the following expression:
\begin{equation}
f(z) = (1+z)^{3 \left ( 1+w_0+w_a\right ) }\exp \left ( -3w_a\frac{z}{1+z} \right ) \,,
\label{eq:fcpl}
\end{equation}
which trivially follows when inserting Eq.~(\ref{eq:w0wa}) into Eq.~(\ref{eq:fz}). For a state-of-the-art critical assessment of current constraints on CPL dynamical DE, see Ref.~\cite{Giare:2025pzu}.

There are various reasons why the CPL parametrization is widely utilized. First and foremost, it has been shown to be especially accurate in recovering observables for physical models of quintessence DE~\cite{Linder:2002et,Linder:2007wa}, as tested against solutions of the Klein-Gordon equation, a characteristic it shares with the 4-parameter Copeland-Corasaniti-Linder-Huterer parametrization~\cite{Corasaniti:2002vg,Linder:2005ne}. In this sense the CPL parametrization is best viewed as a physics-based encapsulation of the 2 full DE EoS phase space functions $w(a)$ and $dw(a)/d\ln a$~\cite{Linder:2024rdj}, rather than a truncated Taylor expansion (see however Ref.~\cite{Nesseris:2025lke}).

In addition, its 2-parameter nature makes the CPL parametrization highly manageable and flexible. Its flexibility is particularly significant because it allows for the DE EoS to cross the phantom divide. This is a powerful probe for new physics because, as mentioned earlier, it is not achievable by the simplest models of DE featuring a single scalar field with a canonical kinetic term, minimally coupled to gravity, and without higher derivative operators, whose EoS is always bounded by $w \geq -1$~\cite{Nesseris:2006er}. Broadly speaking, there are two theoretical avenues to realize a phantom divide crossing. The first involves multi-component dark energy models (``quintom'' models) that include at least one phantom component with a negative kinetic term~\cite{Cai:2009zp}. However, this approach is often considered theoretically problematic due to potential catastrophic vacuum instabilities~\cite{Nesseris:2006er}. The second, more theoretically motivated possibility is to consider extensions to General Relativity, e.g.\ scalar-tensor theories. The dynamics of these models can naturally accommodate a smooth transition to the phantom regime ($w<-1$) without violating observational constraints~\cite{Vikman:2004dc,Perivolaropoulos:2005yv,Nesseris:2006hp}. On general grounds, an observational confirmation of a phantom divide crossing would provide a compelling signature of new physics beyond standard DE models.

It is worth noting that recent data from the DESI survey provide compelling new evidence pointing towards such a crossing. When combined with various CMB and SNeIa datasets, DESI DR2 BAO data show a strong preference for a dynamical DE model over $\Lambda$CDM, with a significance of up to $4.2\sigma$~\cite{DESI:2025zgx}. The favored solution consistently lies in the quadrant with a present-day EoS $w_0>-1$ and a negative evolution parameter $w_a<0$. This implies a DE EoS that was phantom ($w<-1$) in the recent past and has evolved to be quintessence-like ($w>-1$) today, necessarily crossing the phantom divide at intermediate redshifts. This finding, supported by various data combinations and robust against different analysis choices~\cite{Giare:2024oil} (see however Refs.~\cite{Sakr:2025daj,Colgain:2025fct,Keeley:2025rlg,Roy:2025cxk,Toomey:2025xyo}), strengthens the motivation for considering theoretical frameworks that permit such behavior.

\subsection{Phenomenologically emergent dark energy}
\label{subsec:pede}

The phenomenologically emergent dark energy (PEDE) model was originally proposed in Ref.~\cite{Li:2019yem} on purely phenomenological grounds as an attempt to alleviate the Hubble tension without introducing additional model parameters beyond those of $\Lambda$CDM. Within the PEDE model, the redshift-dependent function $f(z)$ appearing in Eq.~(\ref{eq:fF_equation}) is given by the following:~\footnote{We stress that it is not straightforward to count the true degrees of freedom of a non-linear model such as PEDE, where a very specific shape for $w(z)$ (the true ``free function'') has been chosen.}
\begin{equation}
f(z)= 1 - \tanh \left [ \log_{10}(1+z) \right ] \,.
\label{eq:fpede}
\end{equation}
The above describes a DE component which is completely subdominant in the past, and starts emerging at redshift $z \sim 10$, before becoming the dominant component at late times. Using Eq.~(\ref{eq:fz}), the EoS of PEDE is found to take the following form:
\begin{equation}
w(z) = -\frac{1 + \tanh \left [ \log_{10}(1+z) \right ] }{3\ln10}-1\,.
\label{eq:wpede}
\end{equation}
We see that the EoS of PEDE starts from $w=-2/(3\ln 10)-1 \sim -1.29$ in the far past, reaches $w=-1/(3\ln 10)-1 \sim -1.14$ at present time, and keeps increasing until it will asymptotically reach $w \to -1$ in the far future. Therefore, the PEDE model corresponds to a DE component which is always in the phantom regime, and asymptotically approaches a cosmological constant.

The PEDE model was originally introduced on purely phenomenological grounds, with the purpose of illustrating the possible phenomenology of a potential minimal (i.e.\ one not introducing additional degrees of freedom in a statistical sense) solution to the Hubble tension, and pave the way to more theoretically motivated models. The model has later been generalized to a broader class of parametrizations where the location and steepness of the transition can be controlled by additional parameters, and which include PEDE as a particular case: examples include the generalized emergent dark energy (GEDE)~\cite{Li:2020ybr} and modified emergent dark energy (MEDE)~\cite{Benaoum:2020qsi} models. PEDE and related models has also received significant attention in the literature, not only in the context of the Hubble tension (see e.g.\ Refs.~\cite{Pan:2019hac,Liu:2020vgn,Hernandez-Almada:2020uyr,Rezaei:2020mrj,Yang:2020ope,Yang:2021eud,Staicova:2021ntm,DiValentino:2021rjj,Liu:2022mpj,John:2023fsy,Yao:2023ybs,Staicova:2023jic,Nelleri:2023bdm,LHuillier:2024rmp,Hernandez-Almada:2024ost,Liu:2025mub,Manoharan:2025uix,Li:2025eqh,Sharma:2025qmv,Chaudhary:2025vzy}): for instance, these models have been included among those tested by the DESI collaboration in their papers devoted to DE models beyond the simplest ones~\cite{DESI:2024kob,DESI:2025fii}. To the best of our knowledge, at the time of writing there is no proposal for a fundamental theoretical origin of PEDE. We note that the phantom nature thereof precludes a microscopical model based on a single canonical, minimally coupled scalar field: however, it should in principle be feasible to reconstruct a scenario featuring multiple coupled scalar fields which, through an appropriate choice of interaction potential, could reproduce the phenomenology of PEDE. Going beyond the specific details of PEDE, the generic feature of a (more or less smooth) transition in the DE EoS has been observed in several constructions (including those based on phase transitions inspired by the physics of critical phenomena), see e.g.\ Refs.~\cite{Khosravi:2017hfi,Banihashemi:2018oxo,Banihashemi:2018has,DiValentino:2019exe,Farhang:2020sij,Banihashemi:2020wtb,Khosravi:2021csn,Banihashemi:2022vfv,Farhang:2023hen,Scherer:2025esj}.

\subsection{Holographic dark energy}
\label{subsec:hde}

The holographic principle, first inspired by Bekenstein's work on the entropy of black holes (BHs)~\cite{Bekenstein:1973ur}, suggests that the information contained within a given volume of space can be fully described by degrees of freedom residing on its boundary. Widely regarded as a potential cornerstone of quantum gravity, this principle implies a deep and non-trivial connection between the ultraviolet and infrared cutoffs of a quantum field theory or, in other words, between its smallest and largest scales. Applying this principle to the Universe as a whole, with typical cosmological length scale $L$, one demands that the (vacuum) energy within a radius $L$ does not exceed the mass of a BH of Schwarzschild radius $L$: else, the energy density would be sufficient to form a BH, leading to gravitational collapse. Saturating this inequality leads to the holographic dark energy (HDE) paradigm, where the vacuum energy density is not a constant, but is dynamically related to a cosmological horizon scale, and therefore evolves in time. More specifically, denoting by $M_{\text{Pl}}$ the Planck mass, and following the arguments of Ref.~\cite{Cohen:1998zx}, one finds that the quantum zero-point energy density is given by the following~\cite{Li:2004rb,Wang:2016och}:
\begin{equation}
\rho_{\text{hde}}=3C^2M_{\text{Pl}}^2L^{-2}\,,
\label{eq:rhodeholography}
\end{equation}
where $C$ is a dimensionless constant parameter, and the standard Bekenstein-Hawking entropy has been explicitly assumed. For what concerns the choice of length scale $L$, various possibilities have been explored in the literature: it turns out that simple choices such as the Hubble scale or the particle horizon do not lead to an accelerating Universe, whereas one of the simplest choices which does is that where $L$ is set to the future event horizon of the Universe $R_h$, given by the following:
\begin{equation}
R_h=a(t)\int_{t}^{\infty}\frac{dt'}{a(t')}=a\int_{a}^{\infty}\frac{da^{\prime}}{Ha^{\prime 2}}\,,
\label{eq:rh}
\end{equation}
where $a$ denotes the scale factor. While several other choices of infrared cutoff/characteristic length scale have been studied in the literature, we will stick to the future event horizon, as in the original HDE model.

With this choice, the evolution of the density parameter of HDE, $\Omega_{\text{hde}} \equiv \rho_{\text{hde}}/\rho_c$, where $\rho_c$ is the critical density, is governed by the following differential equation (setting $M_{\text{Pl}}=1$):
\begin{equation}
\frac{d\Omega_{\text{hde}}(z)}{dz}=-\frac{[1-\Omega_{\text{hde}}(z)]\Omega_{\text{hde}}(z)}{1+z} \left ( \frac{\sqrt{\Omega_{\text{hde}}(z)}}{2C}+1 \right ) \,,
\label{eq:omegadeholography}
\end{equation}
where we have explicitly neglected the contribution from radiation, which is completely subdominant at the late times we are interested in. The above equation has an analytic solution in implicit form, at least for the simplest case $C=1$, but is otherwise easy to solve numerically. One can also write down the effective EoS associated to the HDE component, which is obviously time-varying, and is given in implicit form by the following:
\begin{equation}
w_{\text{hde}}(z) = -\frac{1}{3}-\frac{2\sqrt{\Omega_{\text{hde}}(z)}}{3C}\,.
\label{eq:wdeholography}
\end{equation}
Clearly, when HDE is dominant, the equation of state is $w \simeq -1/3-2/3C$, and HDE can therefore drive accelerated expansion as long as $C>0$. Moreover, if $0<C<1$ the HDE component exhibits phantom behaviour, whereas if $C>1$ the behaviour is quintessence-like (in the limiting case $C=1$, HDE behaves as a cosmological constant): in any case, regardless of the value of $C$, in the far future HDE approaches a de Sitter-like phase of expansion, where $\Omega_{\text{hde}} \to 1$ and $w_{\text{hde}} \to -1$. In addition, HDE has been argued to solve the coincidence problem, provided the primordial phase of cosmic inflation lasted at least 60 $e$-folds. This combination of appealing theoretical properties and phenomenological flexibility has led HDE (including generalizations of the original model based on entropies other than the Bekenstein-Hawking one) to receive considerable attention in the literature, not only in the context of cosmological tensions (see e.g.\ Refs.~\cite{Zhao:2017urm,Saridakis:2018unr,Dai:2020rfo,Saridakis:2020zol,Anagnostopoulos:2020ctz,Nojiri:2020wmh,Colgain:2021beg,Drepanou:2021jiv,Telali:2021jju,Luciano:2022pzg,Oliveros:2022biu,Landim:2022jgr,Cardona:2022pwm,Basilakos:2023kvk,Nakarachinda:2023jko,Tang:2024gtq,Trivedi:2024inb,Li:2024qus,Yarahmadi:2024afr,Bidlan:2025pzi,Cimdiker:2025vfn,Luciano:2025hjn,Luciano:2025elo,Guin:2025xki,Plaza:2025nip,Zapata:2025ngr,Wu:2025vfs}).

Two final comments are in order regarding HDE. While one can translate the implicit form for the HDE EoS in Eq.~(\ref{eq:wdeholography}) to a corresponding form for $f(z)$ in Eq.~(\ref{eq:fF_equation}), this is not particularly illuminating nor useful. In fact, in our subsequent analysis, we will explicitly (numerically) solve the differential equation given in Eq.~(\ref{eq:omegadeholography}). Finally, the value of the parameter $C$ cannot be determined from first principles, and instead has to be fit to observations. From our earlier discussions, it is clear that values of $0<C<1$ are of interest in the Hubble tension discussion, due to the effective phantom behaviour of HDE in this regime.

\subsection{Sign-switching cosmological constant}
\label{subsec:lscdm}

The $\Lambda_s$CDM model replaces the positive cosmological constant of $\Lambda$CDM with a sign-switching cosmological constant~\cite{Akarsu:2021fol,Akarsu:2023mfb}. Specifically, the cosmological constant abruptly changes sign at a transition redshift $z=z_{\dagger} \lesssim 2.5$, mimicking a rapid phase transition from an anti-de Sitter (AdS) vacuum $\Lambda_{\text{AdS}}<0$ at $z>z_{\dagger}$, to a de Sitter (dS) vacuum with $\Lambda_{\text{dS}}=-\Lambda_{\text{AdS}}>0$ at $z<z_{\dagger}$. The $\Lambda_s$CDM model was originally inspired by the so-called graduated DE model, where the abrupt transition in $\Lambda_s$CDM is smoother~\cite{Akarsu:2019hmw}: however, cosmological data was shown to prefer a rather rapid transition~\cite{Akarsu:2022typ}. Alongside related models, the resulting $\Lambda_s$CDM model has been the subject of several studies (see e.g.\ Refs.~\cite{Paraskevas:2023itu,Anchordoqui:2023woo,Paraskevas:2024ytz,Akarsu:2024qsi,Anchordoqui:2024gfa,Akarsu:2024eoo,Yadav:2024duq,Anchordoqui:2024dqc,Akarsu:2024nas,Souza:2024qwd,Soriano:2025gxd,Akarsu:2025ijk,Bouhmadi-Lopez:2025ggl,Bouhmadi-Lopez:2025spo,Yadav:2025vpx,Akarsu:2025nns}). It has also been shown that the sudden AdS-to-dS transition does not lead to pathological consequences for astrophysical systems such as bound cosmic structures~\cite{Paraskevas:2023itu,Paraskevas:2024ytz}. Recent works have also demonstrated that the $\Lambda_s$CDM model can emerge from a variety of microphysical scenarios, ranging from Casimir forces and thin-wall vacuum decay within string-inspired extra dimensional scenarios~\cite{Anchordoqui:2023woo,Anchordoqui:2024gfa,Anchordoqui:2024dqc}, to type-II minimally modified theories of gravity~\cite{Akarsu:2024qsi,Akarsu:2024eoo}, to teleparallel $f(T)$ gravity~\cite{Akarsu:2024nas,Souza:2024qwd}, providing further theoretical grounding to what was initially considered a purely phenomenological model.

Within the $\Lambda_s$CDM model, the redshift-dependent function $f(z)$ which appears in Eq.~(\ref{eq:fF_equation}) can be written in the following form:
\begin{equation}
f(z) = \text{sgn} \left [ z^{\dagger}-z \right ] \,,
\label{eq:flscdm}
\end{equation}
where ${\text{sgn}}(x)$ is the \textit{signum} function, which takes the following values:
\begin{eqnarray}
{\text{sgn}}[x] =
\begin{cases}
-1 \quad \quad &(x<0)\,, \\
0 \quad \quad &(x=0)\,, \\
+1 \quad \quad &(x>0)\,. \\
\end{cases}
\label{eq:sgn}
\end{eqnarray}
Although not particularly informative, one can also compute the EoS of the sign-switching cosmological constant of $\Lambda_s$CDM. This is given by $w_{\text{DE}}(z)=-1$ as long as $z \neq z_{\dagger}$, and features a pole at the transition redshift $z=z_{\dagger}$, with $\lim_{z \to z_{\dagger}^{\pm}}=\pm \infty$. As discussed earlier as well as in the literature~\cite{Ozulker:2022slu,Adil:2023ara,Menci:2024rbq}, the existence of this pole is inevitable given that the effective DE density crosses zero, and is not pathological, as $w_{\text{DE}}(z)$ is not a directly observable quantity.

We note that, in the limit $z_{\dagger} \to \infty$, $\Lambda_s$CDM approaches the $\Lambda$CDM model. On the other hand, for sufficiently low $z_{\dagger} \lesssim 2$, the model is able to partially alleviate the Hubble tension, as well as the now much less significant discrepancy in $S_8$~\cite{DiValentino:2018gcu,DiValentino:2020vvd,Nunes:2021ipq}. The mechanism is essentially the one discussed at the start of Sec.~\ref{sec:models} in the context of phantom DE: the decrease in energy density relative to $\Lambda$CDM due to the AdS phase would result in a lower $\theta_s$, which is then compensated by a higher $H_0$. This, however, is achieved without invoking any (theoretically problematic) phantom DE component. Moreover, for $z<z_{\dagger}$ the shape of the expansion history is identical to that of $\Lambda$CDM (up to possible shifts in parameters such as $\Omega_m$), automatically ensuring consistency with probes which are only sensitive to the shape of the expansion history, e.g.\ unanchored SNeIa or uncalibrated BAO. This is completely different from the case of phantom DE, where the entire shape of the expansion history can be drastically modified, quickly leading to disagreement with the same datasets. For these reasons, compared to other late-time modifications, $\Lambda_s$CDM stands as a more competitive contender to partially alleviate cosmological tensions.

\subsection{Negative cosmological constant}
\label{subsec:ncc}

The $\Lambda_s$CDM model discussed earlier is an example of DE sector featuring negative energy densities. Although this may appear strange at first, it is worth recalling that negative energy densities arise frequently in string theory-inspired frameworks. This is mostly because AdS vacua are common within string theory (due in part to reasons connected to the AdS/CFT correspondence), whereas it is conjectured that stable dS vacua cannot be constructed within string theory~\cite{Danielsson:2018ztv}: this conjecture, and more generally the swampland program, is taken seriously in the theory community, and has important cosmological consequences~\cite{Obied:2018sgi,Agrawal:2018own,Achucarro:2018vey,Garg:2018reu,Kehagias:2018uem,OColgain:2018czj,Kinney:2018nny,Ooguri:2018wrx,Palti:2019pca,Colgain:2019joh,Odintsov:2020zkl,Banerjee:2020xcn,Trivedi:2020wxf,Oikonomou:2020oex,Lehnert:2025izp}. Motivated by these insights, numerous studies in recent years have considered the possibility that the DE sector may include a negative cosmological constant (nCC) $\Lambda<0$, corresponding to an AdS vacuum. By itself, such a component obviously cannot drive cosmic acceleration, so an additional evolving component with positive energy density on top is required: such a string-inspired combination can be viewed as a quintessence field rolling down a potential whose minimum is negative (i.e.\ an AdS vacuum). This class of models, and more generally models featuring negative energy densities in the DE sector, have been extensively studied over the past years in light of a wide variety of cosmological and astrophysical datasets (see for instance Refs.~\cite{Poulin:2018zxs,Dutta:2018vmq,Ruchika:2020avj,Calderon:2020hoc,Hogas:2021fmr,Malekjani:2023ple,VanRaamsdonk:2023ion,Colgain:2024clf,VanRaamsdonk:2024sdp,Wang:2024hwd,Menci:2024hop,Mukherjee:2025myk,VanRaamsdonk:2025wvj,Wang:2025dtk,Chakraborty:2025yuo}).~\footnote{A related yet distinct scenario studied in the context of the Hubble tension involves an AdS phase in the pre-recombination Universe~\cite{Ye:2020btb,Ye:2020oix,Ye:2021nej,Jiang:2021bab,Ye:2021iwa,Jiang:2022uyg,Peng:2025tqt}.}

To be concrete, we now need to specify further details regarding the evolving DE component with positive energy density (on top of the nCC), which in what follows we generically refer to as ``\textit{x}''. For simplicity, we assume that the EoS of this component is a constant $w_x \neq -1$.~\footnote{We note that the case $w_x=-1$ is obviously trivial, since it leads to an overall positive cosmological constant, given by the sum of the negative cosmological constant and the positive cosmological constant on top, and is therefore equivalent to $\Lambda$CDM.} With these assumptions, the Friedmann equation is modified to the following:
\begin{align}
H^2(z) &= H_0^2\left [ \Omega_r(1+z)^4+\Omega_m(1+z)^3 \right.\\
&\left. +\, \Omega_{\Lambda}+\Omega_x(1+z)^{3(1+w_x)} \right ] \,,
\label{eq:friedmannncc}
\end{align}
where $\Omega_{\Lambda}<0$ and $\Omega_x>0$ are the density parameters of the nCC and the evolving DE component respectively. Since we are assuming a spatially flat Universe, $\Omega_r+\Omega_m+\Omega_{\Lambda}+\Omega_x=1$ holds, and it is natural to identify the density parameter of the whole DE sector as being $\Omega_{\text{DE}}=\Omega_x+\Omega_{\Lambda}$. We note that $\Omega_{\text{DE}} \simeq 1-\Omega_m$ is positive (the approximation follows from having neglected the radiation component, which is subdominant at late times), and is in fact $\Omega_{\text{DE}} \simeq 0.7$ in order to agree with background cosmological observations. Therefore, provided $\Omega_{\Lambda}$ is sufficiently negative, $\Omega_x$ can actually exceed $1$. A sufficiently negative $\Omega_{\Lambda}$ needs to be compensated by an EoS of the $x$-component $w_x$ which moves towards the phantom regime, in order for cosmic acceleration to take place at present time. Late-time background cosmological observations alone set a rather loose limit on $\vert \Omega_{\Lambda} \vert \lesssim {\cal O}(10)$~\cite{Visinelli:2019qqu}. However, the long lever arm which results from including geometrical information from the CMB tightens the previous limit by about an order of magnitude, to $\vert \Omega_{\Lambda} \vert \lesssim {\cal O}(1)$~\cite{Sen:2021wld}.

The effective EoS of the DE sector ($\Lambda$ plus $x$-component) can easily be calculated, and is given by the following expression:
\begin{equation}
w_{\text{ncc}}(z)=\frac{\Omega_x(1+z)^{3(1+w_x)}w_x-\Omega_{\Lambda}}{\Omega_x(1+z)^{3(1+w_x)}+\Omega_{\Lambda}}\,.
\label{eq:wncc}
\end{equation}
From Eq.~(\ref{eq:wncc}) we notice that, for sufficiently negative $\Omega_{\Lambda}$, the effective EoS can be phantom even if $w_x$ itself is quintessence-like. Following a similar argument as that provided for $\Lambda_s$CDM, the fact that the nCC model lowers the $z \lesssim 2$ redshift rate relative to $\Lambda$CDM (as is evident from its potential effective phantom behaviour) is the reason why such a model has been regarded of interest in the Hubble tension literature. Finally, while we have considered an $x$-component with constant EoS $w_x$, various works have explored the case where the $x$-component is dynamical, for instance modelled by the CPL parametrization: while this constitutes a natural extension of our scenario, it does not introduce significant conceptual differences. In what follows, we refer to the model defined by Eq.~(\ref{eq:friedmannncc}) as nCC model.

\section{Datasets and methodology}
\label{sec:datasets}

To test whether a possible BAO miscalibration could rescue late-time solutions to the Hubble tension, we compare the late-time models discussed in Sec.~\ref{sec:models} against various combinations of late-time datasets and/or priors, which we now discuss.
\begin{itemize}
\item \textbf{Standard BAO measurements} -- BAO measurements from the completed Baryon Oscillation Spectroscopic Survey (BOSS) extended BOSS (eBOSS) survey programs of the Sloan Digital Sky Survey (SDSS)~\cite{eBOSS:2020yzd}. In particular, we adopt measurements from the Main Galaxy Sample (MGS) at effective redshift $z_{\text{eff}}=0.15$, the BOSS DR12 galaxy samples at $z_{\text{eff}}=0.38$ and $0.51$, the eBOSS Luminous Red Galaxies (LRG) sample at $z_{\text{eff}}=0.70$, the eBOSS Emission Line Galaxies (ELG) sample at $z_{\text{eff}}=0.85$, and the eBOSS quasar (QSO) sample at $z_{\text{eff}}=1.48$. For reasons we will elaborate on later in Sec.~\ref{subsubsec:bao}, in order to be as conservative as possible, we have opted not to include BAO measurements from the Ly-$\alpha$ forest (Ly$\alpha$) and the Ly$\alpha$-QSO cross-correlation. This combination of measurements, which constitutes our baseline BAO dataset, is referred to as \textit{\textbf{BAO}}.
\item \textbf{Rescaled BAO measurements} -- Recall that, playing devil's advocate, we adopt the guiding hypothesis that fiducial cosmology assumptions in the standard BAO pipeline bias the above measurements such that the resulting $H_0$ is low. We undo this bias and therefore ``correct'' the above standard BAO measurements by rescaling them as indicated in Eqs.~(\ref{eq:lambda}). We set the scaling parameter $\lambda$ appearing in Eqs.~(\ref{eq:lambda}) to $\lambda=73.0/68.5 \approx 1.06$, for the reasons outlined at the end of Sec.~\ref{sec:bao}. The actual BAO measurements of $D_V/r_d$, $D_M/r_d$, and $D_H/r_d$ are therefore rescaled by $1/\lambda \approx 0.94$, see Eq.~(\ref{eq:lambdad}). We make the conservative assumption of not modifying the covariance matrix, i.e.\ our working hypothesis is that fiducial cosmology assumptions to not lead to significant biases in the uncertainties of the BAO measurements, but only in their central values. We refer to this set of rescaled BAO measurements as \textit{\textbf{BAOr}}.
\item \textbf{Unanchored SNeIa} -- We adopt the \textit{PantheonPlus} SNeIa catalog, which consists of 1701 light curves for 1550 unique SNeIa~\cite{Brout:2022vxf}. We only use SNeIa in the redshift range $0.01<z<2.26$, and do not include the SH0ES calibration, making these SNeIa effectively relative distance indicators, sensitive to $E(z)$. The reason for choosing the \textit{PantheonPlus} catalog over other available ones (such as the \textit{DES-Y5} and/or \textit{Union3} compilations) is discussed in Sec.~\ref{subsubsec:sneia}. We refer to this dataset as \textbf{\textit{PP}}.
\item \textbf{Compressed CMB Likelihood} -- Since the models we consider only affect the late-time expansion history, their impact on the CMB anisotropies is indirect and manifests mostly at the geometrical level, through changes in distance measures, rather than altering the perturbation dynamics or the physics of recombination. These considerations lead us to adopt a compressed likelihood approach for CMB data which, while not the standard in all cosmological analyses, is well-motivated and not uncommon in this context, given that it greatly simplifies the required computations (see e.g.\ Refs.~\cite{Lin:2021sfs,Lynch:2025ine} for examples of recent works adopting this approach). In our case, we compress the CMB information in a $2 \times 2$ likelihood for the parameter vector $\mathbf{v}=\{\theta_s^{-1}(z_{\star}),\omega_b\}$, where $\theta_s^{-1}(z_{\star})$ is the angular size of the sound horizon at recombination, controlling the location of the acoustic peaks in the CMB. We model the likelihood as a bivariate Gaussian with mean $\mathbf{\mu}$ and covariance matrix $\mathbf{\mathcal{C}}$ given by the following:
\begin{equation}
\begin{split}
\textbf{v} &=
\begin{pmatrix}
94.3342 \\ 0.0224
\end{pmatrix}\,,\\
C_\textbf{v} &= 10^{-8} \times
\begin{pmatrix}
2.2280 & -0.0151\\-0.0151 & 0.0007931
\end{pmatrix}\,.
\end{split}
\label{eq:compressed}
\end{equation}
In practice, the $\theta_s^{-1}(z_{\star})$ part of the likelihood treats the CMB as a BAO measurement at $z_{\star}=1100$. The above mean vector and covariance matrix have been obtained by analyzing the public \textit{Planck} legacy chains, considering the \texttt{TTEEE}+\texttt{lowl}+\texttt{lowE}+\texttt{lensing} dataset combination, within the $\Lambda$CDM model.~\footnote{This is referred to as \texttt{base\_plikHM\_TTTEEE\_lowl\_lowE\_lensing} in the \textit{Planck} parameter tables.} In practice, these values are very stable against choices of different full CMB dataset combinations and/or cosmological models, given that they are related to the position of the first acoustic peak for $\theta_s^{-1}(z_{\star})$, and to the relative height of the odd and even peaks for $\omega_b$, features which are very well measured in \textit{Planck} independently of the assumed cosmological model and choice of subset of \textit{Planck} CMB data. This treatment of the compressed CMB likelihood differs from other choices in the literature, a difference which we will comment on in Sec.~\ref{subsubsec:cmb}. We refer to this likelihood as $\boldsymbol{P18^{\star}}$.

\item \textbf{Prior on $\boldsymbol{\Omega_m}$} -- We consider a Gaussian prior on the matter density parameter $\Omega_m = 0.30 \pm 0.03$. The rationale behind this prior will be discussed in detail in Sec.~\ref{subsubsec:omegam}. We refer to this prior as $\boldsymbol{\Omega_m^{\boldsymbol{\mathit{P}}}}$.
\end{itemize}
Since the aforementioned datasets constrain background quantities, we restrict our analysis to background parameters. In particular we ignore the $\Lambda$CDM parameters $A_s$, $n_s$, and $\tau$, which play no role in our discussion, and consider the reduced parameter basis spanned by $\Omega_m$, $\omega_b$, $h$, and extra degrees of freedom in models beyond $\Lambda$CDM.

We sample the posterior distributions of our cosmological parameters using Monte Carlo Markov Chain (MCMC) methods, adopting the cosmological MCMC sampler \texttt{SimpleMC}.~\footnote{This code was first presented and used in the seminal Ref.~\cite{BOSS:2014hhw} by the BOSS collaboration, and is available at \url{https://github.com/ja-vazquez/SimpleMC}.} The code can be used to perform parameter estimation against datasets for which only the background expansion history matters, which is precisely the case for the measurements used here. We set wide, flat priors on all the cosmological parameters, verifying a posteriori that our posteriors are not affected by the choice of lower and upper prior boundaries. We assess the convergence of our MCMC chains using the Gelman-Rubin $R-1$ parameter~\cite{Gelman:1992zz}, with $R-1<0.01$ required for our chains to be considered converged. Our chains are subsequently analyzed via the \texttt{GetDist} package~\cite{Lewis:2019xzd}.

\subsection{Discussion on choice, treatment, and combinations of datasets}
\label{subsec:treatment}

Some clarifications are in order for what concerns the choice and treatment of certain datasets and likelihoods used in our analysis, especially regarding the compressed CMB likelihood and $\Omega_m$ prior. We now turn to a brief discussion of these aspects, before further discussing the logic behind the dataset combinations we adopt, with each combination selected to highlight a specific aspect of the problem, particularly the key role of unanchored SNeIa measurements.

\subsubsection{(Standard or rescaled) BAO}
\label{subsubsec:bao}

As mentioned earlier, our (standard and rescaled) BAO datasets do not include Ly$\alpha$ and Ly$\alpha$-QSO cross-correlation-based distance measurements~\cite{BOSS:2017fdr,BOSS:2017uab}. One key reason behind this choice is that these measurements indicate slight deviations from $\Lambda$CDM which are not clearly related to the Hubble tension, and could therefore complicate our interpretation. This is especially true since these measurements tend to indicate a lower expansion rate compared to what is expected within $\Lambda$CDM, effectively easing the challenge for late-time modifications attempting to solve the Hubble tension (as per our earlier discussion in Sec.~\ref{sec:models}). By excluding these datapoints we set a harder, but much more conservative, task for our late-time modifications, requiring them to resolve the Hubble tension without relying on potentially favorable features in these measurements.

Similar considerations apply to the DESI BAO measurements~\cite{DESI:2025zgx}, which appear to indicate $2$-$3\sigma$ deviations from $\Lambda$CDM at $z \lesssim 2$, as extensively studied in the literature, see e.g.\ Refs.~\cite{Luongo:2024fww,Cortes:2024lgw,Colgain:2024xqj,Carloni:2024zpl,Giare:2024smz,Yang:2024kdo,Dinda:2024kjf,Ramadan:2024kmn,Mukherjee:2024ryz,Notari:2024rti,Li:2024qso,Ye:2024ywg,Giare:2024gpk,Dinda:2024ktd,Jiang:2024viw,Wolf:2024eph,RoyChoudhury:2024wri,Giani:2024nnv,Wolf:2025jlc,Colgain:2025nzf,Wolf:2025jed,RoyChoudhury:2025dhe,Giani:2025hhs,Mukherjee:2025ytj,Ling:2025lmw,Ozulker:2025ehg,Chaudhary:2025pcc,Chaudhary:2025uzr,Jia:2025poj,RoyChoudhury:2025iis,Chaudhary:2025bfs,Capozziello:2025lor,Li:2025vuh,Capozziello:2025qmh}. Although these deviations are not directly tied to the Hubble tension (at least not in an obvious way), they potentially provide more room for late-time modifications to help alleviate it (see for instance the discussion in Ref.~\cite{Poulin:2024ken}). Additionally, DESI BAO data indicate a higher value of $r_dH_0$ compared to SDSS measurements~\cite{Jiang:2024xnu}, generally making it easier for models to address the Hubble tension. Following the same logic as above, we therefore exclude these measurements in order not to mix the ``DESI tension'' (also referred to as CMB-BAO tension in the literature) with the Hubble tension, which could obscure our interpretation. This choice also imposes a stricter and more conservative challenge for our models. Nevertheless, in Appendix~\ref{sec:appendix} we provide a brief assessment of the impact of DESI data, focusing on the $w$CDM model.

\subsubsection{Unanchored SNeIa}
\label{subsubsec:sneia}

As explained earlier, we make use of unanchored SNeIa from the \textit{PantheonPlus} sample. However, the more recent \textit{DESY5} and \textit{Union3} samples are also available. The choice to not adopt these samples follows the same reasoning as above: to avoid mixing potential tensions or systematics that could complicate the interpretation of our analysis. As extensively studied in the literature,  combining these SNeIa samples with BAO measurements tends to strengthen indications of late-time deviations from $\Lambda$CDM, especially when DESI BAO data are included. Furthermore, in the case of \textit{DESY5}, potential calibration offsets (particularly evident when comparing SNeIa common to \textit{PantheonPlus}) have been reported~\cite{Efstathiou:2024xcq,Huang:2025som,Ormondroyd:2025phk,Dinda:2025hiu,Lopez-Hernandez:2025lbj}, and shown to favor deviations from $\Lambda$CDM (however, see also Ref.~\cite{DES:2025tir}). Until these issues are better understood, and at the same time impose a harder challenge on our models, we therefore opt to only include the \textit{PantheonPlus} SNeIa sample, which shows no compelling evidence for late-time deviations from $\Lambda$CDM, and remains consistent with the best-fit \textit{Planck} $\Lambda$CDM cosmology.

Our choice of not including the SH0ES calibration for the \textit{PantheonPlus} sample is guided by a similar reasoning. Including this calibration would automatically raise the inferred value $H_0$ (typically by $1\,{\text{km}}/{\text{s}}/{\text{Mpc}}$ or sometimes even more, although the details depend on the specific model), seemingly easing the challenge for any late-time modification, \textit{regardless of the model's intrinsic ability to address the Hubble tension}. To remain as conservative as possible, we therefore opt to not calibrate our SNeIa data.

\subsubsection{Compressed CMB likelihood}
\label{subsubsec:cmb}

As argued earlier, we use a compressed CMB likelihood since the models considered affect only the late-time expansion history, with primarily geometrical effects on the CMB. Moreover, the full CMB spectrum exhibits several minor anomalies that are not obviously geometrical in origin (e.g.\ lensing anomaly, lack of power on large angular scales), and which may partly reflect residual systematics. Nevertheless, these features may affect the inferred values of parameters related to the DE sector, potentially suggesting non-standard behaviour for the latter even when none is truly present. An example is the apparent $>2\sigma$ preference for phantom DE from \textit{Planck} data alone, extensively investigated by some of us in Ref.~\cite{Escamilla:2023oce}, and traced at least partly to the anomalous lack of large-scale power, which phantom DE can help explain via changes to the latest ISW effect. To avoid confounding such effects with the much cleaner geometrical information, we restrict our analysis to the latter, making use of the compressed CMB likelihood.

Our approach to compress the CMB measurements is slightly different compared to other approaches in the literature. The standard approach is to make use of the two so-called shift parameters, $R$ and $\ell_a$, which are defined as follows~\cite{Wang:2007mza}:
\begin{align}
R &\equiv \sqrt{\Omega_mH_0^2}D_M(z_{\star})\,,
\label{eq:cmbr} \\
\ell_a &\equiv \pi\frac{D_M(z_{\star})}{r_s(z_{\star})}\,,
\label{eq:cmbla}
\end{align}
where we recall that $D_M(z_{\star})$ is the transverse comoving distance to the last-scattering surface, and $r_s(z_{\star})$ is the comoving sound horizon at recombination. The CMB information is then usually compressed in a $3 \times 3$ multivariate Gaussian likelihood for $R$, $\ell_a$, and $\omega_b$. We see from Eq.~(\ref{eq:cmbla}) that, up to a factor of $\pi$, $\ell_a$ is essentially $\theta_s^{-1}(z_{\star})$. On the other hand, once $r_s(z_{\star})$ is calibrated (as we will do, assuming that $\Lambda$CDM holds prior to recombination), and therefore $D_M(z_{\star})$ is determined from $\ell_a$, $R$ essentially sets a prior on $\omega_m$. Including information on $\omega_b$, and leaving aside neutrinos, this essentially maps $R$ onto $\omega_c$. This $3 \times 3$ compression strategy for the parameter vector $\mathbf{v}=\{R,\ell_a,\omega_b\}$ has been widely adopted in the literature, and is referred to as ``Wang-Wang'' parametrization in \texttt{SimpleMC}~\footnote{This comes from the initials of the authors of Ref.~\cite{Wang:2013mha}, although we note that the terminology ``Wang-Mukherjee'' would be more appropriate~\cite{Wang:2007mza}.}. We will discuss in a moment why such a compression strategy, while widely adopted in the literature, is potentially problematic in the context of the Hubble tension.

A different but conceptually similar compression strategy has been adopted in other works, most notably in the seminal Ref.~\cite{BOSS:2014hhw}. Here the CMB information is also compressed in a $3 \times 3$ multivariate Gaussian likelihood, this time for the parameter vector $\mathbf{v}=\{\theta_s^{-1}(z_{\star}),\omega_b,\omega_c\}$. This is in fact the compressed CMB likelihood recommended in \texttt{SimpleMC}, and has also found widespread use. As per our above discussion, the information content of the $\mathbf{v}=\{\theta_s^{-1}(z_{\star}),\omega_b,\omega_c\}$ compression is similar to the $\mathbf{v}=\{R,\ell_a,\omega_b\}$ one, since $\ell_a$ and $R$ can be mapped onto $\theta_s^{-1}(z_{\star})$ and $\omega_c$ respectively.

We take issue with these compression strategies primarily because they incorporate information on $\omega_c$. The standard reasoning behind this choice is that $\omega_c$ can be determined from the CMB temperature power spectrum through its impact on the early integrated Sachs-Wolfe (eISW) effect, the ``potential envelope'' effect, and the gravitational lensing-induced smoothing of the acoustic peaks, with limited sensitivity to post-recombination departures from $\Lambda$CDM. This is only partially true (see e.g.\ Ref.~\cite{Lynch:2025ine} for a recent discussion), and becomes problematic in the context of the Hubble tension. In fact it has recently been argued, both independently~\cite{Poulin:2024ken} and in work involving some of the present authors~\cite{Pedrotti:2024kpn}, that once $\Omega_m$ and $\omega_b$ are well calibrated (e.g.\ from unanchored SNeIa and BBN respectively), any model attempting to solve the Hubble tension and therefore raise $H_0$ will necessarily raise $\omega_c$ as well. This conclusion is true regardless of whether the model in question operates before or after recombination.~\footnote{In principle these concerns hold also for the compressed likelihood of \href{https://github.com/cmbant/PlanckEarlyLCDM}{github.com/cmbant/PlanckEarlyLCDM}~\cite{Lemos:2023xhs}. This provides constraints on early-Universe parameters, including $\omega_c$, (almost) independently of the late-time Universe, but explicitly assumes an early-$\Lambda$CDM Universe.} As shown by some of us in Ref.~\cite{Pedrotti:2024kpn}, a fractional increase in $\delta H_0/H_0$ needs to be accompanied by a fractional increase in $\omega_c$ approximately given by $\delta \omega_c/\omega_c \approx 2.38\,\delta H_0/H_0$. Taking the hypothetical case where a model is able to fully solve the Hubble tension, i.e.\ bring $H_0$ from $\approx 68\,{\text{km}}/{\text{s}}/{\text{Mpc}}$ to $\approx 74\,{\text{km}}/{\text{s}}/{\text{Mpc}}$, one finds $\delta H_0/H_0 \approx 9\%$, and therefore $\delta \omega_c/\omega_c \approx 21\%$. This behaviour has been explicitly observed in a number of early-time new physics models, including early dark energy, and can help partially absorb the excess eISW power these models would otherwise predict~\cite{Hill:2020osr,Vagnozzi:2021gjh}.

The above arguments suggest that the value of $\omega_c$ is not stable against precisely those modifications of $\Lambda$CDM which attempt to address the Hubble tension. Within this context, imposing a strong prior on $\omega_c$ is therefore not an appropriate compression choice. We therefore opt for reducing the standard $3 \times 3$ compressed CMB likelihood to a $2 \times 2$ version, as in Eq.~(\ref{eq:compressed}). We restrict the likelihood to retain information only on quantities which are robustly and (quasi-)model-independently constrained, choosing as compression parameter vector $\mathbf{v}=\{\theta_s^{-1}(z_{\star}),\omega_b\}$ -- note that, as per our earlier discussion, this would be entirely equivalent to a $\mathbf{v}=\{\ell_a,\omega_b\}$ compression. As argued earlier, both parameters are inferred from the CMB temperature power spectrum to very high accuracy, relying primarily on features inherent in the data (such as the position of the first acoustic peak and the relative heights of odd and even peaks) -- as a result, their determination displays very minimal dependence on the assumed cosmological model.~\footnote{The only potential exception for what concerns $\theta_s^{-1}(z_{\star})$ is related to models such as the strongly interacting neutrino one, where the phase shift induced by free-streaming neutrinos is partially suppressed~\cite{Kreisch:2019yzn}. However, despite hints for potential self-interacting neutrino modes from galaxy clustering, these models are now too tightly constrained from CMB polarization data to be able to significantly alleviate the Hubble tension~\cite{RoyChoudhury:2020dmd,Brinckmann:2020bcn,RoyChoudhury:2022rva,Camarena:2023cku,Bostan:2023ped,Camarena:2024daj}.}

The physical baryon density $\omega_b$ is particularly robust in this regard, for two reasons. Firstly, as demonstrated in Ref.~\cite{Motloch:2020lhu}, it can be determined from CMB data through at least eight independent physical effects: each of these effects constrains $\omega_b$ separately, and all eight determinations are found to be in excellent agreement. Next, $\omega_b$ is tightly constrained by BBN, through its impact on the predicted abundances of light elements (especially deuterium), with the BBN determination being competitive and in excellent agreement with the CMB inference. Taken together, these considerations support the robustness of the $2 \times 2$ compression strategy we adopt, given the aims and context of this work. A comparative study on different compression strategies and their robustness when testing late-time cosmological models is the subject of work in progress by some of us, which we hope to report on soon~\cite{compressedinpreparation}.

\subsubsection{$\Omega_m$ prior}
\label{subsubsec:omegam}

We have chosen to give up on $\omega_c$ information because its dependence on the assumed cosmological model becomes especially significant for models attempting to solve the Hubble tension. As discussed in Sec.~\ref{sec:bao}, this is not a problem for calibrating the sound horizon once $\omega_b$ is known, and therefore extracting $H_0$ from BAO measurements, if information on $\Omega_m$ (which is much more stable than $\omega_c$) is provided, see Eq.~(\ref{eq:h0rd}). In particular from Eq.~(\ref{eq:soundhorizonomegam}) we clearly see why \textit{a)} $r_dH_0$ is not perfectly degenerate with $\Omega_m$, but there is a small indirect dependence on $h$ which is reflected in Eq.~(\ref{eq:h0rd}), and especially \textit{b)} trading (strongly model-dependent) information on $\omega_c$ for more robust information on $\Omega_m$ still allows for $H_0$ to be determined from BAO data, once $\omega_b$ (and of course $\omega_r$) is known. In fact, when BAO and BBN are combined, excluding any form of CMB data (full or compressed), this is essentially the way $H_0$ is inferred from this combination (see Refs.~\cite{Addison:2013haa,Addison:2017fdm,Cuceu:2019for,Schoneberg:2019wmt,Schoneberg:2022ggi} for further discussions).  While Eq.~(\ref{eq:h0rd}) holds for $\Lambda$CDM, it can be suitably adapted for late-time modifications to $\Lambda$CDM to account for the impact of the beyond-$\Lambda$CDM parameters on $D_M(z_{\star})$, as long as the sound horizon is unmodified from Eq.~(\ref{eq:rd}). With these considerations in mind, we have therefore chosen to complement the $2 \times 2$ compressed CMB likelihood by adding a prior on $\Omega_m$. Aside from allowing us to calibrate $r_dH_0$ as shown in Eqs.~(\ref{eq:h0rd},\ref{eq:soundhorizonomegam}), information on $\Omega_m$ is particularly important as it controls the shape of the late-time expansion history, which is especially helpful to stabilize constraints on beyond-$\Lambda$CDM parameters.

Two final comments are in order concerning the choice of adopting a prior on $\Omega_m$. First of all, we choose to use the same prior (which we have yet to specify) both when adopting the \textit{BAO} (standard) and \textit{BAOr} (rescaled) measurements. Is this consistent? To put it differently, would one expect rescaled BAO measurements to favor a different value of $\Omega_m$? The answer is negative, i.e.\ the two sets of BAO measurements should in fact indicate the same values of $\Omega_m$. To understand why, let us recall what makes BAO measurements sensitive to $\Omega_m$: at different redshifts $z_{\text{eff}}$, these measurements give different degeneracy directions in the $\Omega_m$-$r_dH_0$ plane. This implies that, with a single BAO measurement at a single $z_{\text{eff}}$, $\Omega_m$ and $r_dH_0$ are completely degenerate and cannot be disentangled. The fact that these degeneracy directions change (and in particular steepen) as $z_{\text{eff}}$ increases, however, implies that multiple BAO measurements over a sufficiently wide range of redshifts can in fact constrain $\Omega_m$, especially if BAO measurements at sufficiently high redshift ($z_{\text{eff}}$) are included.~\footnote{The BAO-only determination of $\Omega_m$ is of course not competitive with the CMB determination. However, including CMB information on $\theta_s(z_{\star})$, i.e.\ treating the CMB as a BAO measurement at $z_{\text{eff}}=z_{\star}$, significantly improves this determination and makes it competitive with that from the full CMB, the reason being the extremely long lever arm provided by the geometrical CMB datapoint~\cite{Lin:2019htv,Jedamzik:2020zmd,Lin:2021sfs}.} This is because $\Omega_m$ determines the \textit{relative} change of the expansion rate with redshift, $dE(z)/dz$: therefore, $\Omega_m$ can be constrained from uncalibrated BAO alone, provided these are measured over a wide redshift range. In other words, the information on $\Omega_m$ comes from the \textit{shape} of the BAO measurements, which are being used as relative distance indicators (analogously to unanchored SNeIa). This makes it clear that the BAO-only determination of $\Omega_m$ is insensitive to the absolute BAO scale (or equivalently to the calibration). In fact, a constant rescaling of the BAO measurements such as the one described by Eqs.~(\ref{eq:lambda}) would not change the overall shape of the BAO measurements, and therefore their properties as relative distance indicators. We would therefore expect the \textit{BAO} and \textit{BAOr} datasets to prefer the same values of $\Omega_m$.

\begin{figure}[!ht]
\centering
\includegraphics[width=0.9\linewidth]{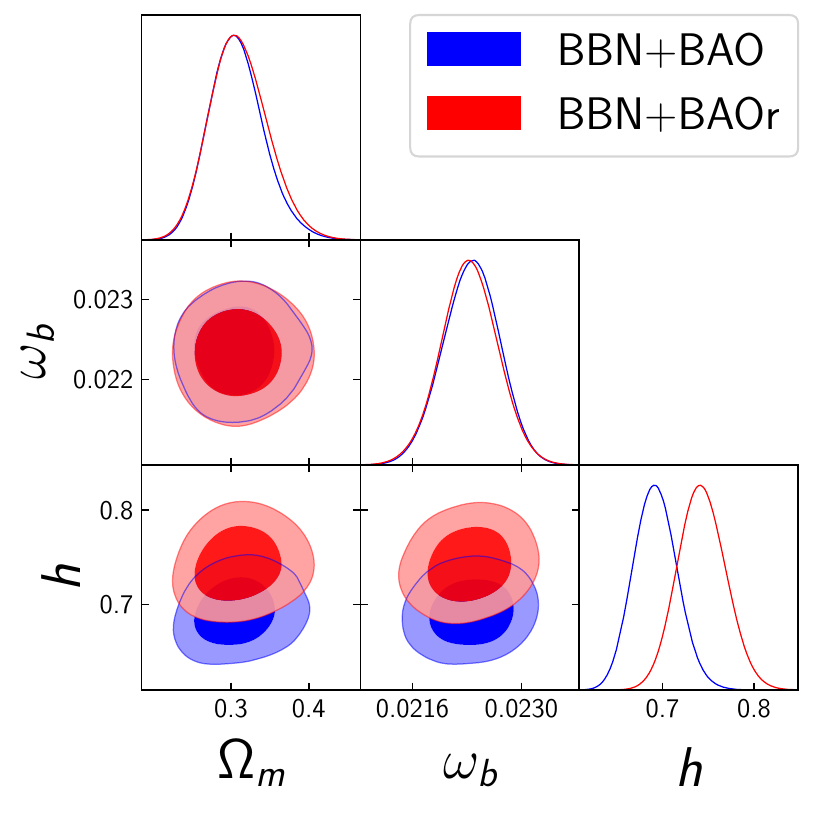}
\caption{Triangular plot showing 2D joint and 1D marginalized posterior probability distributions for the fractional matter density parameter $\Omega_m$, the physical baryon density $\omega_b$, and the reduced Hubble constant $h \equiv H_0/(100\,{\text{km}}/{\text{s}}/{\text{Mpc}})$, obtained within the baseline $\Lambda$CDM model by combining a BBN prior on $\omega_b$ with standard (blue curves) or rescaled (red curves) BAO measurements. We clearly see that the effect of rescaling the BAO measurements transfers entirely to a rescaling of $H_0$, which controls the overall BAO calibration, whereas the inferred value of $\Omega_m$ is unchanged, since this parameter controls the (loosely constrained) shape of BAO measurements.}
\label{fig:baobaor}
\end{figure}

We demonstrate the above point explicitly in Fig.~\ref{fig:baobaor}. Working within the same set of parameters discussed earlier ($\Omega_m$,$\omega_b$,$h$), we consider the combination of a BBN prior on $\omega_b = 0.02233 \pm 0.00036$ (resulting from an improved determination of the deuterium burning rate from the LUNA experiment~\cite{Mossa:2020gjc}, see Ref.~\cite{Schoneberg:2024ifp} for a more recent determination) and BAO measurements, in the form of either the standard \textit{BAO} dataset (blue contours) or the rescaled \textit{BAOr} one (red contours). We see that in both cases the inferred value of $\Omega_m$ is identical, the reason being that the shape of the \textit{BAO} and \textit{BAOr} measurements is the same. On the other hand, the value of $h$ shifts significantly, increasing in the \textit{BAOr}+\textit{BBN} case as expected. While we could have made this point considering BAO measurements alone, our choice to calibrate them with the BBN prior on $\omega_b$ was deliberately made to stress the significant difference between parameters which are sensitive to the overall calibration ($H_0$) versus shape ($\Omega_m$).

Having established that setting a prior on $\Omega_m$ is reasonable in this context, our final comment concerns the actual value of this (Gaussian) prior. While it is true that $\Omega_m$ is well constrained by late-time probes, there is a mild level of disagreement for what concerns its actual value (see e.g.\ Refs.~\cite{Sakr:2023hrl,Baryakhtar:2024rky,Colgain:2024ksa,Loverde:2024nfi,Colgain:2024mtg,Lee:2025hjw,Lee:2025kbn}). For instance, while the earlier \textit{Pantheon} SNeIa dataset preferred $\Omega_m \sim 0.3$, this increased to $\Omega_m \sim 0.33$ in \textit{PantheonPlus}, whereas \textit{DES-Y5} and \textit{Union3} appear to indicate $\Omega_m \sim 0.35$~\cite{DES:2024jxu} and $\Omega_m \sim 0.36$~\cite{Rubin:2023ovl} respectively. On the other hand, \textit{DESI} BAO measurements indicate systematically lower values of $\Omega_m \sim 0.29$~\cite{DESI:2024mwx,DESI:2025zgx}, or potentially as low as $0.27$, as reported in the recent external independent reanalysis of DESI data of Ref.~\cite{Chudaykin:2025aux}. Other less standard probes (e.g.\ cosmic chronometers, galaxy number counts, and so on) also exhibit some scatter in the inferred value of $\Omega_m$, although all these inferences fall within the range $0.25 \lesssim \Omega_m \lesssim 0.35$. While some tension exists among the values of $\Omega_m$ inferred from different late-time probes, it is nonetheless clear that $\Omega_m \sim 0.3$ provides a broadly consistent central value. Extreme and unreasonable values such as $\Omega_m \lesssim 0.25$ or $\Omega_m \gtrsim 0.35$ are strongly disfavored by essentially all available datasets. To remain conservative yet reasonable, while reflecting the scatter among different determinations, we adopt a Gaussian prior on $\Omega_m=0.30 \pm 0.03$. This choice is not anchored to any specific dataset, but rather reflects the collective trend from a broad range of late-time cosmological measurements, including the less standard ones mentioned earlier. It also reflects the values of $\Omega_m$ obtained in the more model-independent ``uncalibrated cosmic standards'' approach (see e.g.\ Fig.~3 of Ref.~\cite{Lin:2019htv} and Tab.~2 of Ref.~\cite{Wang:2025mqz}). We note that the standard deviation of our prior is deliberately conservatively large (e.g.\ up to a factor of $\approx 5$ larger compared to the uncertainty on $\Omega_m$ from combined BAO and CMB measurements), in order to reflect the scatter mentioned above, and at the same avoid over-constraining models that genuinely predict modest shifts in $\Omega_m$.

\subsubsection{Dataset combinations}
\label{subsubsec:combinations}

We now discuss the different dataset combinations we consider, emphasizing the rationale behind our choices. Our baseline dataset combination includes the \textit{P18}$^{\star}$ compressed CMB likelihood and the $\Omega_m^{P}$ prior on $\Omega_m$. This reflects the crucial role of both \textit{a)} the long lever arm provided by the geometrical CMB information treated as a BAO measurement at $z=z_{\star}$, as well as \textit{b)} a consistent calibration for $\Omega_m$: these points become particularly important when considering late-time deviations from $\Lambda$CDM although, as we will see later, the \textit{PP} unanchored SNeIa dataset places extremely stringent constraints on these deviations. We refer to the \textit{P18}$^{\star}$+$\Omega_m^{P}$ combination as \textit{base}.

For each of the models considered, we essentially want to address the following three points/questions (deliberately phrased in a more informal tone):
\begin{enumerate}
\item \textit{\textbf{Focusing on a minimal dataset combination, what is the effect of the BAO rescaling? Is it enough to save late-time solutions to the Hubble tension?}} \\
This question is most naturally addressed by considering and comparing the \textit{base}+\textit{BAO} and \textit{base}+\textit{BAOr} dataset combinations.
\item \textit{\textbf{Say we ignore the role of BAO in the Hubble tension ``no-go theorem'', what is the status of late-time solutions? Is there a BAO-free benchmark combination which strongly constrains these?}} \\
This question is easily addressed by considering unanchored SNeIa, whose role in the Hubble tension is clearly underappreciated. A natural dataset combination to consider in this case is \textit{base}+\textit{PP}, which we will show places extremely strong constraints on late-time solutions.
\item \textit{\textbf{It looks like unanchored SNeIa very strongly constrain late-time solutions, perhaps even more so than BAO. But what happens if we consider SNeIa and BAO together? Is their impact still so strong? What is the effect of the BAO rescaling now?}} \\
This question touches a scenario somewhat intermediate between 1.\ and 2., and is most naturally addressed by considering and comparing the \textit{base}+\textit{BAO}+\textit{PP} and \textit{base}+\textit{BAOr}+\textit{PP} dataset combinations. As we will see, the results reinforce the crucial role played by unanchored SNeIa measurements, regardless of the BAO rescaling.
\end{enumerate}
In summary, each of the models presented in Sec.~\ref{sec:models} is confronted against the following five combinations:
\begin{itemize}
\item \textit{base}+\textit{BAO}
\item \textit{base}+\textit{BAOr}
\item \textit{base}+\textit{PP}
\item \textit{base}+\textit{BAO}+\textit{PP}
\item \textit{base}+\textit{BAOr}+\textit{PP}
\end{itemize}
Our expectation, which our analysis will explicitly confirm, is that including unanchored SNeIa (the \textit{PP} dataset) imposes such strong constraints on the shape of the expansion history that it effectively overwhelms any shift in $H_0$ arising from the use of rescaled BAO measurements (see also Ref.~\cite{Zhou:2025kws}). Since there is no compelling a priori reason to exclude unanchored SNeIa data we conclude that, even setting aside BAO and their potential fiducial cosmology-related systematics, the SNeIa themselves place sufficiently tight constraints on the late-time expansion to rule out a viable loophole to the ``no-go theorem'', thereby precluding a late-time solution to the Hubble tension.

\section{Results}
\label{sec:results}

\begin{table*}[!t]
\centering
\renewcommand{\arraystretch}{1.2}
\scalebox{0.8}{
\begin{tabular}{?c?ccccc?}
\thickhline
\multicolumn{1}{?c?}{\textbf{Model / parameters}} & \multicolumn{5}{c?}{\textbf{Dataset: \textit{base}+}} \\
\thickhline
$\mathbf{\Lambda}$\textbf{CDM} & \textbf{\textit{BAO}} & \textbf{\textit{BAO}}+\textbf{\textit{PP}} & \textbf{\textit{PP}} & \textbf{\textit{BAOr}} & \textbf{\textit{BAOr}}+\textbf{\textit{PP}} \\ \hline
$H_0 \left [ {\text{km}}/{\text{s}}/{\text{Mpc}} \right ] $ & $68.42 \!\pm\! 1.28$ (2.8$\sigma$, 1.9\%) & $67.45 \!\pm\! 0.83$ (4.2$\sigma$, 1.2\%) & $66.96 \!\pm\! 0.99$ (4.2$\sigma$, 1.5\%) & $71.68 \!\pm\! 1.43$ (0.8$\sigma$, 1.9\%) & $68.80 \!\pm\! 0.86$ (3.1$\sigma$, 1.2\%) \\
$\Omega_m$ & $0.30\pm 0.02$ & $0.31\pm 0.01$ & $0.32\pm 0.02$ & $0.26 \pm 0.02$ & $0.29 \pm 0.01$ \\
$\omega_b$ & $0.02236\pm 0.00015$ & $0.02236\pm 0.00015$ & $0.02237\pm 0.00014$ & $0.02237\pm 0.00015$ & $0.02237\pm 0.00015$ \\ \hline
$\chi^2_{\min}$ & $7.61$ & $1412.28$ & $1403.91$ & $16.75$ & $1429.13$ \\
\thickhline

$\mathbf{w}$\textbf{CDM} & \textbf{\textit{BAO}} & \textbf{\textit{BAO}}+\textbf{\textit{PP}} & \textbf{\textit{PP}} & \textbf{\textit{BAOr}} & \textbf{\textit{BAOr}}+\textbf{\textit{PP}} \\ \hline
$w$ & $-0.97\pm 0.11$ & $-0.94\pm 0.04$ & $-0.93\pm 0.08$ & $-0.73 \pm 0.09$ & $-0.84 \pm 0.04$ \\ \hline
$H_0 \left [ {\text{km}}/{\text{s}}/{\text{Mpc}} \right ] $ & $68.51 \!\pm\! 2.28$ (1.8$\sigma$, 3.4\%) & $68.00 \!\pm\! 0.87$ (3.7$\sigma$, 1.3\%) & $68.22 \!\pm\! 1.12$ (3.2$\sigma$, 1.6\%) & $66.01 \!\pm\! 2.37$ (2.7$\sigma$, 3.7\%) & $69.11 \!\pm\! 0.91$ (2.8$\sigma$, 1.3\%) \\
$\Omega_m$ & $0.31\pm 0.02$ & $0.31\pm 0.01$ & $0.30\pm 0.03$ & $0.28 \pm 0.02$ & $0.27 \pm 0.01$ \\
$\omega_b$ & $0.02236\pm 0.00015$ & $0.02236\pm 0.00015$ & $0.02236\pm 0.00015$ & $0.02236\pm 0.00015$ & $0.02236\pm 0.00015$ \\ \hline
$\chi^2_{\min}$ & $7.38$ & $1410.27$ & $1402.84$ & $6.51$ & $1411.73$ \\
\thickhline

\textbf{CPL} & \textbf{\textit{BAO}} & \textbf{\textit{BAO}}+\textbf{\textit{PP}} & \textbf{\textit{PP}} & \textbf{\textit{BAOr}} & \textbf{\textit{BAOr}}+\textbf{\textit{PP}} \\ \hline
$w_0$ & $-0.93^{+0.19}_{-0.22}$ & $-0.92\pm 0.08$ & $-0.92\pm 0.11$ & $-0.54 \pm 0.17$ & $-0.85 \pm 0.05$ \\
$w_a$ & $-0.08^{+0.91}_{-0.66}$ & $-0.09^{+0.52}_{-0.45}$ & $-0.20^{+0.66}_{-0.50}$ & $-0.71 \pm 0.59$ & $ 0.18 \pm 0.31$ \\
\hline
$H_0 \left [ {\text{km}}/{\text{s}}/{\text{Mpc}} \right ] $ & $67.01 \!\pm\! 2.58$ (2.2$\sigma$, 3.9\%) & $67.01^{+1.22}_{-1.03}$ (3.8$\sigma$, 1.8\%) & $67.18 \!\pm\! 1.43$ (3.3$\sigma$, 2.1\%) & $64.72  \!\pm\! 2.11$ (3.6$\sigma$, 3.2\%) & $67.37 \!\pm\! 1.41$ (3.2$\sigma$, 2.1\%) \\
$\Omega_m$ & $0.30\pm 0.02$ & $0.31\pm 0.02$ & $0.31\pm 0.03$ & $0.29 \pm 0.02$ & $0.26 \pm 0.01$ \\
$\omega_b$ & $0.02236\pm 0.00015$ & $0.02236\pm 0.00015$ & $0.02236\pm 0.00015$ & $0.02236\pm 0.00015$ & $0.02236\pm 0.00015$ \\ \hline
$\chi^2_{\min}$ & $7.25$ & $1410.14$ & $1402.87$ & $6.12$ & $1411.71$ \\
\thickhline

\textbf{PEDE} & \textbf{\textit{BAO}} & \textbf{\textit{BAO}}+\textbf{\textit{PP}} & \textbf{\textit{PP}} & \textbf{\textit{BAOr}} & \textbf{\textit{BAOr}}+\textbf{\textit{PP}} \\ \hline
$H_0 \left [ {\text{km}}/{\text{s}}/{\text{Mpc}} \right ] $ & $71.22 \!\pm\! 1.43$ (1.1$\sigma$, 2.0\%) & $67.32 \!\pm\! 0.87$ (4.2$\sigma$, 1.3\%) & $66.09 \!\pm\! 0.87$ (5.1$\sigma$, 1.3\%) & $74.91 \!\pm\! 1.63$ (1.0$\sigma$, 2.1\%) & $68.31 \!\pm\! 0.88$ (3.5$\sigma$, 1.3\%) \\
$\Omega_m$ & $0.29\pm 0.02$ & $0.34\pm 0.01$ & $0.36\pm 0.01$ & $0.25 \pm 0.02$ & $0.33 \pm 0.01$ \\
$\omega_b$ & $0.02236\pm 0.00015$ & $0.02237\pm 0.00015$ & $0.02237\pm 0.00015$ & $0.02237\pm 0.00015$ & $0.02237\pm 0.00015$ \\ \hline
$\chi^2_{\min}$ & $11.36$ & $1429.55$ & $1410.18$ & $29.13$ & $1463.44$ \\
\thickhline  

\textbf{HDE} & \textbf{\textit{BAO}} & \textbf{\textit{BAO}}+\textbf{\textit{PP}} & \textbf{\textit{PP}} & \textbf{\textit{BAOr}} & \textbf{\textit{BAOr}}+\textbf{\textit{PP}} \\ \hline
$C$ & $0.82^{+0.12}_{-0.15}$ & $0.87^{+0.07}_{-0.08}$ & $0.91^{+0.12}_{-0.11}$ & $1.01^{+0.04}_{-0.12}$ & $1.03^{+0.03}_{-0.11}$ \\
\hline
$H_0 \left [ {\text{km}}/{\text{s}}/{\text{Mpc}} \right ] $ & $67.61^{+2.03}_{-2.52}$ (2.2$\sigma$, 3.3\%) & $66.93 \!\pm\! 0.87$ (4.5$\sigma$, 1.3\%) & $67.33^{+1.02}_{-1.21}$ (3.8$\sigma$, 1.6\%) & $68.61 \!\pm\! 1.49$ (2.4$\sigma$, 2.2\%) & $69.02 \!\pm\! 0.83$ (3.0$\sigma$, 1.2\%) \\
$\Omega_m$ & $0.30\pm 0.02$ & $0.30\pm 0.01$ & $0.29\pm 0.02$ & $0.26 \pm 0.02$ & $0.27 \pm 0.02$ \\
$\omega_b$ & $0.02236\pm 0.00015$ & $0.02236\pm 0.00015$ & $0.02236\pm 0.00015$ & $0.02236\pm 0.00015$ & $0.02235\pm 0.00015$ \\ \hline
$\chi^2_{\min}$ & $7.53$ & $1411.41$ & $1403.11$ & $7.61$ & $1410.06$ \\
\thickhline 

$\mathbf{\Lambda_s}$\textbf{CDM} & \textbf{\textit{BAO}} & \textbf{\textit{BAO}}+\textbf{\textit{PP}} & \textbf{\textit{PP}} & \textbf{\textit{BAOr}} & \textbf{\textit{BAOr}}+\textbf{\textit{PP}} \\ \hline
$z_\dagger$ & $2.25\pm 0.43$ & $2.12^{+0.55}_{-0.49}$ & $2.08^{+0.66}_{-0.56}$ & $2.02 \pm 0.42$ & $1.91^{+0.55}_{-0.24}$ \\
\hline
$H_0 \left [ {\text{km}}/{\text{s}}/{\text{Mpc}} \right ] $ & $69.12 \!\pm\! 1.31$ (2.4$\sigma$, 1.9\%) & $68.27 \!\pm\! 0.99$ (3.3$\sigma$, 1.5\%) & $68.11^{+1.21}_{-1.53}$ (2.9$\sigma$, 2.0\%) & $72.78 \!\pm\! 1.51$ (0.1$\sigma$, 2.1\%) & $70.04 \!\pm\! 1.03$ (2.0$\sigma$, 1.5\%) \\
$\Omega_m$ & $0.30\pm 0.02$ & $0.32\pm0.01$ & $0.32\pm 0.01$ & $0.26 \pm 0.02$ & $0.29 \pm 0.01$ \\
$\omega_b$ & $0.02236\pm 0.00015$ & $0.02236\pm 0.00015$ & $0.02236\pm 0.00015$ & $0.02236\pm 0.00015$ & $0.02236\pm 0.00015$ \\ \hline
$\chi^2_{\min}$ & $7.48$ & $1411.66$ & $1403.61$ & $12.01$ & $1422.54$ \\
\thickhline  

\textbf{nCC} & \textbf{\textit{BAO}} & \textbf{\textit{BAO}}+\textbf{\textit{PP}} & \textbf{\textit{PP}} & \textbf{\textit{BAOr}} & \textbf{\textit{BAOr}}+\textbf{\textit{PP}} \\ \hline
$w_x$ & $-0.99^{+0.04}_{-0.02}$ & $-0.99^{+0.02}_{-0.01}$ & $-0.99^{+0.03}_{-0.02}$ & $-0.96^{+0.03}_{-0.02}$ & $-0.98^{+0.02}_{-0.01}$ \\
$\Omega_x$ & $4.14^{+2.72}_{-2.88}$ & $3.59^{+2.62}_{-2.44}$ & $4.16^{+2.28}_{-2.47}$ & $4.68^{+2.37}_{-2.02}$ & $4.76^{+3.14}_{-2.11}$ \\
\hline
$H_0 \left [ {\text{km}}/{\text{s}}/{\text{Mpc}} \right ] $ & $67.32 \!\pm\! 2.08$ (2.4$\sigma$, 3.1\%) & $67.23 \!\pm\! 0.84$ (4.3$\sigma$, 1.2\%) & $67.31^{+1.04}_{-1.25}$ (3.8$\sigma$, 1.6\%) & $64.56 \!\pm\! 2.23$ (3.4$\sigma$, 3.5\%) & $68.29 \!\pm\! 0.86$ (3.5$\sigma$, 1.3\%) \\
$\Omega_m$ & $0.30\pm 0.02$ & $0.31\pm 0.01$ & $0.30\pm 0.03$ & $0.28 \pm 0.02$ & $0.26 \pm 0.01$ \\
$\omega_b$ & $0.02236\pm 0.00015$ & $0.02236\pm 0.00015$ & $0.02236\pm 0.00015$ & $0.02236\pm 0.00015$ & $0.02236\pm 0.00015$ \\ \hline
$\chi^2_{\min}$ & $7.24$ & $1410.06$ & $1402.82$ & $6.26$ & $1412.08$ \\
\thickhline
\end{tabular}}
\caption{68\% credible intervals on the fractional matter density parameter $\Omega_m$, physical baryon density $\omega_b$, Hubble constant $H_0$, and other model-specific parameters, for the seven models (separated by thick horizontal lines) and five dataset combinations (different columns) considered in this work. In all dataset combinations, the \textit{base} dataset combination which consists of the \textit{P18}$^{\star}$ $2 \times 2$ compressed CMB likelihood as well as the $\Omega_m^{P}$ prior on $\Omega_m$ is always included. For all models and dataset combinations, we report the best-fit $\chi^2$, and in brackets the residual tension with the representative Cepheid-calibrated SNeIa measurement $H_0=(73.04 \pm 1.04)\,{\text{km}}/{\text{s}}/{\text{Mpc}}$~\cite{Riess:2021jrx} (in excellent agreement with the community consensus measurement from the Local Distance Network reported recently in Ref.~\cite{H0DN:2025lyy}), as well as the relative precision with which $H_0$ is inferred (given that for some dataset combinations the tension is formally alleviated mostly due to significantly larger uncertainties). For the nCC model, given that the posteriors for $w_x$ and $\Omega_x$ are highly non-Gaussian, we do not report the posterior mean and corresponding 68\% credible interval uncertainties, but the median alongside the 16th and 84th percentiles.}
\label{tab:parameters}
\end{table*}

We now discuss the results obtained using the methodology and datasets presented in Sec.~\ref{sec:datasets}, applied to the different late-time modifications discussed in Sec.~\ref{sec:models}. Our results are compactly summarized in Tab.~\ref{tab:parameters}. Before turning to the late-time modifications, we begin by examining the impact of rescaled BAO data on the $\Lambda$CDM model. This preliminary step serves to highlight a number of generic features that will reappear in our subsequent analyses.

\subsection{$\Lambda$CDM}
\label{subsec:resultslcdm}

Before explicitly testing the $\Lambda$CDM model against rescaled BAO data, it is useful to reason about what should be expected from such a comparison. It is important to recall that, within $\Lambda$CDM, the CMB power spectrum on its own acts as a \textit{self-calibrated standard ruler}, i.e.\ one where there is sufficient internal information to self-calibrate the absolute size of the ruler ($r_d$) and determine $H_0$ from the ruler's apparent size, without relying on external calibrations.~\footnote{See e.g.\ Sec.~IIA and~IIB of Ref.~\cite{Knox:2019rjx}, as well as Slides~12-15 of~\href{https://tinyurl.com/tonalevagnozzilecture2}{tinyurl.com/tonalevagnozzilecture2}, for a detailed explanation of this aspect.} This implies that within $\Lambda$CDM there is in principle no need to add BAO data to CMB data to reliably infer $H_0$. BAO can complement the CMB-only determination of $H_0$, most notably by tightening the associated uncertainties. Perhaps more importantly, once calibrated, BAO provide an important consistency check for the CMB-only determination. Stated differently, within $\Lambda$CDM the number of model degrees of freedom is effectively matched by the number of physical constraints from the CMB alone. The addition of BAO, or for that matter any geometrical dataset, renders the system \textit{over-constrained} (see e.g.\ Refs.~\cite{Bernal:2021yli,Vagnozzi:2021tjv} for related discussions): at best, one can hope that the additional constraints are consistent with the CMB ones, which is fortunately the case with BAO and unanchored SNeIa within $\Lambda$CDM. Importantly, these considerations remain valid under the compression strategy adopted here: the combination of the $\Omega_m$ prior with our $2 \times 2$ compressed CMB information allows the latter to retain its status as self-calibrated standard ruler (by virtue of the same arguments presented in Sec.~\ref{sec:bao}). Before moving on we note that, for late-time extensions to $\Lambda$CDM that introduce additional DE-related parameters, BAO play a central role due to their ability to break the geometrical degeneracy, and stabilize constraints on both the additional parameters and $H_0$, whereas the CMB in general is no longer a self-calibrated standard ruler.

Based on the previous arguments we expect that the rescaled BAO measurements, recalibrated in such a way to favor SH0ES-level values of $H_0$, should exhibit a strong tension with the self-calibrated CMB information alone. This internal tension should prevent the value of $H_0$ inferred from the CMB+BAO combination from reaching the high SH0ES-level values which the rescaling was meant to favor. Unlike the case of standard BAO measurements, we would therefore expect rescaled BAO measurements to fail the ``consistency test'' discussed above.

\begin{figure}[!ht]
\centering
\includegraphics[width=0.9\linewidth]{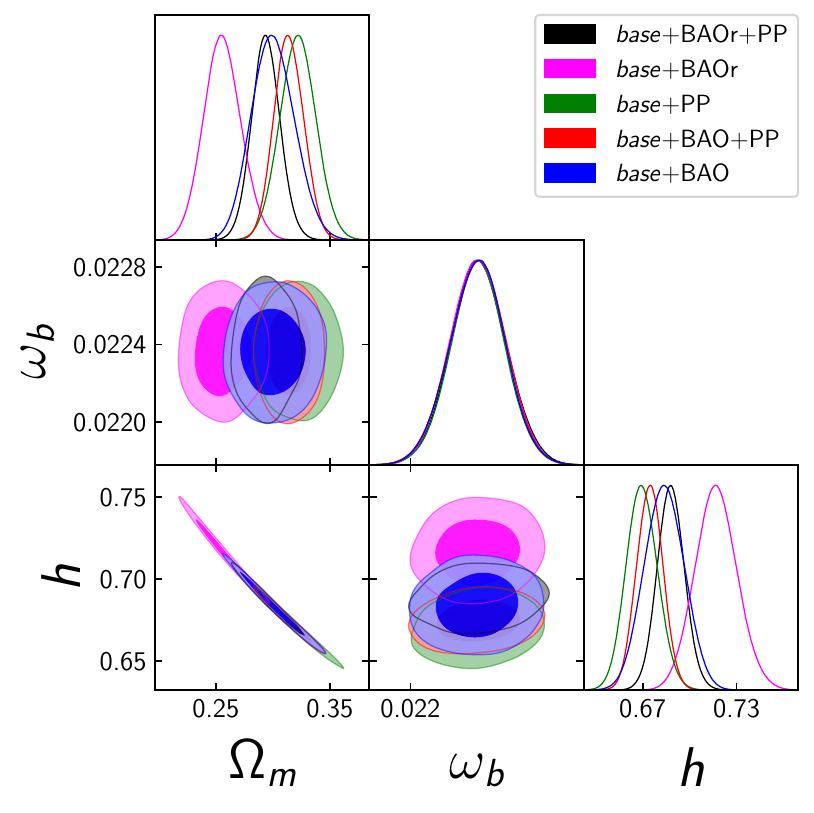}
\caption{Triangular plot showing 2D joint and 1D marginalized posterior probability distributions for the fractional matter density $\Omega_m$, the physical baryon density $\omega_b$, and the reduced Hubble constant $h \equiv H_0/(100\,{\text{km}}/{\text{s}}/{\text{Mpc}})$, obtained within the baseline $\Lambda$CDM model (see Sec.~\ref{subsec:lcdm}) in light of the \textit{base}+\textit{BAO} (blue contours), \textit{base}+\textit{BAOr} (magenta contours), \textit{base}+\textit{BAO}+\textit{PP} (red contours), \textit{base}+\textit{PP} (green contours), and \textit{base}+\textit{BAOr}+\textit{PP} (black contours) dataset combinations. We recall that the \textit{base} combination includes the \textit{P18}$^{\star}$ $2 \times 2$ compressed CMB likelihood as well as the $\Omega_m^{P}$ prior on $\Omega_m$.}
\label{fig:lcdm}
\end{figure}

Our expectations are explicitly confirmed in the corner plot of Fig.~\ref{fig:lcdm}, showing 2D joint and 1D marginalized posterior probability distributions for $\Omega_m$, $\omega_b$, and $h$ in light of the 5 dataset combinations discussed in Sec.~\ref{subsubsec:combinations}. Recall that by \textit{base} we denote the \textit{P18}$^{\star}$+$\Omega_m^{P}$ combination. For the standard \textit{base}+\textit{BAO} dataset combination (blue contours) we find $H_0=(68.40 \pm 1.30)\,{\text{km}}/{\text{s}}/{\text{Mpc}}$, in line with expectations from the literature. In contrast, when adopting the \textit{base}+\textit{BAOr} dataset combination (magenta contours), we infer $H_0=(71.69 \pm 1.35)\,{\text{km}}/{\text{s}}/{\text{Mpc}}$. This confirms our earlier expectation that even the rescaled BAO dataset is not able to drive $H_0$ all the way up to SH0ES-level values: what is preventing this is precisely the BAO-CMB internal tension, which despite the comparatively large uncertainties is quite evident when comparing the blue and magenta contours in Fig.~\ref{fig:lcdm}. This internal tension is also evident when comparing the quality of fit to the \textit{base}+\textit{BAO} versus \textit{base}+\textit{BAOr} dataset combinations, with the best-fit $\chi^2$ of the latter exceeding the former by $\Delta \chi^2=+9.14$.

A further noteworthy feature is the shift in the matter density parameter. From the \textit{base}+\textit{BAOr} dataset combination we infer $\Omega_m=0.26 \pm 0.02$, which is at odds by almost $2\sigma$ with $\Omega_m=0.300 \pm 0.018$ as inferred from \textit{base}+\textit{BAO} dataset combination, despite the generous $\Omega_m=0.30 \pm 0.03$ prior being included in the \textit{base} dataset combination. This behaviour is readily understood: increasing $H_0$ requires a compensating decrease in $\Omega_m$ to preserve consistency with the CMB acoustic angular scale $\theta_s(z_{\star})$, since in $\Lambda$CDM there is no other remaining degree of freedom which one can vary to adjust $\theta_s(z_{\star})$. However, reaching SH0ES-level values of $H_0$ would force $\Omega_m$ to decrease to unrealistically low values. The outcome therefore ends up being an intermediate one, with \textit{BAOr} pulling $H_0$ upwards, but with a limit to how far $\Omega_m$ can fall, resulting in $H_0 \approx 71.5\,{\text{km}}/{\text{s}}/{\text{Mpc}}$ and $\Omega_m \approx 0.26$. This illustrates the remarkable constraining power of the CMB geometrical information, which provides an extremely long lever arm up to the last scattering surface at redshift $z_{\star} \approx 1100$, well beyond the reach of BAO (see e.g.\ Ref.~\cite{Lin:2021sfs} for further discussions). It is worth stressing that the $\Delta \chi^2=+9.14$ is at least in part driven by the lower matter density, given our Gaussian (but generous) prior on $\Omega_m$. For instance, at the best-fit $\Omega_m=0.26$, the $\Omega_m$ prior contributes to the $\Delta \chi^2$ by $\Delta \chi^2_{\Omega_m}=((0.30-0.26)/0.03)^2 \approx +1.77$.

\begin{figure}[ht!]
\centering
\includegraphics[width=0.9\linewidth]{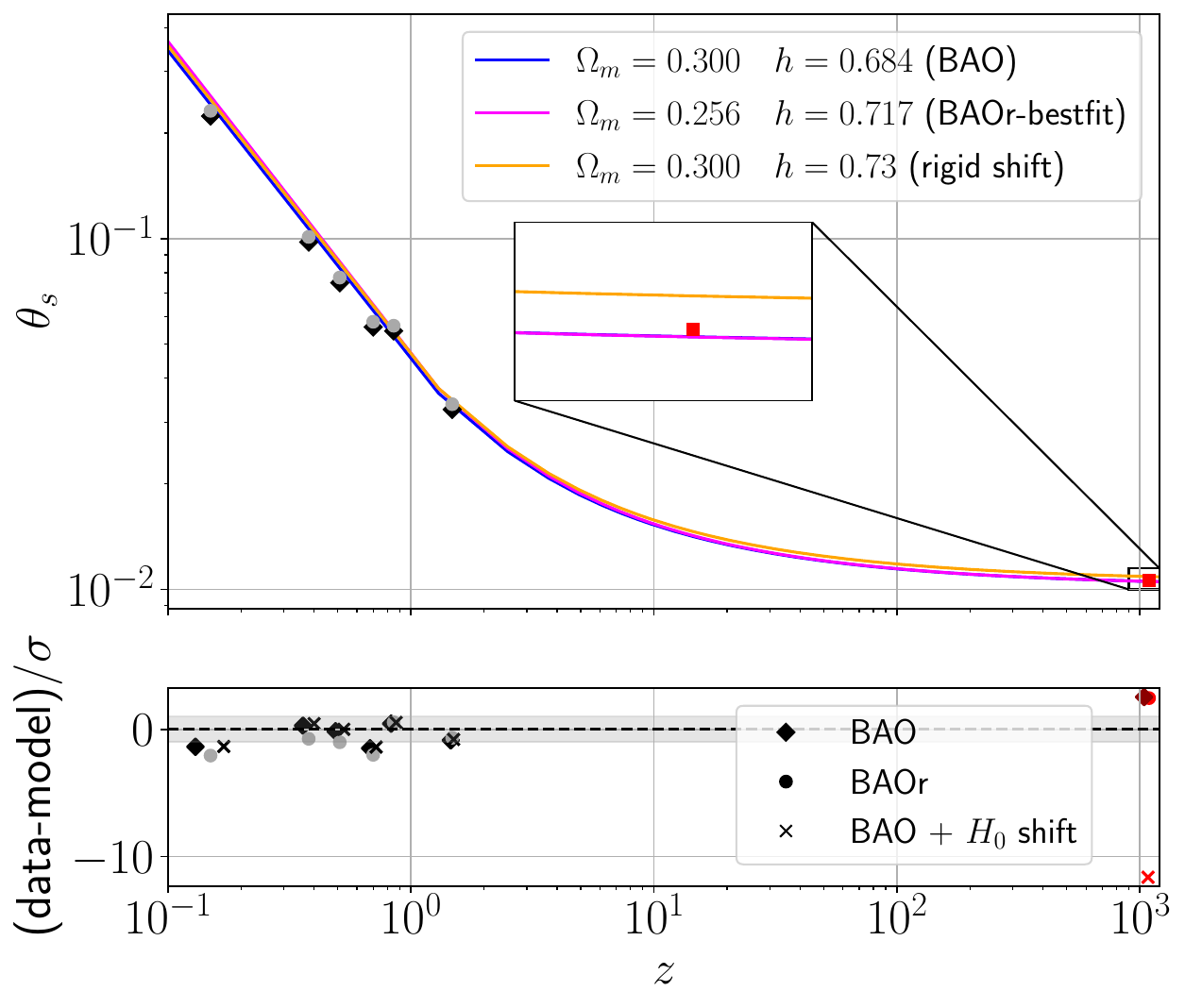}
\caption{\textit{Upper panel}: recombination sound horizon $\theta_s$ observed, inferred, or predicted at different redshifts. The black diamonds correspond to $\theta_s$ inferred from a selection of BAO measurements (the difference between sound horizon at recombination and at baryon drag has been taken into account when deducing these measurements), whereas the grey circles are the same values but after rescaling the BAO measurements. The red square at $z=z_{\star}$ corresponds to the CMB acoustic scale measurement. All datapoints come with uncertainties, but these are too small to be appreciated by the eye. The three curves correspond to the predictions for $\theta_s(z)$ within the three cosmological models discussed in the text: the best-fit cosmologies inferred from the \textit{base}+\textit{BAO} (blue curve) and \textit{base}+\textit{BAOr} (magenta curve) dataset combinations, and the ``rigid shift'' cosmology which is identical to best-fit \textit{base}+\textit{BAO} one, except for a shift in $H_0=73.0\,{\text{km}}/{\text{s}}/{\text{Mpc}}$ (yellow curve). \textit{Lower panel}: normalized residuals. What is plotted are the differences between datapoints and model prediction for each of the three cosmologies shown in the upper panel, with the differences normalized by the uncertainty of the associated datapoint. For the diamond [circle] residuals the datapoints are the standard [rescaled] BAO measurements, and the model is the best-fit cosmology inferred from the \textit{base}+\textit{BAO} [\textit{base}+\textit{BAOr}] dataset combination; for the cross residuals the datapoints are the rescaled BAO measurements, whereas the model is the ``rigid shift'' cosmology.}
\label{fig:fit_cmb_acoustic_lcdm}
\end{figure}

The importance of the CMB geometrical datapoint is clearly illustrated in Fig.~\ref{fig:fit_cmb_acoustic_lcdm}, where in the upper panel we plot the evolution of the acoustic angular scale $\theta_s(z)$ up to $z_{\star}$, for three different models (to be defined soon), alongside measurements of $\theta_s$ from transverse and volume-averaged BAO data, as well as from the CMB at $z_{\star}$.~\footnote{This plot is analogous to Fig.~1 of Ref.~\cite{Lin:2021sfs} and Fig. 2 of Ref.~\cite{eBOSS:2020yzd}, and for the same reasons discussed there (see e.g.\ footnote~5 of Ref.~\cite{Lin:2021sfs}) we only plot angular scales and not the redshift span $\Delta zr_d$. Therefore, only transverse and volume-averaged BAO measurements are included, but not line-of-sight ones, although this in no way alters the message of the plot.} The three models are as follows:
\begin{enumerate}
\item the cosmology inferred from the \textit{base}+\textit{BAO} dataset combination, with $\Omega_m=0.296$ and $H_0=68.40\,{\text{km}}/{\text{s}}/{\text{Mpc}}$ (blue curve);
\item the cosmology inferred from the \textit{base}+\textit{BAOr} dataset combination, with $\Omega_m=0.256$ and $H_0=71.69\,{\text{km}}/{\text{s}}/{\text{Mpc}}$ (magenta curve);
\item a \textit{rigid} rescaling of the first (\textit{base}+\textit{BAO}) cosmology, with $\Omega_m=0.296$ and $H_0=73.0\,{\text{km}}/{\text{s}}/{\text{Mpc}}$ (yellow curve): we refer to this as rigid rescaling since it is performed shifting only $H_0$ but not $\Omega_m$, leading to a rigid shift in the amplitude of the expansion history $H(z)$, without altering its shape.
\end{enumerate}
Hereafter, we shall refer to these three cosmologies as cosmologies 1., 2., and 3. Since the uncertainties in the measurements are too small to be appreciated by the naked eye (see also Fig.~1 of Ref.~\cite{Lin:2021sfs} for a similar plot), in the lower panel we instead plot uncertainty-normalized residuals. Specifically, we plot the difference between datapoints and model predictions, divided by the associated uncertainty: the data are the standard BAO data for cosmology 1.\ (diamond residuals), and the rescaled BAO data for cosmologies 2.\ (circle residuals) and 3.\ (cross residuals) respectively.

Fig.~\ref{fig:fit_cmb_acoustic_lcdm} allows us to draw a few important conclusions. Firstly, we note that fit of cosmology 2.\ to rescaled BAO data is only slightly worse than the fit of cosmology 1.\ to standard BAO data. Perhaps more surprising is the fact that the fit of cosmology 3.\ (the ``rigid shift'' cosmology) to rescaled BAO data is almost identical to, and in fact even slightly better than, that of cosmology 2.\ to the same dataset. This directly reflects the fact that the sensitivity of BAO measurements to $\Omega_m$, discussed in Sec.~\ref{sec:bao}, is limited. It is only once BAO measurements are combined with geometrical CMB information that the sensitivity to $\Omega_m$ is significantly sharpened, due to the resulting long lever arm. This is very clear when comparing the fit of the three cosmologies to the geometrical CMB datapoint at $z=z_{\star}$. Again, the difference in fit is hardly visible to the naked eye, but is made very clear by the inset in the upper panel, and even more clearly by the residuals in the lower panel. We see that cosmology 2.\ fits the geometrical CMB datapoint only slightly worse than cosmology 1.  On the other hand, the fit of the rigid shift cosmology 3.\ to the same datapoint is disastrous, being off by over $10\sigma$. The conclusion we draw is that within the range of values for $H_0$ and $\Omega_m$ considered here, BAO data alone have limited ability to distinguish between the three cosmologies, and in particular cannot rule out a rigid $H_0$-only rescaling of the best-fit \textit{base}+\textit{BAO} cosmology. What truly exposes the inadequacy of the rigid shift in $H_0$ is the inclusion of geometrical CMB information, whose long lever arm up to $z_{\star}$ makes manifest the BAO-CMB internal tension discussed earlier. The rigidly rescaled cosmology can only be brought into agreement with the geometrical CMB information if $\Omega_m$ is decreased to values that are strongly penalized by our (already rather generous) $\Omega_m$ prior, precisely because such values are unrealistic.

So far we have focused on the impact of (rescaled or standard) BAO data on the inferred values of $H_0$ and $\Omega_m$. However, an arguably even more important role in shaping the overall cosmological constraints is that played by unanchored SNeIa. Let us begin from the standard \textit{base}+\textit{BAO}+\textit{PP} dataset combination (red contours), from which we infer $H_0=(67.45 \pm 0.83)\,{\text{km}}/{\text{s}}/{\text{Mpc}}$ and $\Omega_m=0.31 \pm 0.01$, perfectly in line with expectations from the literature. As discussed earlier, the \textit{base}+\textit{PP} combination provides an extremely useful BAO-free benchmark to constrain late-time modifications to $\Lambda$CDM, and will therefore play an important role in the later subsections. From this dataset combination (green contours) we infer $H_0=(66.96 \pm 0.99)\,{\text{km}}/{\text{s}}/{\text{Mpc}}$ and $\Omega_m=0.32 \pm 0.02$, also in excellent agreement with the known fact that \textit{PP} appears to prefer slightly higher values of $\Omega_m$. Finally, from the \textit{base}+\textit{BAOr}+\textit{PP} dataset combination (black contours) we find $H_0=(68.80 \pm 0.86)\,{\text{km}}/{\text{s}}/{\text{Mpc}}$ and $\Omega_m=0.29 \pm 0.01$. The value of $H_0$, in particular, is far from the SH0ES-level values the BAO rescaling was in principle designed to reach. In fact, it is even lower than the value $H_0=(71.69 \pm 1.35)\,{\text{km}}/{\text{s}}/{\text{Mpc}}$ inferred from the \textit{base}+\textit{BAOr} dataset combination. In addition, the internal BAO-CMB tension is again reflected in a $\Delta \chi^2=+16.85$ compared to the \textit{base}+\textit{BAO}+\textit{PP} case.

The addition of unanchored SNeIa has therefore further strengthened the conclusions reached previously. The reason once more has to do with $\Omega_m$. High-redshift SNeIa, whether anchored or unanchored, carry significant sensitivity to $\Omega_m$, reflecting their sensitivity to the shape of the expansion history $E(z)$, which in the $\Lambda$CDM model only depends on $\Omega_m$. In fact, the \textit{PP} dataset alone is able to provide a constraint on $\Omega_m=0.33 \pm 0.02$~\cite{Brout:2022vxf}, significantly tighter than the constraint which can be achieved from BAO measurements alone. As noted earlier, in order to fit the geometrical CMB point while increasing $H_0$, a decrease in $\Omega_m$ which is already penalized by our $\Omega_m$ prior is required. The inclusion of unanchored SNeIa data significantly limit the extent to which this decrease is possible. For instance, regardless of the value of $H_0$, the shape of the unnormalized expansion history $E(z)$ in a cosmology with $\Omega_m \sim 0.26$, as indicated by the \textit{base}+\textit{BAOr} combination, would be completely inconsistent with unanchored SNeIa data. This is particularly evident since the \textit{PP} dataset prefers values of $\Omega_m$ slightly higher than $0.3$, going exactly in the opposite direction of what would be required to fit the geometrical CMB point. The fact that $\Omega_m$ cannot be lowered substantially prevents $H_0$ from reaching higher values. As a closing remark, here we have focused on the $\Lambda$CDM model, where $E(z)$ is characterized exclusively by $\Omega_m$: however, we will see that unanchored SNeIa play an even more critical role in the case of late-time extensions to $\Lambda$CDM, where the shape of the expansion history is characterized by additional parameters beyond $\Omega_m$ (such as the DE equation of state), typically strongly degenerate with $H_0$.

\subsection{$w$CDM}
\label{subsec:resultswcdm}

\begin{figure}[!ht]
\centering
\includegraphics[width=0.9\linewidth]{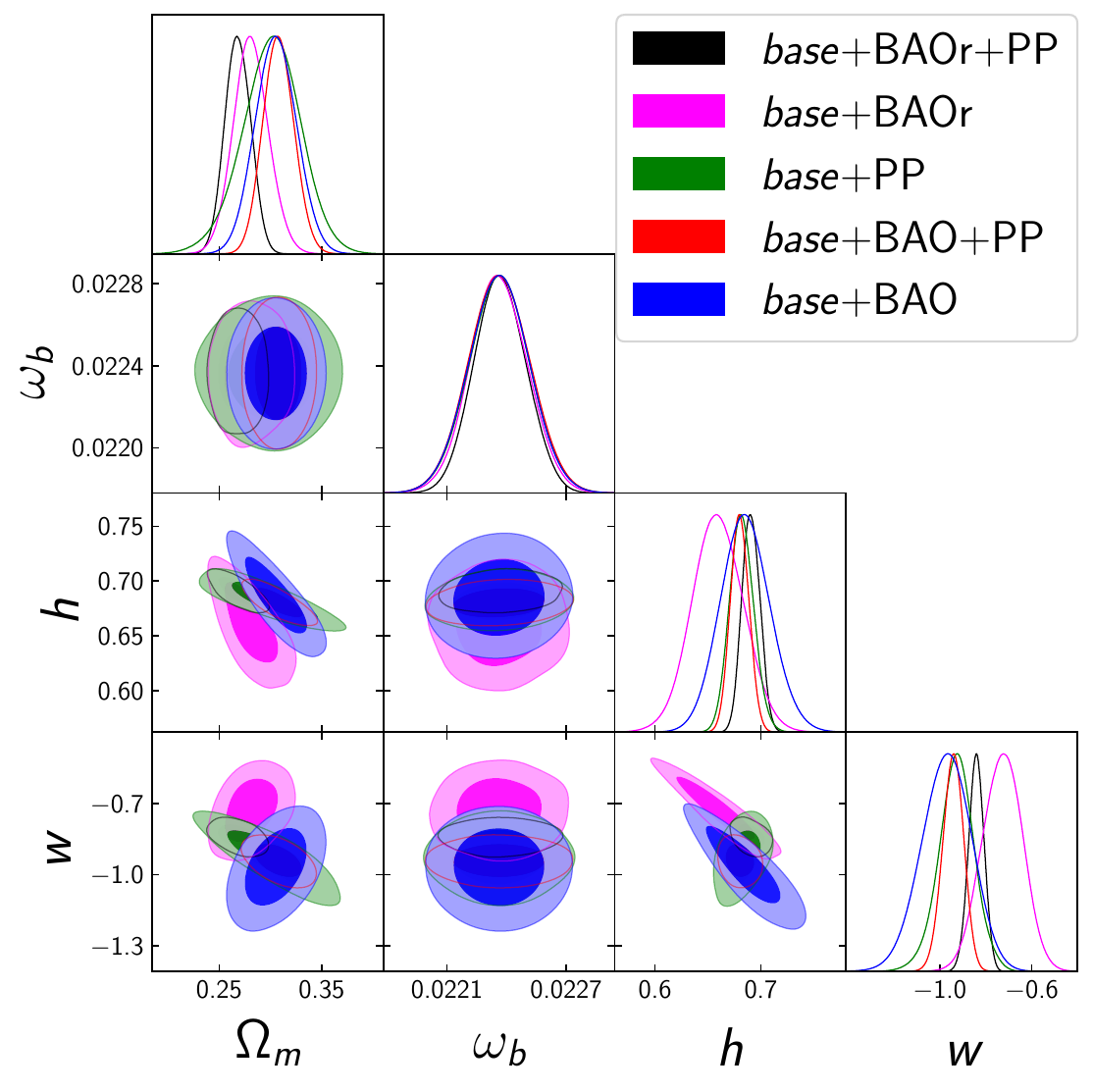}
\caption{As in Fig.~\ref{fig:lcdm}, but in light of the $w$CDM model (see Sec.~\ref{subsec:wcdm}), therefore including also the dark energy equation of state $w$ among the parameters.}
\label{fig:wcdm}
\end{figure}

With the general features discussed previously in mind, we now turn to the $w$CDM model, arguably one of the simplest late-time extensions of $\Lambda$CDM. A corner plot for $\Omega_m$, $\omega_b$, $h$, and $w$ in light of the 5 dataset combinations discussed in Sec.~\ref{subsubsec:combinations} is given in Fig.~\ref{fig:wcdm}. In particular, for the standard \textit{base}+\textit{BAO} dataset combination we infer $H_0=(68.50 \pm 2.30)\,{\text{km}}/{\text{s}}/{\text{Mpc}}$ and $w=-0.97 \pm 0.11$, i.e.\ an EoS slightly in the quintessence-like regime. Considering instead the \textit{base}+\textit{BAO}+\textit{PP} dataset combination further tightens these findings, leading to $H_0=(68.00 \pm 0.87)\,{\text{km}}/{\text{s}}/{\text{Mpc}}$ and $w=-0.94 \pm 0.05$. This confirms that, within the $w$CDM model and using standard BAO measurements, the $H_0$ tension remains unsolved, in agreement with the ``no-go theorem''. For both the \textit{base}+\textit{BAO} and \textit{base}+\textit{BAO}+\textit{PP} dataset combinations, the improvement in fit over $\Lambda$CDM is very mild, with $\Delta \chi^2=-0.25$ and $-2.01$ respectively.

Turning instead to the rescaled BAO measurements, in particular considering the \textit{base}+\textit{BAOr} dataset combination, our results are somewhat unexpected. In particular, we infer an even lower value for $H_0=(65.95 \pm 2.42)\,{\text{km}}/{\text{s}}/{\text{Mpc}}$. At the same time, following the well-known negative correlation between $w$ and $H_0$, itself driven by the geometrical degeneracy, the DE EoS moves further into the quintessence-like regime, to $w=-0.73 \pm 0.09$. Although the outcome is unexpected, it reflects the well-known fact that the late-time expansion history is jointly determined by $H_0$, $\Omega_m$, and $w$, rather than by $H_0$ alone. A uniform rescaling of the BAO measurements should therefore not be interpreted as merely shifting $H_0$. Once the rescaling is applied, certain features of the rescaled BAO data may be better accommodated by a lower $H_0$ together with a quintessence-like value of $w$, which appears to be precisely the case here.

\begin{figure}[ht!]
\centering
\includegraphics[width=0.9\linewidth]{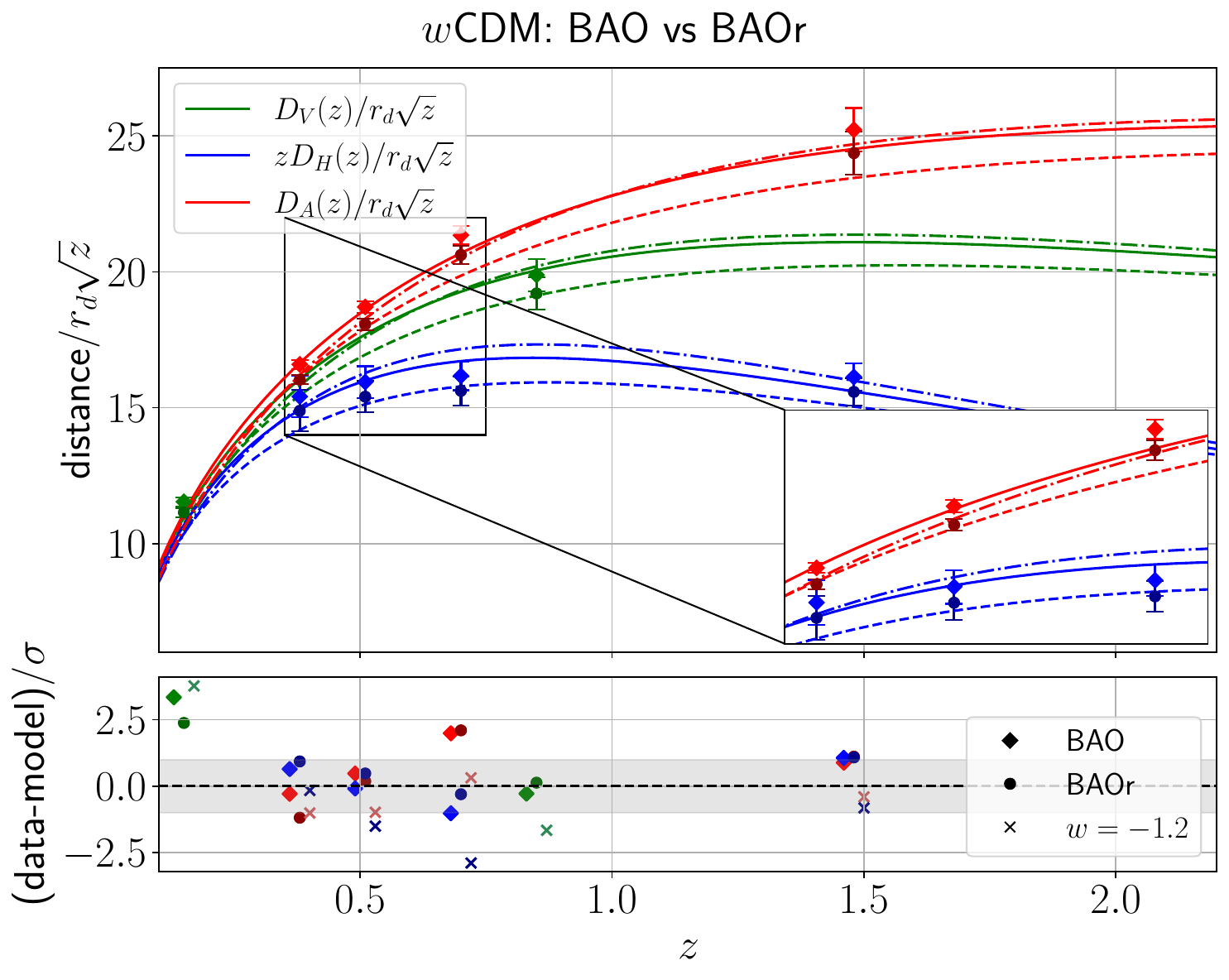}
\caption{\textit{Upper panel}: best-fit predictions for (appropriately normalized) distance-redshift relations within $w$CDM, compared to volume-averaged (green curves and datapoints), line-of-sight (blue curves and datapoints), and transverse (red curves and datapoints) BAO measurements. Both standard and rescaled BAO measurements are included, with the former [latter] associated to diamond [circle] datapoints and being lighter [darker] in color. The solid [dashed] curves correspond to the best-fit predictions obtained from the \textit{base}+\textit{BAO} [\textit{base}+\textit{BAOr}] dataset combination, whereas the dash-dotted curves correspond to the best-fit predictions from the improved ``rigid shift'' cosmology where $H_0$, $w$, and $\Omega_m$ are varied simultaneously to preserve the fit to the CMB acoustic scale. \textit{Lower panel}: normalized residuals, as in the lower panel of Fig.~\ref{fig:fit_cmb_acoustic_lcdm} (we stress again that for the cross residuals the datapoints are the rescaled BAO measurements).}
\label{fig:fit_wCDM}
\end{figure}

The residuals in the lower panel of Fig.~\ref{fig:fit_wCDM} allow us to identify precisely those datapoints which drive this unexpected result. In particular, it is useful to compare the performance of the diamond (model=best-fit to standard BAO, data=standard BAO) versus circle (model=best-fit to rescaled BAO, data=rescaled BAO) residuals. Two datapoints immediately stand out: the MGS volume-averaged $D_V/r_d$ measurement at $z_{\text{eff}}=0.15$, and the eBOSS LRG line-of-sight $D_H/r_d$ measurement at $z_{\text{eff}}=0.72$. These two datapoints are better fit by a slightly lower value of $H_0$, compensated by a larger value of $w$, a combination which does not significantly alter the quality of the fit to the other datapoints, as is evident from the lower panel of Fig.~\ref{fig:fit_wCDM}. In fact, the overall fit compared to that of the \textit{base}+\textit{BAO} dataset combination actually slightly improves by $\Delta \chi^2=-0.87$.

In the $\Lambda$CDM case studied in Sec.~\ref{subsec:resultslcdm}, we introduced a ``rigid shift'' cosmology, where $H_0$ was increased while keeping all other parameters fixed, as a pedagogical exercise to illustrate why high values of $H_0$ were not favored despite the BAO rescaling: such a shift is immediately ruled out by the geometrical CMB datapoint, unless $\Omega_m$ is simultaneously reduced to implausibly low values. In the present case, the natural question is why the rescaled BAO do not favor a higher $H_0$ together with a phantom $w$. To address this, we consider an improved ``rigid shift'' in which $H_0$, $w$, and $\Omega_m$ are varied simultaneously so as to preserve the distance to the CMB, thereby alleviating the shortcomings of the earlier rigid shift.~\footnote{We note that within the $w$CDM model the sound horizon $r_d$ remains unaffected, given that DE is completely subdominant at early times. Ensuring that the distance to the CMB is unaltered is therefore sufficient to maintain $\theta_s$ unchanged, and preserve the fit to the geometrical CMB datapoint.} To do so, we exploit the geometrical degeneracy to move along the $H_0$-$w$ correlation up to $H_0=73.50\,{\text{km}}/{\text{s}}/{\text{Mpc}}$, $w=-1.2$, and $\Omega_m=0.265$, following Tab.~II of Ref.~\cite{Alestas:2020mvb}, see also Eq.~(2.12) of the same work. The resulting theoretical distance-redshift relations are given by the dash-dotted curves in the upper panel of Fig.~\ref{fig:fit_wCDM}, and the corresponding residuals (of course computed against the rescaled BAO data) are given by crosses in the lower panel of the same Figure. We see that the fit to the rescaled BAO datapoints of the resulting cosmology with high $H_0$ and phantom $w$ (crosses) is significantly worse compared to that of the cosmology with lower $H_0$ but quintessence-like $w$ (circles), particularly for what concerns the MGS and eBOSS ELG volume-averaged $D_V/r_d$ measurements at $z_{\text{eff}}=0.15$ and $z_{\text{eff}}=0.85$ respectively, and the BOSS DR12 and eBOSS LRG line-of-sight $D_H/r_d$ measurements at $z_{\text{eff}}=0.51$ and $z_{\text{eff}}=0.72$ respectively.

Our results highlight how, in models introducing late-time modifications to $\Lambda$CDM, one cannot focus on $H_0$ in isolation when analyzing BAO data, even though their sensitivity to the shape of the expansion history $E(z)$ is relatively limited. For the specific case of $w$CDM, $E(z)$ is \textit{directly} controlled by $\Omega_m$ and $w$, but $H_0$ still enters \textit{indirectly} once the CMB geometrical constraint is imposed, through the geometrical degeneracy affecting $\Omega_m$ and $w$, and required to preserve consistency with the CMB acoustic scale. It is therefore essential to analyze the full parameter space rather than expecting the BAO rescaling to lead to a simple shift in $H_0$.

Nevertheless, all these issues become largely irrelevant once unanchored SNeIa measurements are taken in consideration, because of the strong constraints imposed on the shape of the unnormalized expansion history. For instance, the BAO-free benchmark provided by the \textit{base}+\textit{PP} dataset combination leads to $H_0=(68.20 \pm 1.10)\,{\text{km}}/{\text{s}}/{\text{Mpc}}$, while the equation of state is constrained to $w=-0.93 \pm 0.08$. This is again consistent with earlier findings in the literature: within the $w$CDM model, the \textit{PP} dataset is known to prefer values of $w$ slightly within the quintessence-like regime, with $w=-0.90 \pm 0.14$ inferred from this dataset alone~\cite{Brout:2022vxf}. When instead considering the \textit{base}+\textit{BAOr}+\textit{PP} dataset combination, we find $H_0=(69.11 \pm 0.91)\,{\text{km}}/{\text{s}}/{\text{Mpc}}$ and $w=-0.84 \pm 0.04$. This outcome, i.e.\ the fact that including unanchored SNeIa slightly raises $H_0$ (although far from SH0ES-like values), is only apparently surprising, and readily understood. While rescaled BAO measurements prefer a more pronounced quintessence-like equation of state ($w \sim -0.7$) as discussed above, unanchored SNeIa favor milder deviations from $\Lambda$CDM, with $w \sim -0.9$. The outcome therefore ends up being an intermediate one, with \textit{BAOr} shifting $w$ upwards and $H_0$ downwards, and \textit{PP} driving the constraints slightly towards the opposite direction. This slight internal tension between rescaled BAO measurements and other cosmological datasets is also reflected by a degradation in fit of $\Delta \chi^2=+1.46$ for the \textit{base}+\textit{BAOr}+\textit{PP} dataset combination with respect to the \textit{base}+\textit{BAO}+\textit{PP} one. At any rate, our results show that even when taking into account rescaled BAO measurements the Hubble tension remains unresolved in the $w$CDM model, which therefore does not provide a loophole to the ``no-go theorem''. This again emphasizes the crucial but thus far somewhat underappreciated role of unanchored SNeIa in restricting the viability of late-time modifications, even on their own, as also recently noted in Ref.~\cite{Zhou:2025kws}.

\subsection{CPL dynamical dark energy}
\label{subsec:resultscpl}

\begin{figure}[!ht]
\centering
\includegraphics[width=0.9\linewidth]{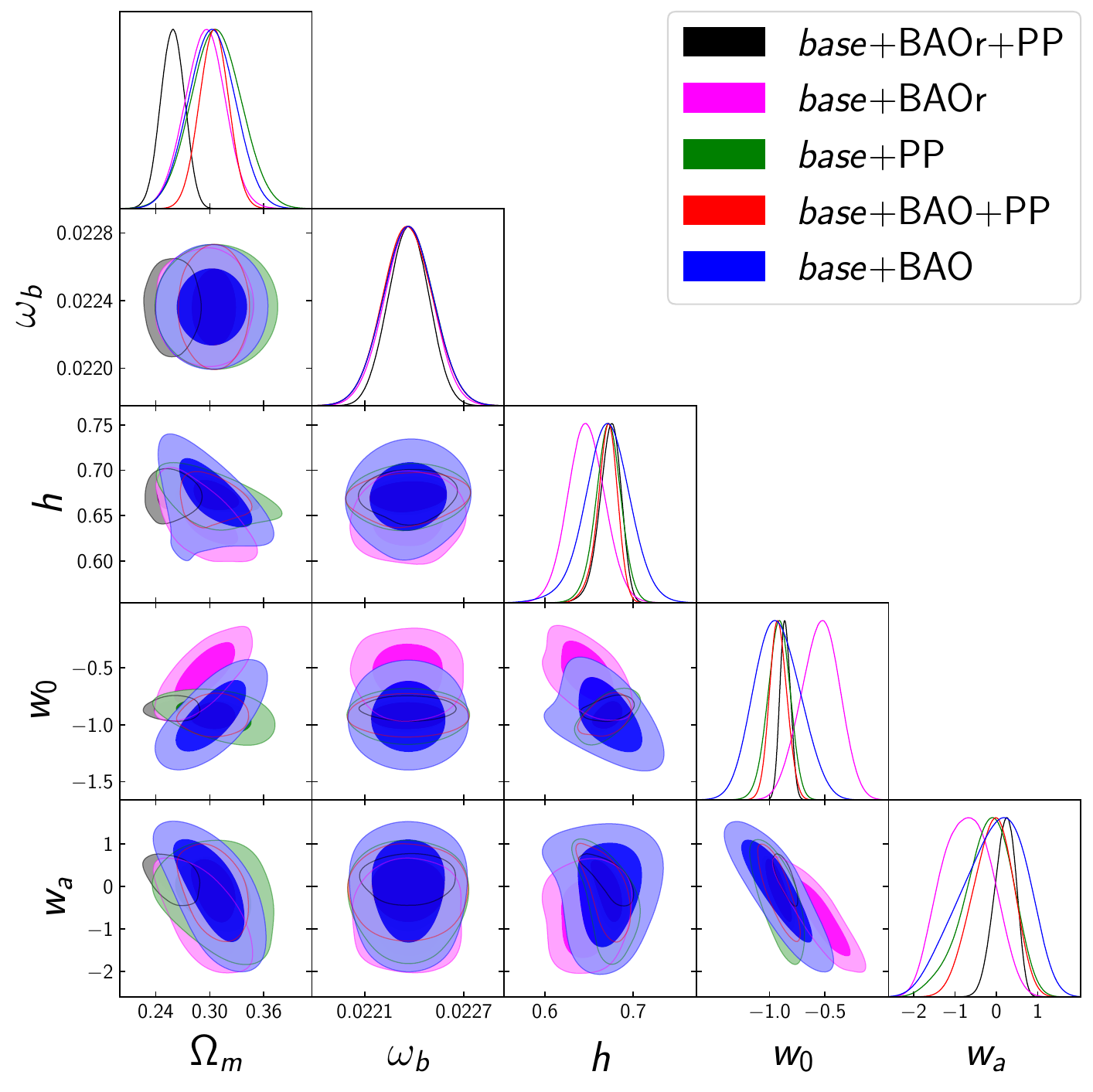}
\caption{As in Fig.~\ref{fig:lcdm}, but in light of the Chevallier-Polarski-Linder dynamical dark energy model (see Sec.~\ref{subsec:cpl}), therefore including also the parameters $w_0$ and $w_a$ which control the evolution of the dark energy equation of state.}
\label{fig:cpl}
\end{figure}

Our discussion now turns to the CPL dynamical DE model, arguably the natural next step beyond $w$CDM. The results obtained for the same 5 dataset combinations discussed earlier are summarized in the corner plot of Fig.~\ref{fig:cpl}. Unsurprisingly, the values of $H_0$ and $w_0$ we infer are very similar to those obtained in the $w$CDM case discussed earlier. Compared to this model, in all cases we observe an extremely slight shift of $w_0$ towards the quintessence-like regime. This shift is compensated by slightly negative values of $w_a$, which are nevertheless consistent with $w_a=0$ at $\lesssim 1.2\sigma$ for all 5 dataset combinations considered here.

More specifically, we find $H_0=(67.00 \pm 2.60)\,{\text{km}}/{\text{s}}/{\text{Mpc}}$ [$67.00^{+1.20}_{-1.0}\,{\text{km}}/{\text{s}}/{\text{Mpc}}$] and $w_0=-0.93^{+0.19}_{-0.22}$ [$w_0=-0.92 \pm 0.08$] for the standard \textit{base}+\textit{BAO} [\textit{base}+\textit{BAO}+\textit{PP}] dataset combination: this confirms that CPL dynamical DE is unable to address the $H_0$ tension when using standard BAO measurements, in agreement with the ``no-go theorem''. For both the \textit{base}+\textit{BAO} and \textit{base}+\textit{BAO}+\textit{PP} dataset combinations, the improvement in fit over $\Lambda$CDM is very mild, with $\Delta \chi^2=-0.38$ and $-2.14$ respectively.

When considering rescaled BAO measurements, the trend is identical to that observed in the $w$CDM case, with $w_0=-0.54 \pm 0.17$ moving deeper into the quintessence-like regime, at the price of a lower value of $H_0=(64.71 \pm 2.09)\,{\text{km}}/{\text{s}}/{\text{Mpc}}$ to maintain the fit to the geometrical CMB datapoint in the case of the \textit{base}+\textit{BAOr} dataset combination. The overall fit compared to that of the \textit{base}+\textit{BAO} dataset combination slightly improves by $\Delta \chi^2=-1.13$. Once more, however, the inclusion of unanchored SNeIa data moves all the constraints back to the standard ones. Concretely, for the BAO-free benchmark \textit{base}+\textit{PP} dataset combination we find $H_0=(67.20 \pm 1.40)\,{\text{km}}/{\text{s}}/{\text{Mpc}}$, consistent with the value inferred within the \textit{base}+\textit{BAOr}+\textit{PP} combination, $H_0=(67.36 \pm 1.43)\,{\text{km}}/{\text{s}}/{\text{Mpc}}$. Again, we note that the slight internal tension when adopting rescaled BAO data alongside unanchored SNeIa is reflected by a degradation in fit of $\Delta \chi^2=+1.57$ for the \textit{base}+\textit{BAOr}+\textit{PP} dataset combination with respect to the \textit{base}+\textit{BAO}+\textit{PP} one.

The reasons behind these observed features are exactly the same as those discussed for the $w$CDM model in Sec.~\ref{subsec:resultswcdm}, and we will therefore refrain from repeating the discussion. We nevertheless take this opportunity to reiterate two important points. Firstly, it is crucial to analyze the full parameter space of late-time modifications to $\Lambda$CDM rather than expecting the BAO rescaling to lead to an upwards shift in $H_0$. We stress once more that, as observed in the $\Lambda$CDM case in Sec.~\ref{subsec:resultslcdm}, an upwards shift to SH0ES-like values would require implausibly low values of $\Omega_m$ to maintain consistency with the geometrical CMB datapoint, even within extended cosmologies. Secondly, and most importantly, such an upwards-shift is anyhow prevented by the extremely tight constraints on the shape of the expansion history from unanchored SNeIa: \textit{even on their own}, these measurements are sufficient to rule out any late-time resolution to the Hubble tension, thereby closing any late-time loophole to the ``no-go theorem'', as explicitly demonstrated in the case of CPL dynamical DE.

\subsection{Phenomenologically emergent dark energy}
\label{subsec:resultspede}

\begin{figure}[!ht]
\centering
\includegraphics[width=0.9\linewidth]{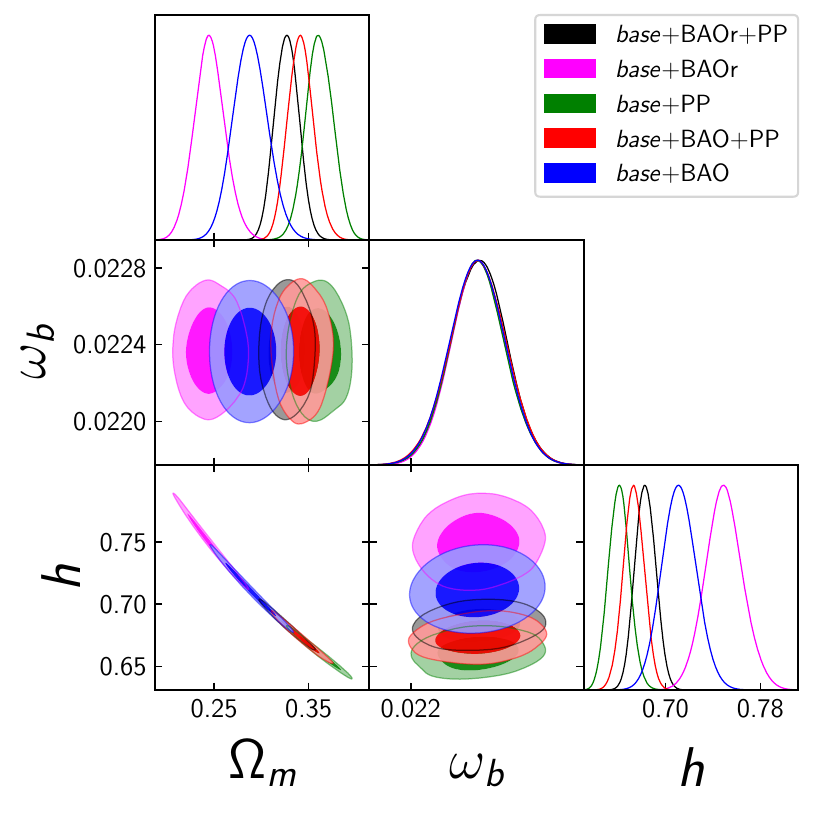}
\caption{As in Fig.~\ref{fig:lcdm}, but in light of the phenomenologically emergent dark energy model (see Sec.~\ref{subsec:pede}), which has no additional parameters beyond those of $\Lambda$CDM.}
\label{fig:pede}
\end{figure}

We now consider the PEDE model, which we recall has no extra degrees of freedom compared to $\Lambda$CDM. In this case, the DE equation of state is forced to take the form given in Eq.~(\ref{eq:wpede}): therefore, the only remaining lever which controls the unnormalized shape of the expansion history $E(z)$ is provided by $\Omega_m$. Our constraints are summarized in the corner plot of Fig.~\ref{fig:pede}. For the standard \textit{base}+\textit{BAO} dataset combination we infer $H_0=(71.20 \pm 1.40)\,{\text{km}}/{\text{s}}/{\text{Mpc}}$, in agreement with earlier findings in the literature~\cite{Pan:2019hac}. However, this comes at the cost of a significantly worse quality of fit compared to $\Lambda$CDM, with $\Delta \chi^2=+3.75$, also in agreement with earlier findings. The reason is essentially that the model ``forces'' the expansion history to take a specific shape regardless of whether such a shape can be accommodated by BAO data or not (see also Refs.~\cite{Vagnozzi:2019ezj,Pedreira:2023qqt,Giare:2024akf} for features observed in similar contexts). The fact that the quality of fit degrades for PEDE relative to $\Lambda$CDM is not in itself a problem, since the two models are not nested.

In addition, including unanchored SNeIa and therefore considering the \textit{base}+\textit{BAO}+\textit{PP} dataset combination pushes the Hubble constant back down to $H_0=(67.32 \pm 0.87)\,{\text{km}}/{\text{s}}/{\text{Mpc}}$, confirming that the model does not evade the ``no-go theorem'' once all available standard datasets are taken into account. Even in this case the quality of fit compared to $\Lambda$CDM is substantially worse, with $\Delta \chi^2=+17.27$. It is interesting to note that this dataset combination pushes the matter density to the rather high value $\Omega_m=0.34 \pm 0.01$. This is required to adjust as much as possible the shape of the (highly non-standard) expansion history, otherwise in strong disagreement with unanchored SNeIa, with $\Omega_m$ being the only available degree of freedom in this sense, as alluded to earlier.

\begin{figure}[ht!]
\centering
\includegraphics[width=0.9\linewidth]{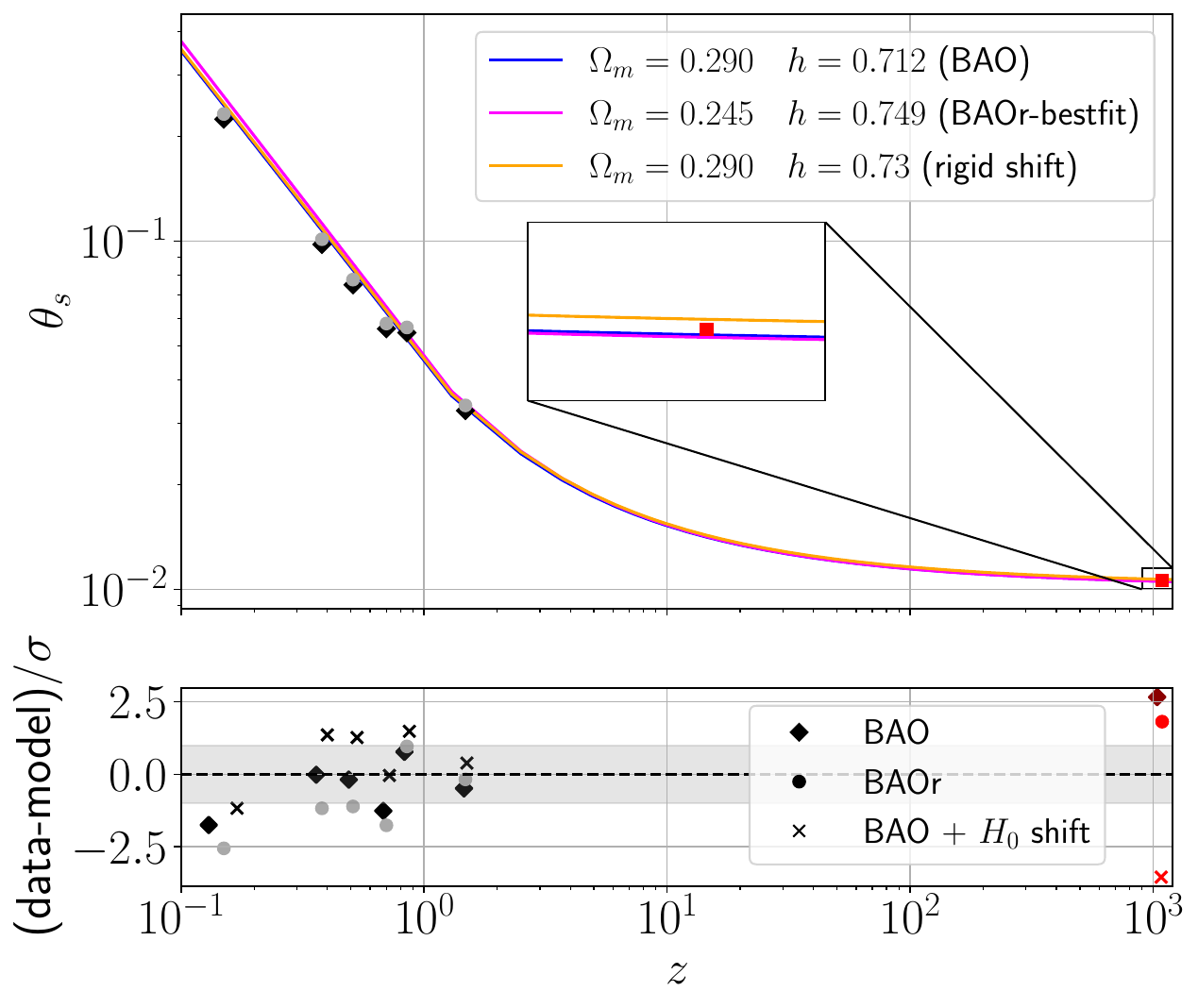}
\caption{As in Fig.~\ref{fig:fit_cmb_acoustic_lcdm}, but for the phenomenologically emergent dark energy model.}
\label{fig:fit_pede}
\end{figure}

When considering the \textit{base}+\textit{BAOr} dataset combination, the Hubble constant shifts significantly upwards to $H_0=(74.94 \pm 1.58)\,{\text{km}}/{\text{s}}/{\text{Mpc}}$, while at the same time the matter density drops to the implausibly low value of $\Omega_m=0.25 \pm 0.02$. However, this comes at the price of an overall degradation in fit compared to the \textit{base}+\textit{BAO} dataset combination of $\Delta \chi^2=+17.77$. The decrease in $\Omega_m$ is required to maintain the fit to the acoustic scale measured from the CMB. To see this explicitly, in Fig.~\ref{fig:fit_pede} we show the evolution of $\theta_s(z)$ up to $z_{\star}$, including a selection of BAO data as well as the geometrical CMB datapoint. Analogously to our $\Lambda$CDM analysis (see Fig.~\ref{fig:fit_cmb_acoustic_lcdm}), we consider not only the best-fit cosmologies inferred from the \textit{base}+\textit{BAO} (blue curve, diamond residuals) and \textit{base}+\textit{BAOr} (magenta curve, circle residuals) dataset combinations, but also a ``rigid shift'' of the first cosmology, with the matter density parameter fixed to $\Omega_m=0.29$, but shifting the Hubble constant from $H_0=71.20\,{\text{km}}/{\text{s}}/{\text{Mpc}}$ to $73.0\,{\text{km}}/{\text{s}}/{\text{Mpc}}$. We immediately see that such a ``rigid shift'' is completely precluded by the geometrical CMB datapoint (see the red cross residual at the bottom right of the lower panel).

At any rate, the shifts discussed above are entirely undone once unanchored SNeIa measurements are included, considering the \textit{base}+\textit{BAOr}+\textit{PP} dataset combination. In this case we find $H_0=(68.31 \pm 0.88)\,{\text{km}}/{\text{s}}/{\text{Mpc}}$ and $\Omega_m=0.33 \pm 0.01$. Even in this case the quality of fit compared to the \textit{base}+\textit{BAO}+\textit{PP} dataset combination is significantly degraded by $\Delta \chi^2=+33.89$, reflecting once more that a very specific expansion history is being ``forced'' upon the data regardless of whether the latter can accommodate it. Finally, considering the BAO-free benchmark \textit{base}+\textit{PP} dataset combination we find $H_0=(66.09 \pm 0.87)\,{\text{km}}/{\text{s}}/{\text{Mpc}}$ and $\Omega_m=0.36 \pm 0.01$. This confirms the inability of the model to address the Hubble tension regardless of any possible miscalibration in the BAO measurements, reinforcing once more the key role of unanchored SNeIa.

\subsection{Holographic dark energy}
\label{subsec:resultshde}

\begin{figure}[!ht]
\centering
\includegraphics[width=0.9\linewidth]{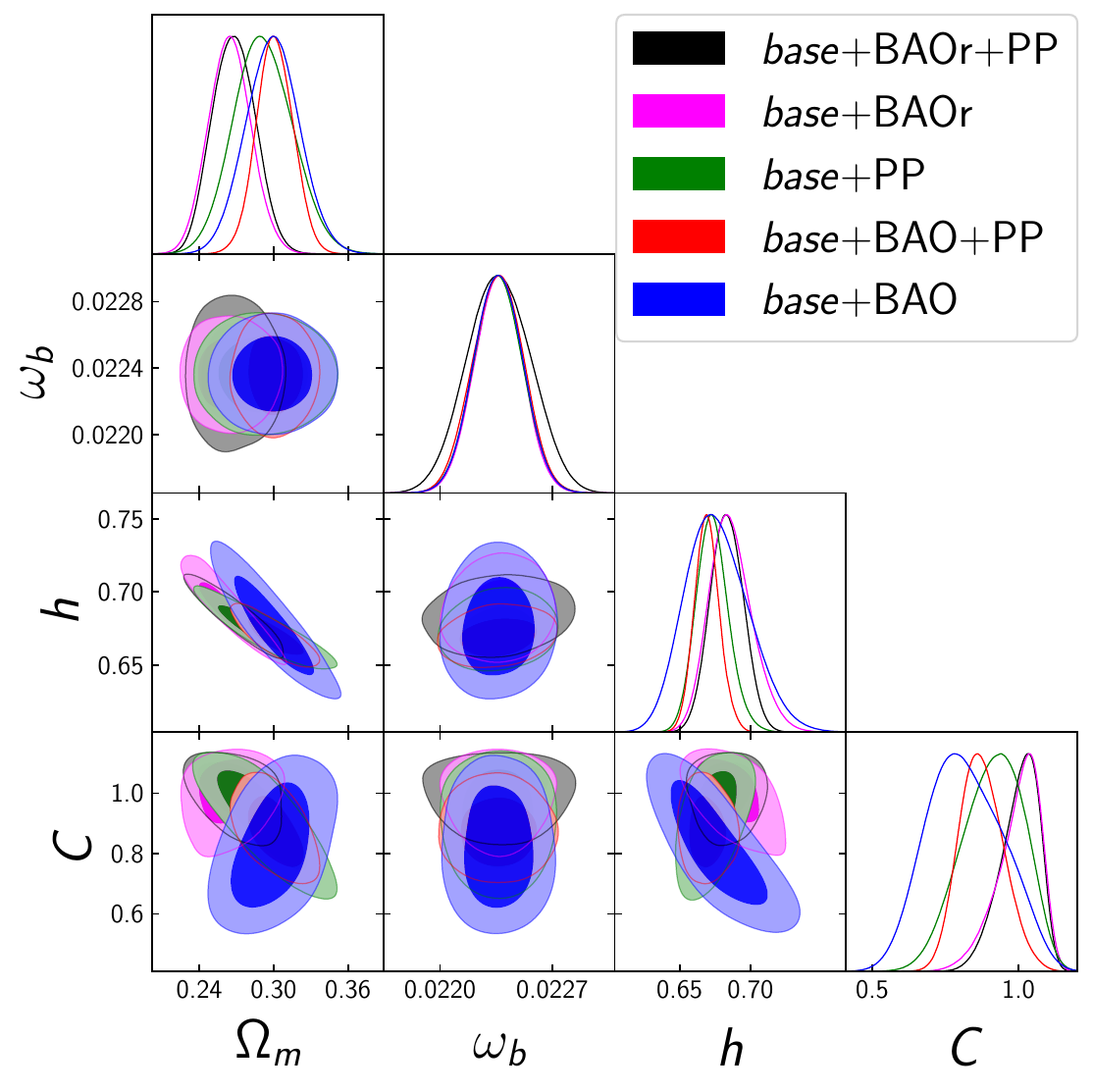}
\caption{As in Fig.~\ref{fig:lcdm}, but in light of the holographic dark energy model (see Sec.~\ref{subsec:hde}), therefore including also dimensionless quantity $C$, which controls the quantum zero-point energy density, among the parameters.}
\label{fig:hde}
\end{figure}

We now turn to the HDE model, with our results shown in the corner plot of Fig.~\ref{fig:hde}. Adopting the standard \textit{base}+\textit{BAO} dataset combination we infer $H_0=67.60^{+2.00}_{-2.50}\,{\text{km}}/{\text{s}}/{\text{Mpc}}$, which is further tightened to $H_0=(66.93 \pm 0.87)\,{\text{km}}/{\text{s}}/{\text{Mpc}}$ when considering the \textit{base}+\textit{BAO}+\textit{PP} dataset combination. This confirms that the HDE model is unable to address the Hubble tension when standard datasets are considered, in spite of its effective phantom nature. In fact, in both cases we infer $C \sim 0.85$, corresponding to an effective present-day equation of state $w_{\text{hde}} \sim -1.10$. For both the \textit{base}+\textit{BAO} and \textit{base}+\textit{BAO}+\textit{PP} dataset combinations, the improvement in fit over $\Lambda$CDM is not significant, with $\Delta \chi^2=-0.08$ and $-0.87$ respectively.

When considering rescaled BAO measurements, we observe only a mild increase in $H_0$. For the \textit{base}+\textit{BAOr} dataset combination we find $H_0=(68.62 \pm 1.53)\,{\text{km}}/{\text{s}}/{\text{Mpc}}$, while the matter density drops to $\Omega_m=0.27 \pm 0.02$ and the HDE parameter $C$ moves slightly into the quintessence-like regime, $C=1.01^{+0.04}_{-0.12}$. The quality of fit compared to the \textit{base}+\textit{BAO} dataset combination is essentially identical, with $\Delta \chi^2=+0.08$. There are two reasons behind the shift in $H_0$ being relatively mild. The first is that, as observed in the $w$CDM (Sec.~\ref{subsec:resultswcdm}) and CPL (Sec.~\ref{subsec:resultscpl}) cases, once the rescaling is applied, the fit to rescaled BAO data (especially for what concerns the MGS datapoint) is improved when the DE equation of state becomes more quintessence-like, limiting the extent to which $H_0$ can increase. In this case, the fact that $H_0$ does not decrease (unlike the $w$CDM and CPL cases) reflects the fact that the HDE expansion history is nevertheless phantom in the past. For conciseness, we do not display a residuals plot analogous to Fig.~\ref{fig:fit_wCDM}, since the observed features are similar.

The second reason why the increase in $H_0$ is mild has to do with the fact that the $\Omega_m$-$H_0$ correlation is much less steep compared to the other models considered: we have explicitly checked that this is consistent with existing constraints on HDE in the literature. This implies that a relatively small increase in $H_0$ needs to be compensated by a comparatively large decrease in $\Omega_m$. In this case, an increase of $\Delta H_0 \sim 1.0\,{\text{km}}/{\text{s}}/{\text{Mpc}}$ requires a decrease in $\Delta \Omega_m \sim 0.035$. Larger value of $H_0$ would require implausibly lower values of $\Omega_m$, which are (exponentially) disfavored by our generous $\Omega_m$ prior.

The addition of unanchored SNeIa data, within the \textit{base}+\textit{BAOr}+\textit{PP} dataset combination, leads to minor shifts in the inferred cosmological parameters, moving $C$ further into the quintessence-like regime, and actually slightly improves the fit by $\Delta \chi^2=-1.35$ compared to the \textit{base}+\textit{BAO}+\textit{PP} dataset combination. Importantly, considering the BAO-free benchmark \textit{base}+\textit{PP} dataset combination we find $H_0=67.30^{+1.00}_{-1.20}\,{\text{km}}/{\text{s}}/{\text{Mpc}}$ and $\Omega_m=0.30 \pm 0.02$. This shows that, regardless of any possible miscalibration in the BAO measurements, the model cannot address the Hubble tension.

\subsection{Sign-switching cosmological constant}
\label{subsec:resultslscdm}

\begin{figure}[!ht]
\centering
\includegraphics[width=0.9\linewidth]{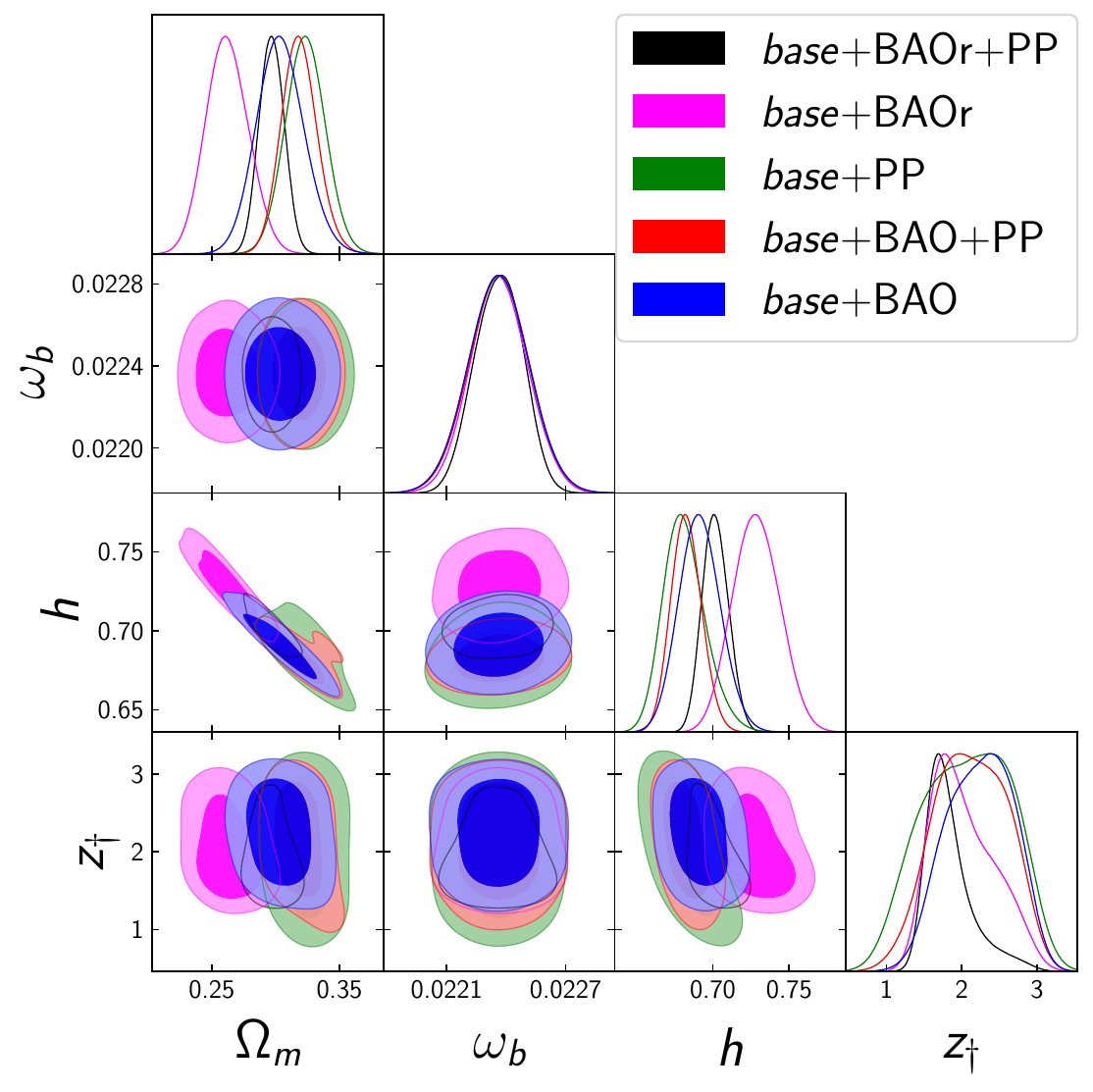}
\caption{As in Fig.~\ref{fig:lcdm}, but in light of the $\Lambda_s$CDM sign-switching cosmological constant model (see Sec.~\ref{subsec:lscdm}), therefore including also the AdS-to-dS transition redshift $z_{\dagger}$ among the parameters.}
\label{fig:lscdm}
\end{figure}

The $\Lambda_s$CDM model turns out to be the most interesting one in the context of our work, for reasons we will discuss shortly. Our results are shown in the corner plot of Fig.~\ref{fig:lscdm}. Starting from the standard \textit{base}+\textit{BAO} and \textit{base}+\textit{BAO}+\textit{PP} dataset combinations, we infer $H_0=(69.10 \pm 1.30)\,{\text{km}}/{\text{s}}/{\text{Mpc}}$ and $(68.27 \pm 0.99)\,{\text{km}}/{\text{s}}/{\text{Mpc}}$ respectively, both in good agreement with recent findings in the literature. Already at this point it is worth noting that the $\Lambda_s$CDM model is the one achieving the highest values of $H_0$ from these dataset combinations (excluding the PEDE model which nevertheless does so -- in the case of the \textit{base}+\textit{BAO} dataset combination -- at the expense of a significantly worse fit compared to $\Lambda$CDM, because the expansion history is essentially ``forced'' upon the data). In both cases the transition redshift is found at $z_{\dagger} \sim 2.1-2.2$, also in agreement with recent findings in the literature. For both the \textit{base}+\textit{BAO} and \textit{base}+\textit{BAO}+\textit{PP} dataset combinations, the improvement in fit over $\Lambda$CDM is extremely mild, with $\Delta \chi^2=-0.13$ and $-0.62$ respectively.

Adopting rescaled BAO measurements, and considering the \textit{base}+\textit{BAOr} dataset combination, we infer $H_0=(72.84 \pm 1.53)\,{\text{km}}/{\text{s}}/{\text{Mpc}}$, which is the highest figure achieved for this dataset combination (again excluding PEDE for the same reasons as previously). Nevertheless, this comes at the price of a significantly lower matter density parameter, $\Omega_m=0.26 \pm 0.02$, with the decrease required to fit the acoustic angular scale in the CMB. This is reflected in a degradation of the quality of fit by $\Delta \chi^2=+4.53$ compared to the \textit{base}+\textit{BAO} dataset combination (although, as noted earlier, this is driven to a significant extent by our $\Omega_m$ prior). When considering the \textit{base}+\textit{BAOr}+\textit{PP} dataset combination, the inclusion of unanchored SNeIa data brings up the matter density parameter to $\Omega_m=0.30 \pm 0.01$, lowering the Hubble constant down to $H_0=(70.05 \pm 1.03)\,{\text{km}}/{\text{s}}/{\text{Mpc}}$. While this is not sufficient for a complete resolution of the Hubble tension, it is worth noting that $\Lambda_s$CDM is the only model (among those we considered) where $H_0>70\,{\text{km}}/{\text{s}}/{\text{Mpc}}$ from this specific dataset combination. Nevertheless, this comes at the cost of a significant degradation in the fit compared to the \textit{base}+\textit{BAO}+\textit{PP} dataset combination, with $\Delta \chi^2=+10.88$, although the same considerations on the $\Omega_m$ prior hold.

Finally, for the BAO-free benchmark provided by the \textit{base}+\textit{PP} dataset combination, we find $H_0=68.10^{+1.20}_{-1.50}\,{\text{km}}/{\text{s}}/{\text{Mpc}}$. This is low enough to draw the conclusion that the model cannot address the Hubble tension regardless of any possible rescaling/miscalibration in the BAO measurements, although together with $w$CDM it represents the only instance where $H_0>68\,{\text{km}}/{\text{s}}/{\text{Mpc}}$ from this specific dataset combination.

Despite its overall inability to fully address the Hubble tension, the above discussion suggests that the $\Lambda_s$CDM model is nonetheless particularly interesting. Indeed, compared to other late-time modifications of $\Lambda$CDM, it is able to partially evade constraints on the unnormalized shape of the expansion history from unanchored SNeIa. The reason is that after the AdS-to-dS transition, for redshifts $z<z_{\dagger}$, the shape is by construction identical to $\Lambda$CDM. Therefore, if $z_{\dagger} \gtrsim 2$, shape constraints from unanchored SNeIa can be automatically satisfied. This is a significant advantage over other models. Of course a lower limit on $z_{\dagger}$ sets an upper limit on the value of $H_0$ which can be achieved, as our results show.

\subsection{Negative cosmological constant}
\label{subsec:resultsncc}

The final model we turn to is that of a nCC component with an evolving $w$CDM fluid on top. Phenomenologically speaking, at the level of shape of the expansion rate, this model is very similar to the $w$CDM model, with the differences that \textit{a)} the energy density of the total DE sector (nCC plus fluid with positive energy density) does not necessarily asymptote to zero at high redshifts, but can reach negative values, and \textit{b)} the effective behavior of the model can be phantom even if $w_x$ itself is in the quintessence-like regime, if $\Omega_{\Lambda}<0$ is sufficiently negative (or equivalently $\Omega_x>0$ sufficiently large). With these considerations in mind, it is more than reasonable to expect that our findings should resemble those of the $w$CDM model.

\begin{figure}[!ht]
\centering
\includegraphics[width=0.9\linewidth]{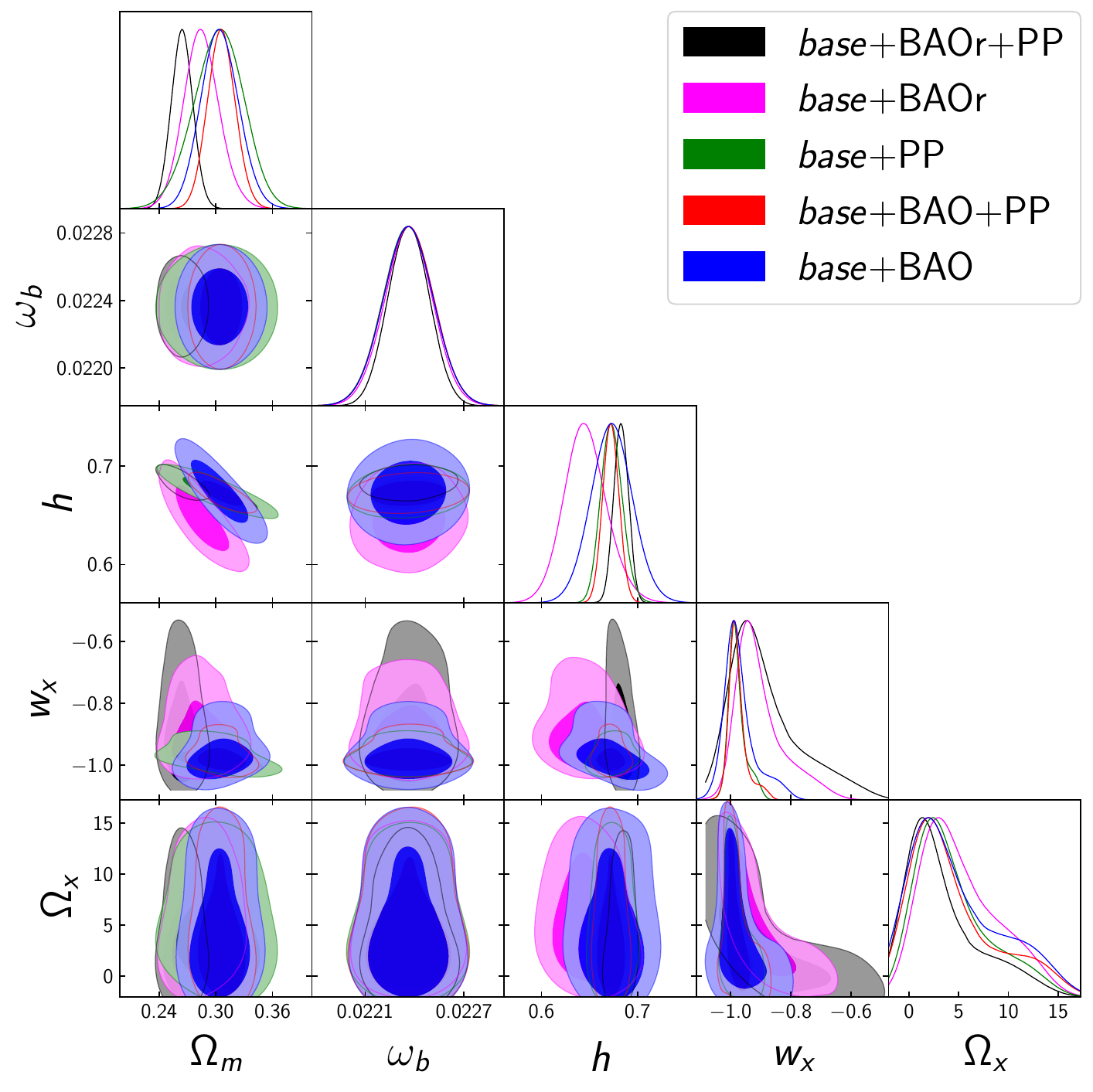}
\caption{As in Fig.~\ref{fig:lcdm}, but in light of the negative cosmological constant model (see Sec.~\ref{subsec:ncc}), therefore including also the equation of state and fractional density parameter of the evolving component with positive energy density, $w_x$ and $\Omega_x$ respectively, among the parameters.}
\label{fig:ncc}
\end{figure}

This expectation is borne out by our analysis, whose results are shown in Fig.~\ref{fig:ncc}. Firstly, adopting the standard \textit{base}+\textit{BAO} and \textit{base}+\textit{BAO}+\textit{PP} dataset combinations, we infer $H_0=(67.30 \pm 2.10)\,{\text{km}}/{\text{s}}/{\text{Mpc}}$ and $(67.23 \pm 0.84)\,{\text{km}}/{\text{s}}/{\text{Mpc}}$ respectively, in agreement with other findings in the literature~\cite{Visinelli:2019qqu,Sen:2021wld}, and confirming that the model does not offer a solution to the Hubble tension. In both cases, the equation of state of the fluid with positive energy density is inferred to be $w_x \sim -0.97$. For both the \textit{base}+\textit{BAO} and \textit{base}+\textit{BAO}+\textit{PP} dataset combinations, we find the improvement in fit over $\Lambda$CDM to be mild, with $\Delta \chi^2=-0.37$ and $-2.22$ respectively.

When we instead consider the \textit{base}+\textit{BAOr} dataset combination, we observe a decrease of the Hubble constant to $H_0=(64.58 \pm 2.23)\,{\text{km}}/{\text{s}}/{\text{Mpc}}$, comparable to the decrease observed in the CPL case. The fluid equation of state also shifts more into the quintessence-like regime, with $w_x \sim -0.90$. The reason behind these shifts is the same as that offered for the $w$CDM and CPL cases, which we will therefore not repeat here. We also note that the fit slightly improves by $\Delta \chi^2=-0.98$ with respect to the \textit{base}+\textit{BAO} dataset combination.

Similar features as for the $w$CDM case are observed when considering the \textit{base}+\textit{BAOr}+\textit{PP} and \textit{base}+\textit{PP} dataset combinations, where we infer $H_0=(68.29 \pm 0.86)\,{\text{km}}/{\text{s}}/{\text{Mpc}}$ and $67.30^{+1.00}_{-1.20}\,{\text{km}}/{\text{s}}/{\text{Mpc}}$ respectively. For the \textit{base}+\textit{BAOr}+\textit{PP} dataset combination, the quality of fit compared to the \textit{base}+\textit{BAO}+\textit{PP} one is slightly worse, by $\Delta \chi^2=+2.02$. We conclude that this model, just as all the other ones studied here, does not provide a loophole to the ``no-go theorem'' regardless of a possible rescaling/miscalibration of BAO measurements. We stress once more the importance of analyzing the full parameter space, rather than expecting the BAO rescaling to lead to a simple shift in $H_0$.

\section{Discussion}
\label{sec:discussion}

Based on our results, the ``no-go theorem'' excluding late-time solutions to the Hubble tension appears to be safe against one of its most frequently invoked loopholes. Even accounting for potential fiducial cosmology-dependent biases in the values of $\alpha_{\perp,\parallel}$ extracted from standard BAO pipelines, none of the post-recombination modifications we have tested is in fact able to address the Hubble tension once we include unanchored SNeIa and geometrical CMB information. Our analysis shows that the latter plays a particularly important role. In fact, rescaled BAO data on their own can tolerate relatively large increases in $H_0$ from $\Lambda$CDM-like values. However, the extremely long lever arm provided by the geometrical CMB datapoint requires that this increase in $H_0$ be achieved at the expense of significantly lower, and therefore highly implausible, values of $\Omega_m$: as we have shown, this conclusion holds even in the presence of the additional freedom provided by parameters controlling the DE sector (such as $w$). These considerations also highlight the remarkable level of internal consistency between the CMB and BAO within $\Lambda$CDM, where the CMB can determine $H_0$ on its own at a competitive level. This internal consistency is somewhat usually taken for granted, but our results show how it can easily be lost, which is perhaps in itself an indirect indication of the overall goodness of standard BAO data.

However, our work highlights the even more important role played by unanchored SNeIa data. These measurements provide very tight constraints on $E(z)$, i.e.\ the shape of the expansion history, irrespective of its normalization. This is because, unlike BAO, which yield sparse determinations of distance measurements at a limited set of effective redshifts, unanchored SNeIa densely sample the unnormalized Hubble diagram over $0.01 \lesssim z \lesssim 2.3$. This dense coverage across redshift, in addition to their overall greater statistical power compared to BAO, is sufficient to preclude large deviations from $\Lambda$CDM-like behaviour at late times. A perfectly legitimate and important question is indeed whether, if one sets aside BAO and their possible systematics aside altogether, late-time new physics which could solve the Hubble tension would then be allowed by other non-BAO datasets. From the perspective of a completely independent dataset, unanchored SNeIa on their own provide a clear negative answer. So far these measurements have played an arguably underappreciated role in the discussion on the Hubble tension and late-time modifications to $\Lambda$CDM which, for obvious reasons, has focused primarily on BAO. Our findings instead show clearly that unanchored SNeIa play a role at least as important as BAO in constraining late-time modifications, and in some respects an even more decisive one, as emphasized in Ref.~\cite{Zhou:2025kws}. Finally, it is worth stressing once more that our results are based on the more ``conservative'' \textit{PantheonPlus} sample of unanchored SNeIa -- although we have chosen not to adopt other samples for the reasons discussed in detail in Sec.~\ref{subsubsec:sneia}, doing so might alter our conclusions at least in part.

Another caveat worth emphasizing is our choice of adopting a deliberately simple redshift-independent rescaling of BAO data, to mimic a possible bias induced by incorrect assumptions on the fiducial cosmology. Although this is a reasonable toy setup, we stress that our results do depend on such an assumption. We can conceive a class of scenarios which could potentially bypass our results, and therefore the ``no-go theorem''. Let us imagine that both BAO and SNeIa are subject to independent, redshift-dependent systematics. Then the expansion history clearly cannot be $\Lambda$CDM, although the three ingredients (modified expansion history, redshift-dependent BAO bias, redshift-dependent SNeIa bias) could \textit{conspire to masquerade as $\Lambda$CDM} if the systematics are not taken into account. We deliberately use the term ``conspire'' to highlight how such a scenario, while mathematically possible, is likely physically implausible, as one would need to explain both the remarkable coincidence of why multiple independent effects combine together to mimic $\Lambda$CDM, as well as the origin of the required systematics in the first place. Variants of this idea have in fact been recently studied in the literature, for instance within scenarios involving redshift-dependent violations of the distance–duality relation~\cite{Tutusaus:2023cms,Teixeira:2025czm} (see also Refs.~\cite{DiValentino:2020evt,Gomez-Valent:2023uof}), and we certainly encourage further research in this direction. We view such (likely finely tuned) scenarios as the only conceivable loophole to the ``no-go theorem'' at this point, albeit one that is probably implausible from a physical standpoint.

We stress that the ``no-go theorem'' precludes post-recombination modifications from fully solving the Hubble tension \textit{on their own}: it does not imply that such modifications cannot play a role in the Hubble tension, nor is such a (erroneous) conclusion implied in any way by our results. In fact, late-time new physics may well play a subdominant role in partially alleviating the Hubble tension, as recently suggested in a number of works~\cite{Vagnozzi:2023nrq,Poulin:2024ken}. In this sense, our results obtained within the $\Lambda_s$CDM model appear somewhat interesting. Recall that, combining geometrical CMB information with rescaled BAO and unanchored SNeIa measurements, we found $H_0 \sim 70\,{\text{km}}/{\text{s}}/{\text{Mpc}}$, representing the largest figure obtained from this specific dataset combination. As explained earlier, the reason is that if the transition occurs at $z_{\dagger} \gtrsim 2$, the shape of the expansion history for $z<z_{\dagger}$ is by construction equivalent to $\Lambda$CDM, and can therefore naturally fit unanchored SNeIa data, while accommodating a larger value of $H_0$, as required by rescaled BAO data. This result has at least two immediate implications. Firstly, given its interesting features with regards to $E(z)$ for $z<z_{\dagger}$, the $\Lambda_s$CDM model could be an interesting candidate for the late-time part of a potential pre-plus-post-recombination solution to the Hubble tension. Moreover, in this context, the $\Lambda_s$CDM model could benefit from a hypothetical miscalibration of BAO data. We stress that obtaining a pre-plus-post-recombination model which can address the Hubble tension is far from a trivial task, see e.g.\ Refs.~\cite{Anchordoqui:2021gji,Clark:2021hlo,Wang:2022jpo,Anchordoqui:2022gmw,daCosta:2023mow,Yao:2023qve,Wang:2024dka,Baryakhtar:2024rky,Toda:2024ncp,Yashiki:2025loj} for examples in this sense. Among the attempts in this direction we find worth mentioning that of Ref.~\cite{Toda:2024ncp}, combining precisely the $\Lambda_s$CDM model with a model where the electron mass is varied, while allowing for a non-zero spatial curvature parameter $\Omega_K$. Nevertheless, for the reasons discussed above, we regard further attempts in this $\Lambda_s$CDM-plus-pre-recombination new physics direction as worthwhile, potentially including possible extensions such as allowing an asymmetric AdS–to-dS transition with different cosmological constants. We recall once more that the \textit{PantheonPlus} unanchored SNeIa sample is the most ``conservative'' one, and the promising features of the $\Lambda_s$CDM model in this context may emerge more prominently when tested against other SNeIa datasets.

Another point worth discussing is that, contrary to expectations, when rescaled BAO are considered without being combined with SNeIa, not all the models exhibit positive $\Delta H_0$ shifts. In fact, the $w$CDM, CPL, and nCC models exhibit precisely the opposite behavior: they appear to prefer a quintessence-like effective equation of state $w>-1$, accompanied by a decrease in $H_0$, i.e.\ $\Delta H_0<0$. This forces us somewhat to rethink the original na\"{i}ve interpretation of the BAO rescaling shown in Fig.~\ref{fig:diagram}, based on which we were expecting to always observe $\Delta H_0>0$. As discussed earlier the reason for such a behavior is that, when we rescale BAO data, models which feature additional cosmological parameters can accommodate at the same time smaller angular diameter distances (as implied by Fig.~\ref{fig:diagram}) \textit{and} lower values of $H_0$ compared to the vanilla $\Lambda$CDM model, thanks to their additional degrees of freedom.

Let us be more specific and assume that $r_d$ is calibrated whereas, following the notation of Eq.~(\ref{eq:lambda}), $\theta_d^{\cal R}>\theta_d$ as implied by Fig.~\ref{fig:diagram}. Now consider a rescaled distance measurement at $z_{\text{eff}}$, $D_M^{\cal R}(z_{\text{eff}})<D_M(z_{\text{eff}})$. Within the context of a given late-time modification to $\Lambda$CDM, whose additional cosmological parameters beyond the 6 $\Lambda$CDM ones we refer to collectively as $X$, we can express $D_M^{\cal R}(z_{\text{eff}})$ as follows:
\begin{equation}
\begin{split}
D_M^{\cal R}(z_{\text{eff}};\Omega_m,X) &= \frac{r_d}{\theta_d^{\cal R} \textcolor{red}{\uparrow}} = \left(\frac{1}{H_0}\int^{z_{eff}}_0\frac{dz'}{E(z';\Omega_m,X)}\right) \\
&\equiv \frac{1}{H_0}F(z_{\text{eff}};\Omega_m,X)\,\textcolor{red}{\big\downarrow}\,,
\label{eq:DM_discussion}
\end{split}
\end{equation}
where $F(z_{\text{eff}};\Omega_m,X)$ is defined as the integral on the right hand side of Eq.~(\ref{eq:DM_discussion}), whereas the red arrows pointing up\textcolor{red}{$\uparrow$} or down\textcolor{red}{$\downarrow$} indicate that a given quantity increases or decreases respectively, as a consequence of the rescaling. Multiple BAO measurements at different redshifts constrain both the normalization, $1/H_0$, and the function $F(z;\Omega_m,X)$ which controls the shape of the expansion history, and in $\Lambda$CDM depends only on $\Omega_m$. In principle, when rescaling BAO measurements, we would expect only the normalization, i.e.\ $1/H_0$, to be affected, since datapoints are rigidly shifted to lower values, each by the same amount. It is indeed interesting to notice that, if $r_d$ is calibrated independently from the CMB (e.g.\ using a BBN-informed prior on the physical baryon density $\omega_b$), such a ``rigid shift'' in the normalization is actually obtained, as shown in Fig.~\ref{fig:baobaor}. However, as discussed in Sec. \ref{sec:results}, the viability of such a scenario is precluded by the geometrical CMB information, responsible for the joint $\Omega_m$-$X$-$H_0$ correlations. This means that the normalization $1/H_0$ and the unnormalized function $F(z_{\text{eff}};\Omega_m,X)$ cannot be treated independently, because of the correlations induced by the geometrical degeneracy, once geometrical CMB data is included. This highlights the importance of analyzing the full parameter space of late-time modifications to $\Lambda$CDM rather than expecting the BAO rescaling to lead to an upwards shift in $H_0$.

In the case of models such as $\Lambda$CDM and PEDE, the shape of the expansion history is fixed once $\Omega_m$ is given. Despite the AdS-to-dS transition, similar considerations hold for the $\Lambda_s$CDM model as well, since before and after the transition $E(z)$ is exactly that of $\Lambda$CDM, up to a potentially different value of $\Omega_m$. These models attempt to fit the rescaled BAO data by increasing $H_0$ along the $\Omega_m$-$H_0$ degeneracy, therefore requiring a decrease in $\Omega_m$ which ultimately affects $E(z)$, explaining the worsened fit observed in Tab.~\ref{tab:parameters}. Models such as $w$CDM, CPL, and nCC have more degrees of freedom to fit the observations, and can potentially fit BAO and geometrical CMB data better than $\Lambda$CDM by varying \textit{all} their parameters ($H_0$, $\Omega_m$, and $X$) in a correlated manner. This explains why these models can potentially accommodate larger angles $\theta_d^{\cal R}>\theta_d$ without necessarily having a larger $H_0$, potentially improving the fit. At any rate, once unanchored SNeIa are included in the dataset, the strong constraints they impose on $E(z)$ at $z \lesssim 2$ strongly limits the freedom to vary $\Omega_m$ and $X$. This, when coupled with the geometrical CMB information, in turn strongly limits the extent to which $H_0$ can be increased, as explicitly observed in all the models.

We close by remarking once more the main result of our work: the ``no-go theorem'' appears to be safe against possible fiducial cosmology-related biases in BAO measurements. This does not imply that it is not of interest to explore these possible systematics, particularly in the context of fiducial cosmologies very far from $\Lambda$CDM. However, unless a contrived and admittedly rather unlikely scenario with multiple intertwined systematics in BAO and unanchored SNeIa are at play, these biases cannot be invoked as loopholes to the ``no-go theorem''. We stress that our conclusions concern cosmological scales at $z \gg 0.01$, and should not be extended to possible local or very low-redshift effects (e.g.\ those studied in Refs.~\cite{Desmond:2019ygn,Ding:2019mmw,Desmond:2020wep,Cai:2020tpy,Camarena:2021jlr,Cai:2021wgv,Marra:2021fvf,Krishnan:2021dyb,Perivolaropoulos:2021bds,Odintsov:2022eqm,Camarena:2022iae,Wojtak:2022bct,Odintsov:2022umu,Perivolaropoulos:2022khd,Oikonomou:2022tjm,Perivolaropoulos:2023iqj,Ruchika:2023ugh,Giani:2023aor,Mazurenko:2023sex,Huang:2024erq,Wojtak:2024mgg,Liu:2024vlt,Ruchika:2024ymt,Wojtak:2025kqo,Perivolaropoulos:2025gzo}, including a possible very late transition in the gravitational constant, see also Refs.~\cite{Perivolaropoulos:2024yxv,Huang:2024gfw}). These still represent a valid and interesting route towards solving the tension, or at least one which is not covered by the ``no-go theorem''.

\section{Conclusions}
\label{sec:conclusions}

The ``no-go theorem'' precluding post-recombination solutions to the Hubble tension is by now a cornerstone of the literature on the subject~\cite{Bernal:2016gxb,Addison:2017fdm,Lemos:2018smw,Aylor:2018drw,Schoneberg:2019wmt,Knox:2019rjx,Arendse:2019hev,Efstathiou:2021ocp,Cai:2021weh,Keeley:2022ojz}, and makes the case for a number of pre-recombination new physics models which are among the targets of next-generation cosmological surveys~\cite{SimonsObservatory:2018koc,SimonsObservatory:2019qwx}. Nevertheless, a concern occasionally raised somewhat loosely is that this conclusion relies on standard BAO analyses, which adopt a fiducial $\Lambda$CDM cosmology at several stages of the pipeline, potentially biasing the resulting measurements. In this work we play devil's advocate and assume, for the sake of argument, that BAO measurements are indeed affected in such a way that the inferred $H_0$ is biased low. We model this via a simple redshift-independent rescaling of BAO measurements, chosen to shift the inferred $H_0$ towards the SH0ES value. This deliberately simplified toy setup is meant to give late-time modifications the most favorable conditions to alleviate the tension. We then test the rescaled BAO dataset against a wide range of late-time dark energy models. We stress that our goal is not merely to restate the ``no-go theorem'', but to directly examine one of its most frequently invoked loopholes: namely, whether the adoption of a fiducial $\Lambda$CDM cosmology in BAO analyses could bias the measurements enough to cast doubts on its validity.

Our results demonstrate that, even under these deliberately favorable (but admittedly implausible) assumptions, late-time modifications to $\Lambda$CDM are unable to solve the Hubble tension, for two main reasons. Firstly, raising $H_0$ to SH0ES-like values while preserving consistency with the CMB acoustic scale, treated here as a purely geometrical constraint, requires lowering $\Omega_m$ to implausibly low values. This explains why, once combined with geometrical CMB information, rescaled BAO data only produce intermediate shifts, falling well short of SH0ES-like values for $H_0$. The situation becomes even clearer once unanchored SNeIa measurements are included, as these tightly constrain the unnormalized shape of the late-time expansion history, $E(z)$ at $z \lesssim 2$. This leaves little room for the deviations from a $\Lambda$CDM-like shape required for a post-recombination solution to the Hubble tension, and therefore for large deviations in $\Omega_m$ and other parameters controlling the dark energy sector (e.g.\ $w$). In fact, the \textit{PantheonPlus} unanchored SNeIa sample on its own drives $H_0$ back towards $\Lambda$CDM-like values, irrespective of whether (standard or rescaled) BAO data are included or not, confirming the key role of unanchored SNeIa recently emphasized by Ref.~\cite{Zhou:2025kws}. Among the models considered, $\Lambda_s$CDM emerges as the least unpromising one, thanks to the fact that after the AdS-to-dS transition at $z_{\dagger}$ the shape of its expansion history is exactly that of $\Lambda$CDM: nevertheless, even this case falls well short of fully bridging the gap to SH0ES-like values for $H_0$.

A clear message from our work is that the ``no-go theorem'' preventing late-time solutions to the Hubble tension holds up against one of its most frequently raised challenges. To be absolutely clear and blunt, our work shows that appeals to such biases are not a \textit{get out of jail free card} for resurrecting late-time models in the context of the Hubble tension, and invoking them amounts to little more than handwavy excuses which do not account for unanchored SNeIa. Of course, this does not imply that late-time new physics is irrelevant to the problem: such modifications may still play a supporting role (as discussed e.g.\ in Refs.~\cite{Vagnozzi:2023nrq,Poulin:2024ken}), but cannot on their own fully address the tension. In closing, it is worth emphasizing a few caveats to our current analysis, which also point to directions worthy of further investigation. First and foremost, as emphasized in Sec.~\ref{sec:discussion}, it is in principle possible that multiple more complex redshift-dependent effects/systematics, entering simultaneously in BAO and unanchored SNeIa, could alter the picture beyond our deliberately simple toy model. It might be interesting to carry out a data-driven joint reconstruction of these corrections, alongside the required modifications to the late-time expansion history, as a proof-of-principle for a potential (mathematically possible but perhaps physically implausible) late-time solution to the Hubble tension. Moreover, it is worth extending the present work on the data side, on the one hand by adopting the full CMB likelihood (which in turn would allow us to reliably test other models, such as interacting dark energy), and on the other hand by exploring the impact of additional BAO and unanchored SNeIa datasets (e.g.\ from DESI, Union3, and DESY5). Last but most certainly not least, a dedicated re-analysis of the standard BAO pipeline itself, assuming fiducial cosmologies extremely far from $\Lambda$CDM, would provide the most direct robustness test for these measurements. We defer these points to follow-up work, which we view as important next steps in further consolidating the case for the Hubble tension ``no-go theorem'', and pre-recombination new physics.

\begin{acknowledgments}
\noindent We are grateful to \"{O}zg\"{u}r Akarsu and Eleonora Di Valentino for several discussions. D.P.\ and S.V.\ acknowledge support from the Istituto Nazionale di Fisica Nucleare (INFN) through the Commissione Scientifica Nazionale 4 (CSN4) Iniziativa Specifica ``Quantum Fields in Gravity, Cosmology and Black Holes'' (FLAG). S.V.\ acknowledges support from the University of Trento and the Provincia Autonoma di Trento (PAT, Autonomous Province of Trento) through the UniTrento Internal Call for Research 2023 grant ``Searching for Dark Energy off the beaten track'' (DARKTRACK, grant agreement no.\ E63C22000500003). L.A.E.\ acknowledges partial financial support from the T\"{u}rkiye Bilimsel ve Teknolojik Ara\c{s}t{\i}rma Kurumu (T\"{U}B\.{I}TAK, Scientific and Technological Research Council of T\"{u}rkiye) through grant no.\ 124N627. V.M.\ acknowledges partial financial support from the Conselho Nacional de Desenvolvimento Cient\`{i}fico e Tecnol\'{o}gico (CNPq, Brazilian National Council for Scientific and Technological Development) and the Funda\c{c}\~{a}o de Amparo \`{a} Pesquisa e Inova\c{c}\~{a}o do Esp\'{i}rito Santo (FAPES, Esp\'{i}rito Santo Research and Innovation Support Foundation). This publication is based upon work from the COST Action CA21136 ``Addressing observational tensions in cosmology with systematics and fundamental physics'' (CosmoVerse), supported by COST (European Cooperation in Science and Technology).
\end{acknowledgments}

\appendix

\section{Impact of DESI data}
\label{sec:appendix}

\begin{table*}[!t]
\centering
\renewcommand{\arraystretch}{1.2}
\scalebox{0.8}{
\begin{tabular}{?c?ccccc?}
\thickhline
\multicolumn{1}{?c?}{\textbf{Model / parameters}} & \multicolumn{5}{c?}{\textbf{Dataset: \textit{base}+}} \\
\thickhline
$\mathbf{w}$\textbf{CDM} & \textbf{\textit{BAO}} & \textbf{\textit{DESI}}+\textbf{\textit{PP}} & \textbf{\textit{PP}} & \textbf{\textit{DESIr}} & \textbf{\textit{DESIr}}+\textbf{\textit{PP}} \\ \hline
$w$ & $-0.95\pm 0.05$ & $-0.93\pm 0.03$ & $-0.93\pm 0.08$ & $-0.70 \pm 0.04$ & $-0.79 \pm 0.02$ \\ \hline
$H_0 \left [ {\text{km}}/{\text{s}}/{\text{Mpc}} \right ] $ & $67.80 \!\pm\! 1.10$ (3.46$\sigma$, 1.6\%) & $67.39 \!\pm\! 0.62$ (4.67$\sigma$, 1\%) & $68.20 \!\pm\! 1.10$ (3.2$\sigma$, 1.6\%) & $64.30 \!\pm\! 0.98$ (6.12$\sigma$, 1.5\%) & $66.75 \!\pm\! 0.61$ (5.22$\sigma$, 0.9\%) \\
$\Omega_m$ & $0.30\pm 0.01$ & $0.30\pm 0.01$ & $0.30\pm 0.03$ & $0.28 \pm 0.01$ & $0.27 \pm 0.06$ \\
$\omega_b$ & $0.02236\pm 0.00015$ & $0.02236\pm 0.00015$ & $0.02237\pm 0.00014$ & $0.02237\pm 0.00015$ & $0.02237\pm 0.00015$ \\ \hline
$\chi^2_{\min}$ & $9.22$ & $1412.29$ & $1402.87$ & $13.49$ & $1427.53$ \\
\thickhline
\end{tabular}}
\vspace{1cm}
\caption{As in Tab.~\ref{tab:parameters}, but focusing only on the $w$CDM model and replacing standard and rescaled SDSS BAO measurements with their DESI counterparts.}
\label{tab:parametersdesi}
\end{table*}

\begin{figure}[!ht]
\centering
\includegraphics[width=0.9\linewidth]{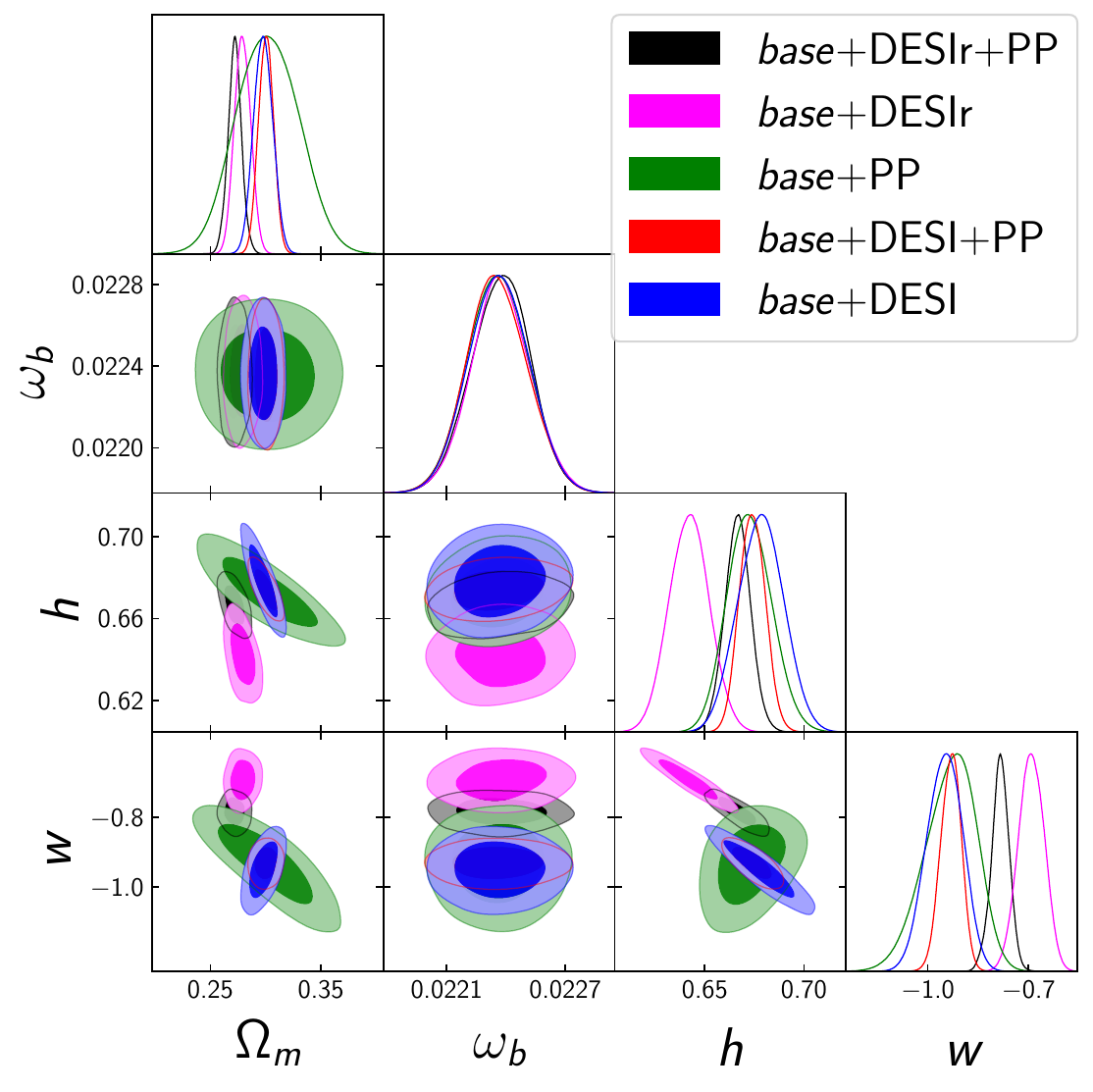}
\caption{As in Fig.~\ref{fig:wcdm}, but in light of standard and rescaled DESI BAO measurements.}
\label{fig:wcdmdesi}
\end{figure}

\begin{figure}
\centering
\includegraphics[width=0.9\linewidth]{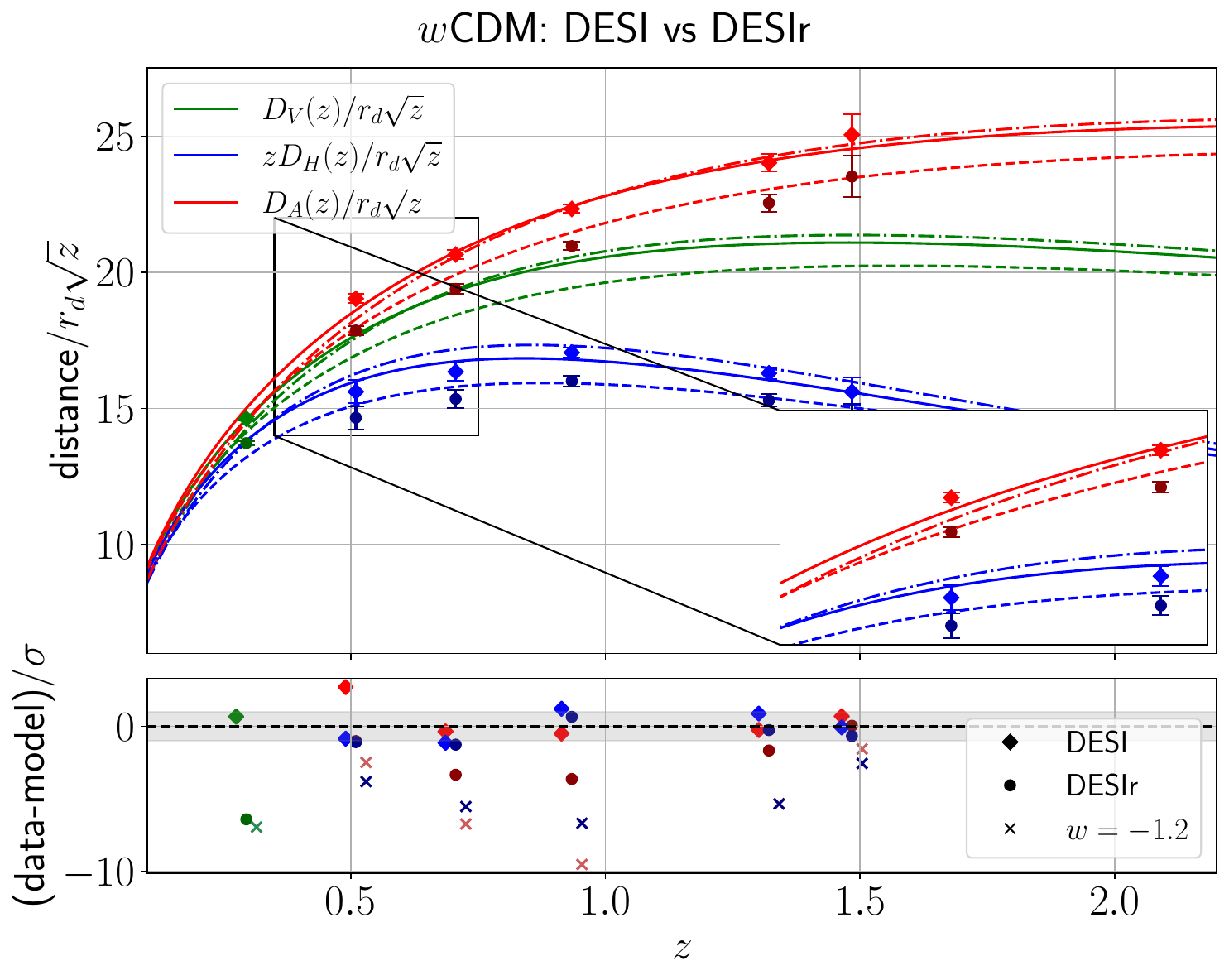}
\caption{As in Fig.~\ref{fig:fit_wCDM}, but in light of standard and rescaled DESI BAO measurements.}
\label{fig:wcdmdesifit}
\end{figure}

Here we provide a brief assessment of the impact of DESI BAO measurements on our results. We recall that the rationale behind our choice of using SDSS rather than DESI BAO data is discussed in detail in Sec.~\ref{subsubsec:bao}. Since our aim is simply that of providing a minimal check, we restrict the analysis to the $w$CDM model.

We proceed as in our standard analysis, considering as baseline BAO dataset the DESI DR2 BAO measurements: we refer the reader to Ref.~\cite{DESI:2025zgx} for a detailed discussion of these measurements, which here we refer to as \textit{\textbf{DESI}}. As we did with the SDSS measurements, we then rescale these as indicated in Eqs.~(\ref{eq:lambda}), setting the scaling parameter $\lambda$ appearing in Eqs.~(\ref{eq:lambda}) to $\lambda=73.0/68.5 \approx 1.06$, as discussed in Sec.~\ref{sec:bao}: the resulting dataset is referred to as \textit{\textbf{DESIr}}. We then consider the same 5 classes of dataset combinations discussed in Sec.~\ref{subsubsec:combinations}, with \textit{BAO} and \textit{BAOr} replaced by \textit{DESI} and \textit{DESIr} respectively.

The resulting corner plot for $\Omega_m$, $\omega_b$, $h$, and $w$ is given in Fig.~\ref{fig:wcdmdesi}, with constraints on these parameters reported in Tab.~\ref{tab:parametersdesi}. For the standard \textit{base}+\textit{DESI} dataset combination we infer $H_0=(67.80 \pm 1.10)\,{\text{km}}/{\text{s}}/{\text{Mpc}}$ and $w=-0.95 \pm 0.05$. These figures change to $(67.39 \pm 0.62)\,{\text{km}}/{\text{s}}/{\text{Mpc}}$ and $-0.93 \pm 0.03$ respectively when instead considering the \textit{base}+\textit{DESI}+\textit{PP} dataset combination. For the BAO-free benchmark provided by the \textit{base}+\textit{PP} dataset combination we find $H_0=(68.20 \pm 1.10)\,{\text{km}}/{\text{s}}/{\text{Mpc}}$ and $w=-0.93 \pm 0.08$, as reported previously in Sec.~\ref{subsec:resultswcdm}. All of these figures confirm that, when considering standard DESI BAO measurements, the $w$CDM model is unable to alleviate the Hubble tension, in agreement with previous findings.

We now consider the rescaled DESI measurements. From the \textit{base}+\textit{DESIr} dataset combination we find $H_0=(64.30 \pm 0.98)\,{\text{km}}/{\text{s}}/{\text{Mpc}}$ and $w=-0.66 \pm 0.04$. This shift towards more quintessence-like values for $w$, and correspondingly lower values of $H_0$ given the direction of the degeneracy between the two parameters, is consistent with the same shift observed with SDSS BAO data and discussed in Sec.~\ref{subsec:resultswcdm}, although more pronounced. The explanation is precisely the same one offered earlier, i.e.\ that once the rescaling is applied, certain features of the rescaled BAO data may be better accommodated by a lower $H_0$ together with a quintessence-like value of $w$. Nevertheless, these shifts come at the cost of a worse quality of fit, with $\Delta \chi^2_{\min}=+4.25$ when compared to the \textit{base}+\textit{DESI} dataset combination.

This point is made clearer by the residuals in Fig.~\ref{fig:wcdmdesifit}. Again, it is useful to compare the circle versus diamond residuals. We see that, for most of the datapoints, the two residuals perform equally well. Nevertheless, the fit of the low-$H_0$/high-$w$ cosmology is clearly better for what concerns the LRG1 $D_A/r_d$ measurement at $z_{\text{eff}}=0.51$, although this comes at the cost of a significantly worse quality of fit for what concerns the BGS $D_V/r_d$ measurement at $z_{\text{eff}}=0.295$. We note that these shifts are more evident in the case of (standard or rescaled) DESI BAO data compared to their (standard or rescaled) SDSS counterparts since the ``shape'' of the expansion history suggested by DESI data is clearly different from $\Lambda$CDM. This highlights once again how the rescaling of BAO measurements should not be interpreted as merely shifting $H_0$ (which only controls the amplitude but not the shape of the expansion history).

Finally, considering the \textit{base}+\textit{DESIr}+\textit{PP} dataset combination, we find $H_0=(66.75 \pm 0.61)\,{\text{km}}/{\text{s}}/{\text{Mpc}}$ and $w=-0.79 \pm 0.02$. The explanation for this upwards shift in $H_0$ is analogous to that offered in the case of SDSS BAO measurements in Sec.~\ref{subsec:resultswcdm}. In particular, it reflects the fact that unanchored SNeIa favor milder deviations from $\Lambda$CDM, with $w \sim -0.9$, which leads to an overall intermediate outcome, with \textit{DESIr} shifting $w$ upwards and $H_0$ downwards, and \textit{PP} driving the constraints slightly towards the opposite direction. However, this comes once again at the cost of a significantly worse quality of fit compared to the \textit{base}+\textit{DESI}+\textit{PP} dataset combination, with $\Delta \chi^2=+15.24$, reflecting an internal tension between rescaled BAO measurements and the other cosmological datasets. These results emphasize once more the extremely important role of unanchored SNeIa data in stabilizing parameter constraints when moving beyond $\Lambda$CDM, thanks to the strong constraints imposed on $E(z)$. Overall, these results indicate that even taking into account rescaled DESI measurements does not offer a solution to the Hubble tension within the $w$CDM model, confirming the trend observed with SDSS BAO measurements. Although, as has been the case here, we expect the interpretation to be somewhat obscured by the ``DESI tension'', we expect a similar trend to hold for all the other models studied earlier.

\bibliography{rescaledbao}

\end{document}